\documentclass[useAMS,usenatbib]{mn2e}
\usepackage{graphicx}
\usepackage{caption}
\usepackage{subfigure}
\usepackage{soul}
\usepackage{fancybox}
\usepackage{amsfonts}
\usepackage{pifont}
\usepackage{caption}
\usepackage{colortbl}
\usepackage{hyperref}
\newcommand \ion[2]{#1$\;${\scshape{#2}}}
\usepackage{amssymb}
\usepackage{lscape}
\usepackage{enumerate}
\usepackage{threeparttable}
\usepackage{afterpage}
\usepackage{longtable}

\usepackage{color}
\usepackage{transparent}
\usepackage{array}
\newcommand{\head}[2]{\multicolumn{1}{>{\centering\arraybackslash}p{#1}}{#2}}

\definecolor{beaublue}{rgb}{0.74, 0.83, 0.9}
\definecolor{bisque}{rgb}{1.0, 0.89, 0.77}
\definecolor{bubblegum}{rgb}{0.99, 0.76, 0.8}
\definecolor{gray}{rgb}{0.75, 0.75, 0.75}

\title[Methane and Ammonia opacity in late T dwarfs]{Methane and Ammonia in the near-infrared spectra of late T dwarfs}
\author[J. I. Canty et al.]
{J. I. Canty$^{1}$\thanks{E-mail:j.canty2@herts.ac.uk}
P.W. Lucas$^{1}$  Sergei N. Yurchenko$^{2}$
Jonathan Tennyson$^{2}$ 
\newauthor S. K. Leggett$^{3}$ C. G. Tinney$^{4}$
H. R. A. Jones$^{1}$ Ben Burningham$^{1}$ \newauthor D. J. Pinfield$^{1}$ 
R. L. Smart$^{5}$\\ 
$^{1}$Centre for Astrophysics Research, University of Hertfordshire, College Lane, Hatfield AL10 9AB, UK\\
$^{2}$Department of Physics and Astronomy, University College London, London WC1E 6BT, UK\\
$^{3}$Gemini Observatory, Northern Operations Center, 670 North A'ohoku Place, Hilo, HI 96720, USA\\
$^{4}$Department of Astrophysics, School of Physics, University of New South Wales, Sydney, NSW 2052, Australia\\
$^{5}$Istituto Nazionale di Astrofisica, Osservatorio Astrofisico di Torino, Strada Osservatorio 20, I-10025 Pino Torinese, Italy}

\begin{document}

\pagerange{\pageref{firstpage}--\pageref{lastpage}} \pubyear{2014}

\maketitle

\label{firstpage}

\begin{abstract}
Analysis of T dwarfs using model atmospheres has been hampered by the absence of reliable line lists 
for methane and ammonia. Newly computed high temperature line lists for both of these important molecules 
are now available, so it is timely to investigate the appearance of the various absorption features in T dwarfs 
in order to better understand their atmospheres and validate the new line lists. We present high quality 
R$\sim$5000 Gemini/NIFS 1.0-2.4 $\mu$m spectra of the T8 standard 2MASS~0415-0935 and the T9 
standard UGPS~0722-0540. We use these spectra to identify numerous methane and ammonia features 
not previously seen and we discuss the implications for our understanding of T dwarf atmospheres.
Among our results, we find that ammonia is the dominant opacity source between 
$\sim$1.233-1.266~$\mu$m in UGPS~0722~0540, and we tentatively identify several absorption features in this 
wavelength range in the T9's spectrum which may be due entirely to ammonia opacity. Our results also suggest that 
water rather than methane is the dominant opacity source 
in the red half of the $J$-band of the T8 dwarf. Water appears to be the main absorber in this wavelength region in the 
T9 dwarf until $\sim$1.31 $\mu$m, when methane starts to dominate.
\end{abstract}

\begin{keywords}
stars: atmospheres $-$ stars: low-mass $-$ stars: brown dwarfs.
\end{keywords}

\section{Introduction}
\label{sec:intro}
Amongst the first brown dwarf discoveries was a T dwarf~\citep{oppenheimer95}. Since then, a large number of 
T dwarfs have been discovered in the local field using surveys such as 2MASS~\citep{skrutskie06}, 
SDSS~\citep{york00}, UKIDSS~\citep{lawrence07}, and now WISE~\citep{wright10}.  T dwarfs provide 
a new arena for studying atmospheric physics at T$_{eff}$ cooler than stars but warmer than gas giant 
planets such as Jupiter. WISE has recently discovered the first Y dwarfs, objects with T$_{eff}$ $\sim$275-450~K,
and has tentatively identified ammonia opacity on the blue wing of the $H$-band flux peak
(\citealt{cushing11}, hereafter C11). However, Y dwarfs are too faint for 
medium resolution spectroscopy of individual narrow molecular absorption features with instruments 
such as Gemini/NIFS~\citep{mcgregor03}. We anticipate that the James Webb Space Telescope 
will provide an ideal platform for spectroscopy of Y dwarfs at a higher  
resolution than is currently possible.

The range of T$_{eff}$ of late T dwarfs is uncertain, owing to the complicated atmospheric microphysics 
of these very cool objects and, until recently, the absence of good methane and ammonia line lists. 
Distance and temperature estimates based on model fits to near-infrared spectra are often found to be 
incorrect, in some instances by a factor of 2 in distance~\citep{liu11}. Improved model atmospheres are 
essential in order to derive reliable temperatures, luminosities, and gravities for cold brown dwarfs, without 
recourse to time-consuming parallax measurements. This is a basic requirement in order to determine the 
substellar mass function in the local field, and to enable brown dwarfs to inform our understanding of the 
many warm gas giant exoplanets, which are hard to study in any detail.

In low resolution late T dwarf spectra, only broad and overlapping absorption bands of water and methane 
are typically observed at 1.0-2.4~$\mu$m. Medium resolution spectroscopy with instruments such as NIFS 
resolves the bands into narrow features produced by blends of individual transition lines
and detects other features such as blends of numerous weak ammonia lines across the near-infrared, 
as well as the temperature sensitive \ion{K}{i} doublet at 1.18~$\mu$m and 1.24~$\mu$m. \citet{bochanski11}, 
hereafter B11, have demonstrated this with the first medium resolution spectrum of a single object (the T9 
standard UGPS~0722-0540, hereafter UGPS~0722), obtained with Magellan/FIRE~\citep{simcoe10} 
during a commissioning run. Our data for UGPS~0722 agree closely with the Magellan/FIRE 
data of B11 (see Figures \ref{fig:jlong},  \ref{fig:hlong},  and  \ref{fig:klong}). 
We note that some molecular features have been resolved in the T6.5 dwarf Gliese 229B at lower resolutions 
(R$\sim$2400$-$2800) \citep{saumon00}.

Such data offer the opportunity to directly test the details of previously inadequate near-infrared model spectra. 
\citet{saumon12}, hereafter S12, have recently published improved model atmospheres that incorporate a 
new high-temperature, synthetic ammonia line list (BYTe)~\citep{yurchenko11}, and new calculations of 
collision-induced absorption of molecular hydrogen (H$_{2}$~CIA)~\citep{richard12}. S12 also had access 
to  an improved, though still incomplete, treatment of methane opacity~\citep{freedman08}. 

The ExoMol project \citep{tennyson12} provides a database (www.exomol.com)
of high temperature line lists for astronomical use.  The database now includes a new high-temperature, 
synthetic methane line list (10to10)~\citep{yurchenko14}.  \citet{yurchenko14} used the 10to10 line list to 
re-identify a number of methane absorption features first identified in the $R\sim$1200 SpeX~\citep{rayner03} 
$H$-band spectrum of the T4.5 brown dwarf 2MASS J0559-1404~\citep{cushing05}. The 10to10 list has 
also been incorporated into the VSTAR model atmosphere code~\citep{bailey12} and used to model the 
$H$- and $K$-band spectrum of 2MASS J0559-1404~\citep{yurchenko14b}. The model spectrum is a 
significantly better fit to the brown dwarf spectrum than an earlier VSTAR model using a methane line list 
computed with the STDS~\citep{wenger98} software. However, medium resolution spectra are needed to 
resolve individual features. It is therefore timely to provide a high quality set of medium resolution spectra 
to compare with the new generation of models.

In this paper, we examine the atmospheres of two late T dwarf standards by comparing absorption 
features in the near-infrared spectra of these objects with synthetic spectra and with absorption cross-sections 
of the most important gas-phase opacity sources at these wavelengths; H$_{2}$O, CH$_{4}$, and NH$_{3}$. 
We also conduct an analysis of the rotational-vibrational (ro-vibrational) transition lines responsible for the 
methane and ammonia absorption features in these objects' near-infrared spectra. 

The structure of this paper is as follows. In Section \ref{sec:obs} we describe how the T dwarfs were observed 
and how we extracted their spectra. Section \ref{sec:methane} contains the results of our analysis of the 
CH$_{4}$ absorption features in the near-infrared spectra of the T dwarfs. Section \ref{sec:ammonia} 
analyses the NH$_{3}$ absorption features in these objects. We discuss our results in Section \ref{sec:discussion}. 
Our conclusions are made in Section \ref{sec:conclusions}.

\section{Observations \& Data Reduction}
\label{sec:obs}
Observations were made with the 8m Gemini Telescope at Gemini North on Mauna Kea, Hawaii, using the 
near-infrared integral field spectrograph (NIFS)~\citep{mcgregor03}.  Observations were made in the $Z$, 
$J$, $H$, and $K$ passbands, (the $Z$ passband includes the $Y$-band flux peak in brown dwarfs). The 
resolution in each passband was R$\sim$5000.
The T8 standard 2MASS~0415-0935 (hereafter 2MASS~0415) was observed over four nights between 
2010 September 30 and 2010 October 12.
Observations of UGPS~0722 were made over seven nights between 2010 October 17 and 2012 October 
29.
Observations in the $H$ and $K$ passbands for UGPS~0722 were made using the Gemini ALTAIR 
adaptive optics system to improve the S/N ratio. Details of the observations and
the physical properties of the two T dwarfs are shown in Tables \ref{tab:observations} and \ref{tab:properties}
respectively.

\begin{table*}
\centering
\begin{minipage}{140mm}
\caption{T Dwarf Observations}  
\label{tab:observations}
\begin{tabular}{cllcccccc} 
\hline\hline
\noalign{\vskip 2mm} 
Object & Observation period  & Grating/centred at & Wavelength Range\\ [0.5ex] 
\hline
\noalign{\vskip 2mm} 
2MASS~0415 &  2010 October 4                                        & $Z$-band/1.05~$\mu$m   & 0.95-1.15~$\mu$m\\ 
                     &  2010 October 12                                      & $J$-band/1.25~$\mu$m    & 1.15-1.35~$\mu$m\\      
                     &  2010 October 1                                        & $H$-band/1.6~$\mu$m     & 1.45-1.75~$\mu$m\\      
                     &  2010 September 30                                  & $K$-band/2.14~$\mu$m    & 1.95-2.37~$\mu$m\\ 
\noalign{\vskip 2mm}                      
\hline
\noalign{\vskip 2mm} 
UGPS~0722 &  2012 September 27 - 2012 September 30   & $Z$-band/1.05~$\mu$m   & 0.95-1.15~$\mu$m\\ 
                   &  2012 October 1 - 2012 October 29             & $J$-band/1.25~$\mu$m    & 1.15-1.35~$\mu$m\\      
                   &  2011 December 11                                      & $H$-band/1.6~$\mu$m     & 1.45-1.75~$\mu$m\\      
                   &  2010 October 17                                         & $K$-band/2.14~$\mu$m    & 1.95-2.37~$\mu$m\\                      
\hline
\end{tabular}
\end{minipage}
\end{table*}

\begin{table*}
\centering
\begin{minipage}{140mm}
\caption{T Dwarf Physical Properties}  
\label{tab:properties}
\begin{tabular}{lllcccccc} 
\hline\hline
\noalign{\vskip 2mm} 
&   UGPS~0722 & 2MASS~0415\\ [0.5ex] 
\hline
\noalign{\vskip 2mm} 
Spectral Type         &     T9~$^{(1)}$                                        &       T8~$^{(2)}$                                \\ [0.5ex]
T$_{eff}$                &      500~K$^{(3),(8)}$                              &       750~K~$^{(4)}$                          \\ [0.5ex]
v sin i                     &      40$\pm$10 kms$^{-1}$~$^{(5)}$       &        33.5  kms$^{-1}$~$^{(6)}$         \\[0.5ex]
RV                         &      46.9$\pm$2.5 kms$^{-1}$~$^{(5)}$    &       49.6~kms$^{-1}$~$^{(6)}$         \\[0.5ex]
Age                        &     1-5~Gyr $^{(5)}$                                 &       1-10~Gyr $^{(2)}$  \\[0.5ex]
log~$g$                 &     4.39-4.90$^{(7)}$                                &       4.64-5.15$^{(7)}$                        \\[0.5ex]
Mass                     &     10.7-25.8~M$_{J}$~$^{(7)}$               &       17.4-40.1~M$_{J}$~$^{(7)}$       \\[0.5ex]
Distance                &     4.12$\pm$0.04 pc$^{(8)}$                  &        5.71$\pm$0.05  pc$^{(9)}$         \\[0.5ex]

$^{(1)}$C11           
$^{(2)}$ \citet{burgasser02}       
$^{(3)}$ \citet{lucas10}\\   
$^{(4)}$ \citet{saumon07} 
$^{(5)}$ B11
$^{(6)}$ \citet{zapatero07}\\
$^{(7)}$ \citet{dupuy13}
$^{(8)}$ \citet{leggett12}\\
$^{(9)}$ \citet{dupuy12}\\

\end{tabular}
\end{minipage}
\end{table*}

Observations were made in an ABBA pattern to facilitate the removal of the sky background and dark 
current. Raw data in each waveband were reduced using the {\sc gemini/nifs} package within {\sc iraf}. 
The reduction was made in three steps.
\begin{enumerate} [(1)]
\item a baseline calibration to produce a reference file to determine the shift between the position of the 
data and the location of the image slices on the detector, a flat field file, a flat bad pixel mask file, a 
wavelength referenced arc file, and a file to correct for spatial distortion of the data;
\item  a telluric calibration reduction to produce a 1D spectrum of the standard star to be used for 
telluric calibration of the science data;
\item a science data reduction to produce a 3D data cube which has been sky subtracted, flat fielded, 
cleaned of bad pixels, and telluric corrected.
\end{enumerate}
The first two steps in the reduction process were completed by editing processing scripts supplied by 
the Gemini Observatory. The science reduction also largely followed a Gemini script. However, 
several additional steps were required to complete the reduction. In particular, any hydrogen absorption 
lines in the spectrum of the standard star chosen for the telluric calibration of each science spectrum 
had to be removed, the modified spectrum then being divided by the star's blackbody spectrum and 
normalised before being divided into the extracted 1D science object spectrum to correct the latter 
for telluric absorption features.

The spectra were extracted using an aperture size of 1.5 times the full width at half maximum (FWHM) 
of each object, as determined from the dispersed images.

Observations in the near-infrared are susceptible to contamination by telluric OH sky lines. For the 
fainter objects observed here, the flux in the sky lines often varied sufficiently during the exposures that 
the lines were poorly subtracted in the reduction. To remove these lines as well as cosmic ray strikes, 
bad pixels, and the general background, the data were processed using our own scripts to subtract the 
residual background along each column and interpolate across isolated pixels with highly anomalous 
counts. Care was taken to ensure that the scripts removed only noise features, using a comparison of 
the many image slices within each dispersed image to distinguish real features from noise. 

\begin{figure*}
\centering
\subfigure{
   \includegraphics[scale=0.4]{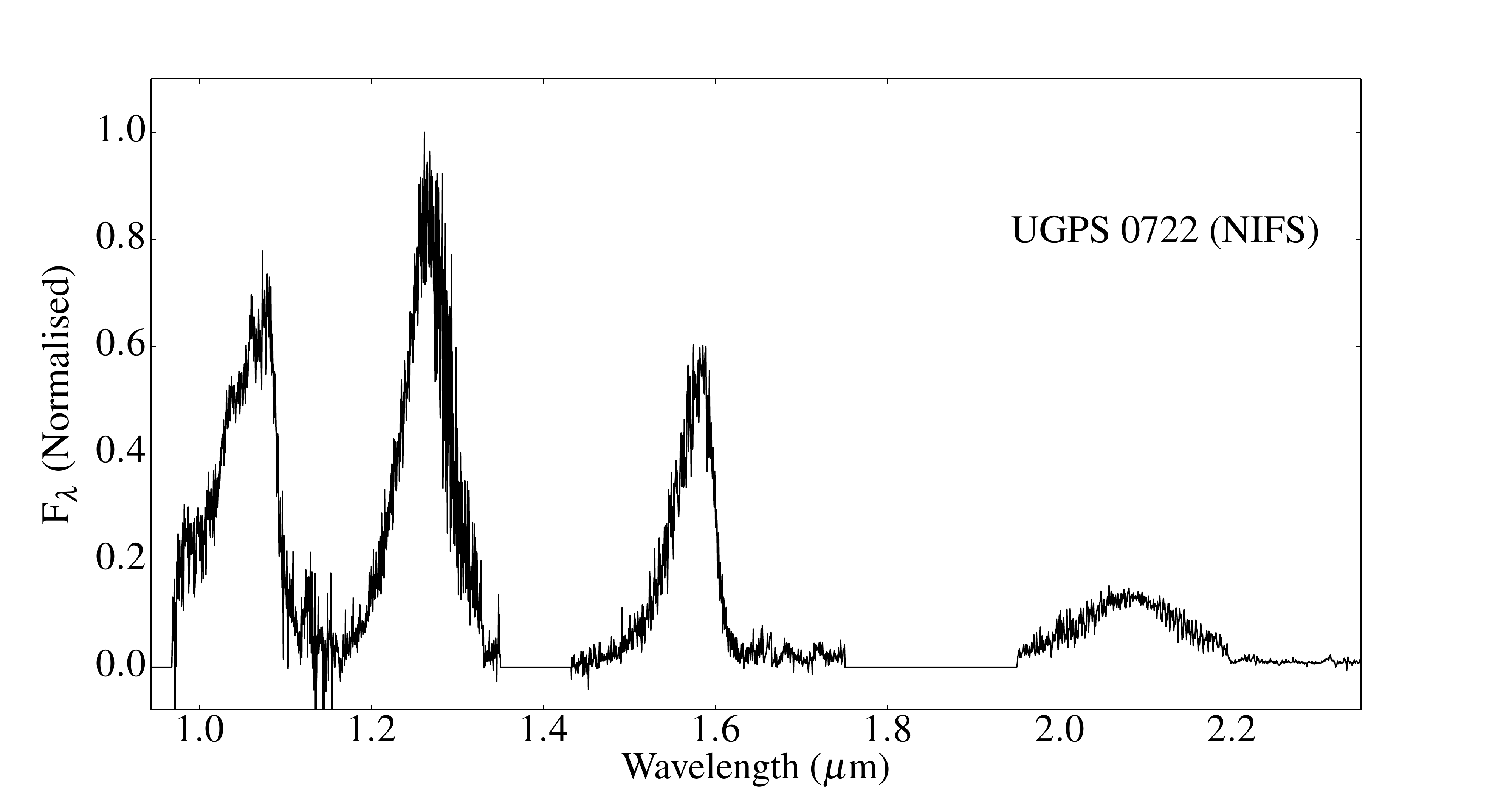}
                 }                                                           
\subfigure{
   \includegraphics[scale=0.4]{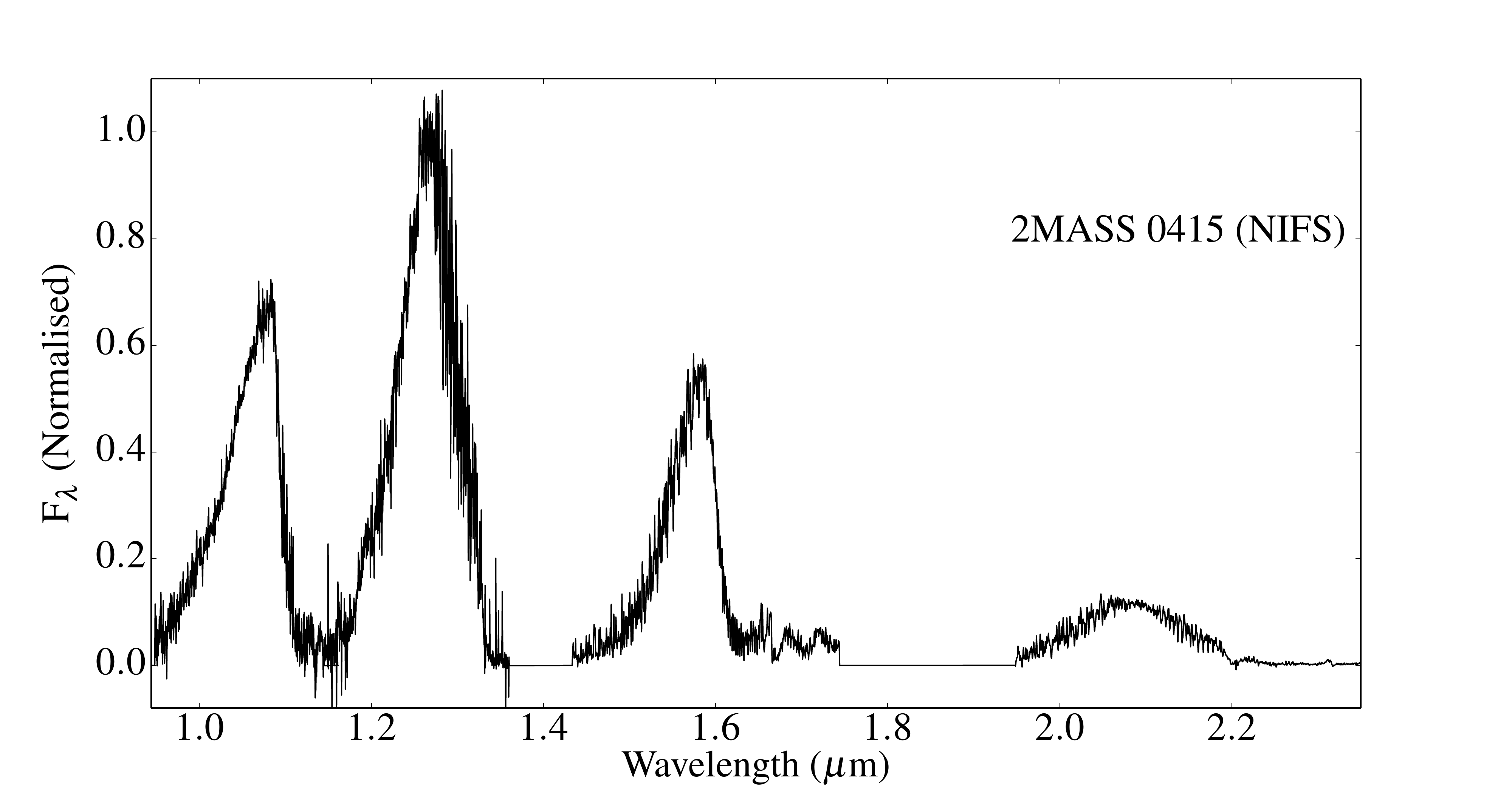}  
                 }
  \caption{\textbf{Top.} The near-infrared spectrum of the T9 standard UGPS~0722.  \textbf{Bottom.} 
  The near-infrared spectrum of the T8 standard 2MASS~0415.}
  \label{fig:0722_2m0415_nir}
\end{figure*}

The science spectra were corrected for the T dwarfs' radial velocities using the {\sc iraf} task {\sc dopcor}. 
The radial velocity for 2MASS~0415 was taken from~\citet{zapatero07}, that for UGPS~0722 from  B11.  
Both T dwarf spectra were corrected for the heliocentric and barycentric velocity components of Earth 
using the  {\sc idl}/{\sc a}{\small stro}{\sc l}{\small ib} task  {\sc baryvel}. In the spectrum of UGPS~0722, 
we estimate the S/N ratio at the peaks of the $Z$-, $J$-, $H$-, and $K$-bands to be $\sim$250, 370, 300, 
and 80 respectively. B11's estimates for the corresponding values in their analysis of UGPS~0722 are 
$\sim$250, 350, 200, and 60. 
We note that Magellan/FIRE data split each passband into multiple orders which can cause more variation 
in the S/N ratio with wavelength. This is not the case with NIFS, where data are collected in separate 
observations in each passband. 
Our T dwarf spectra are shown in Figure \ref{fig:0722_2m0415_nir}

\begin{figure*}
\centering
\begin{minipage}{\linewidth}
   \includegraphics[scale=0.3]{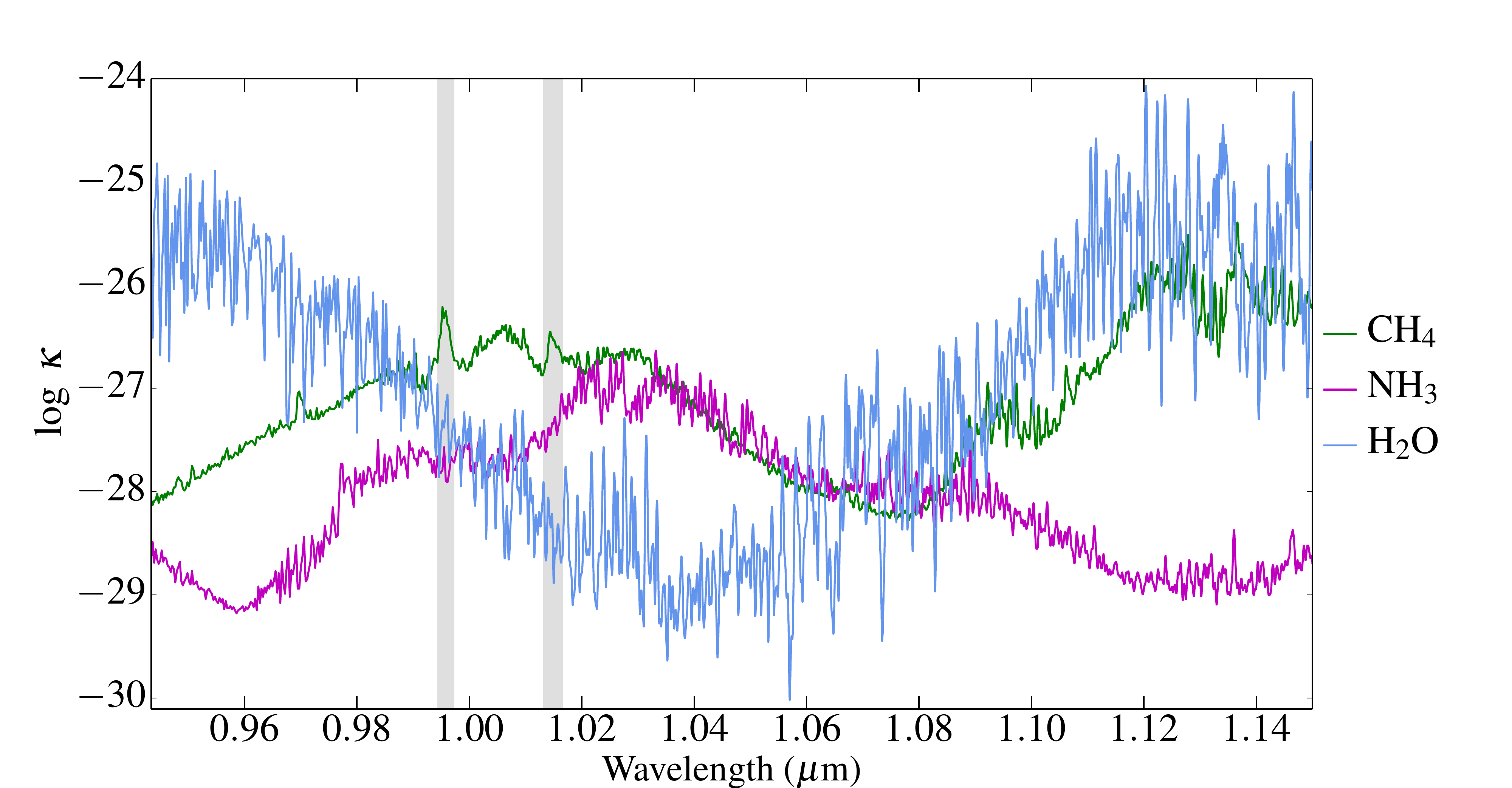} 
  \caption{Molecular opacities in the $Z$-band. The graph shows the log of the scaled absorption cross-sections 
  at 500~K for CH$_{4}$ (green), NH$_{3}$ (magenta) and H$_{2}$O (blue). By scaled, 
  we mean that the cross-sections have been corrected for the relative molecular abundances expected for each 
  molecule according to Figure 3 in~\citealt{saumon06}. Methane opacity is the dominant opacity source between 
  $\sim$0.995-1.032~$\mu$m where the continuum contains two prominent Q-branch regions (shaded).  Cross-sections are measured in units of 
  square centimetres per molecule (see Equation 4 in \citealt{hill13}). These units are used in all similar graphs in this paper.}
  \label{fig:ch4z}
\end{minipage}
\end{figure*}

\section{Methane}
\label{sec:methane}
In the following discussion, the terms CH$_{4}$ and methane refer to the main isotopologue of methane, 
$^{12}$CH$_{4}$. We examined a number of absorption features in the $H$-band of UGPS~0722, which 
we suspected may have been produced by the $^{13}$CH$_{4}$ isotopologue, but none of these features 
coincided with $^{13}$CH$_{4}$ lines in the HITRAN~\citep{rothman09} molecular spectroscopic database.

Carbon in the photospheres of early to mid L type brown dwarfs is predominantly found as carbon monoxide 
(CO) due to the molecule's high dissociation energy~\citep{geballe09}. By spectral type L5, the photosphere 
is cool enough to allow the hydrogenation of CO to CH$_{4}$~\citep{noll00}. As the photosphere continues 
to cool, CH$_{4}$ becomes the dominant carbon-bearing species~\citep{fegley96}. 

We have used absorption cross-sections at 500~K and 750~K, derived from the 10to10 line list 
to identify methane absorption features in the near-infrared spectra of the two T dwarfs. 
Absorption cross-sections were calculated at zero-pressure and do not consider collisional broadening effects. 
At the highest resolutions, this could lead to differences between opacity plots and model spectra, 
but is not a concern at the resolution of our T dwarf spectra \citep{hill13}.
These cross-sections were compared with cross-sections at the same temperatures for H$_{2}$O and NH$_{3}$, calculated 
respectively from the BT2~\citep{barber06} and BYTe line lists. Each cross-section 
was scaled by the relative abundances of these molecules according to Figure 3 in \citet{saumon06}. 
We have adopted the mole fractions of log $-$3.129 (H$_{2}$O), log $-$3.312 (CH$_{4}$), and log $-$4.907 (NH$_{3}$). 
The same values were used for the 500~K and 750~K opacity plots.
We note that the molecular abundances along the profiles in Figure 3 in \citet{saumon06}, and
also in Figure 5 of \citet{geballe09} are constant with depth to $\sim$0.1 dex (except well below the photosphere). 
The abundance for NH$_{3}$ included non-equilibrium effects in the model atmosphere.

\subsection{Ro-vibrational spectroscopy of non-linear molecules}
\label{ch4ro}
CH$_{4}$ electric dipole transitions are due to changes in a manifold of rotational energies, $\Delta$J, 
superimposed on larger vibrational energy changes,  $\Delta\nu$.
In any particular waveband, R-branch ($\Delta$J$=+$1), Q-branch ($\Delta$J$=$0), and P-branch 
($\Delta$J$=-$1) transitions form a sequence in wavelength. We have not detected electric quadrupole 
transitions from the O-branch ($\Delta$J$=-$2), and S-branch ($\Delta$J$=+$2) in our analysis of the T dwarf spectra. 

If CH$_{4}$ was a simple harmonic oscillator, absorption features could only be produced by transitions 
between adjacent vibrational energy levels. As this is not the case, transitions involving $\Delta\nu=\pm$2, 
$\pm$4, $\pm$6,...etc. are possible. We show in Section \ref{sec:ch4h} that the strongest features in the 
$H$-band absorption spectrum of CH$_{4}$ are produced by overtones. 

Molecules may be classified according to their symmetry, and assigned to certain point groups.  CH$_{4}$ 
is a tetrahedral molecule and belongs to the T$_{d}$ point group. It has four fundamental vibrational modes 
or states, described by four quantum numbers. The modes and their properties are shown in Table \ref{tab:ch4modes}.

\begin{table*}
\centering
\begin{minipage}{140mm}
\caption{The Fundamental Vibrational Modes $(\Delta \nu=1)$ of Methane}  
\label{tab:ch4modes}
\begin{tabular}{c c c c c c c c} 
\hline\hline
\noalign{\vskip 2mm} 
Mode & Type & Wavenumber (cm$^{-1}$) & Symmetry & Components\\ [0.5ex] 
\hline
\noalign{\vskip 2mm} 
$\nu_{1}$ & Symmetric Stretch & 2917  & A$_{1}$ & 1\\ 
$\nu_{2}$ & Symmetric Bend & 1534 & E & 2 \\ 
$\nu_{3}$ & Asymmetric Stretch & 3019 & F$_{1}$ & 3\\ 
$\nu_{4}$ & Asymmetric Bend & 1306 & F$_{2}$ & 3\\ 
\hline
\end{tabular}
\end{minipage}
\end{table*}

It can be seen from Table \ref{tab:ch4modes}  that three of the modes are degenerate. $\nu_{2}$ is doubly 
degenerate, while $\nu_{3}$ and $\nu_{4}$ each have three components sharing the same energy. Note that 
the symmetric modes impart no motion to the carbon atom. In most cases, excited stretching states are 
stronger than excited bending states. In the $H$-band, for example, the strongest features belong to the 
2$\nu_{3}$ vibrational band, which is a stretching band.



\subsection{The $Z$-Band}
\label{sec:ch4z} 
In the $Z$-band, CH$_{4}$ is the major opacity source between $\sim$0.995$-$1.032~$\mu$m, 
where the cross-section is marked by two Q-branch regions, centred at $\sim$0.9957 $\mu$m and $\sim$1.0152 $\mu$m (see Figure  \ref{fig:ch4z}). We have not identified 
any methane features in the $Z$-band spectra of either T dwarf. The non-detection of the Q-branch CH$_{4}$ features can be attributed to the low flux level in the $Z$-band at these wavelengths.

\begin{figure*}
\centering
\begin{minipage}{\linewidth}
   \includegraphics[scale=0.45]{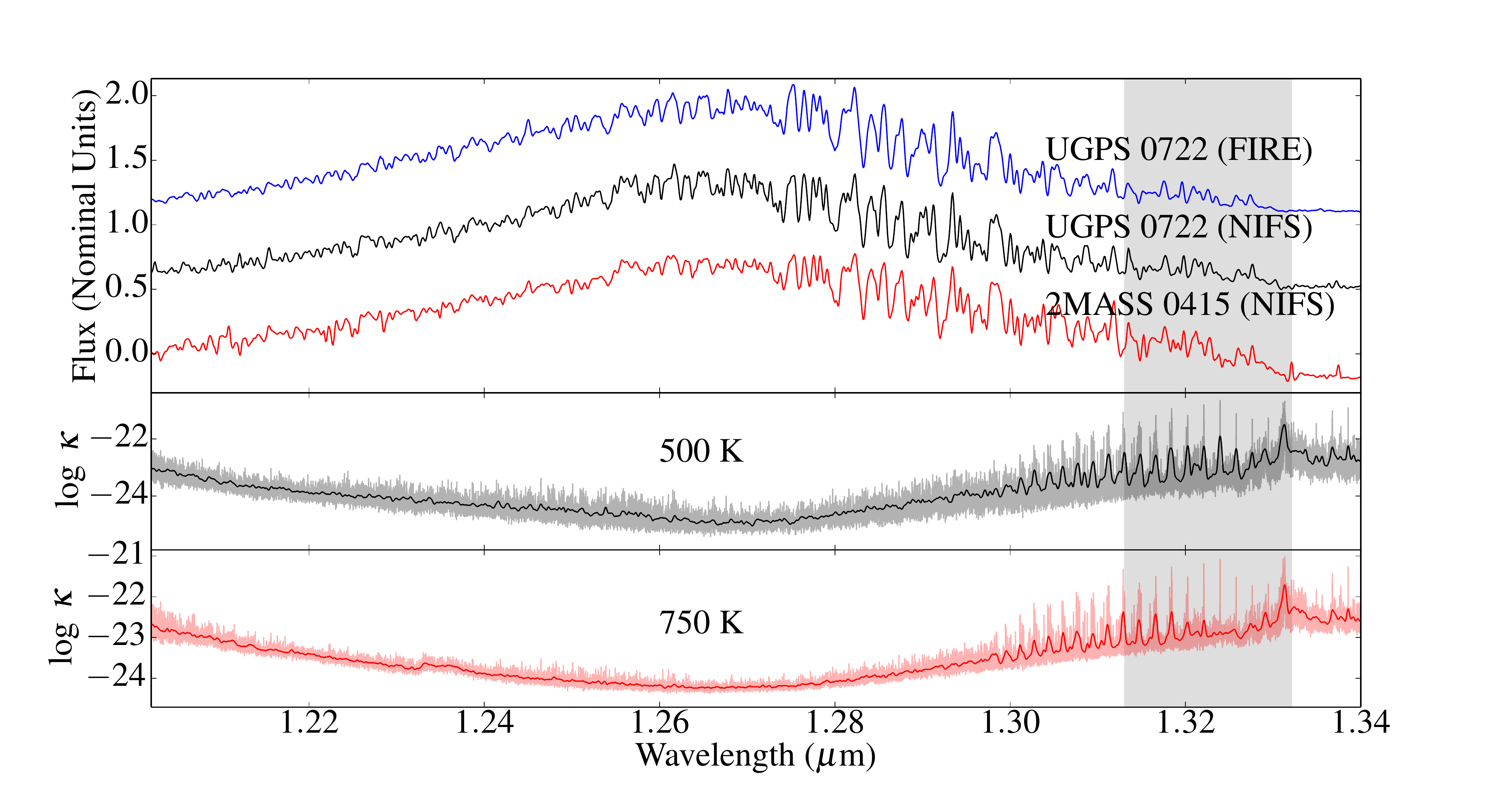} 
  \caption{CH$_{4}$ absorption in the $J$-band spectra of 2MASS~0415 (red) and UGPS~0722 (black).  
  The Magellan/FIRE spectrum of UGPS 0722 (blue) is shown for comparison.
  The middle and lower graphs show the unscaled absorption cross-sections at 500~K (black) and 750~K (red), 
  smoothed to the same resolution as the T dwarf spectra, overplotted on the unsmoothed cross-sections. 
  The shaded region indicates the Q-branch starting at $\sim$1.33~$\mu$m. (T dwarf spectra have been 
  offset to aid identification of spectral features.)}
  \label{fig:jlong}
\end{minipage}
\end{figure*}

\begin{figure*}
\centering
\begin{minipage}{\linewidth}
   \includegraphics[scale=0.45]{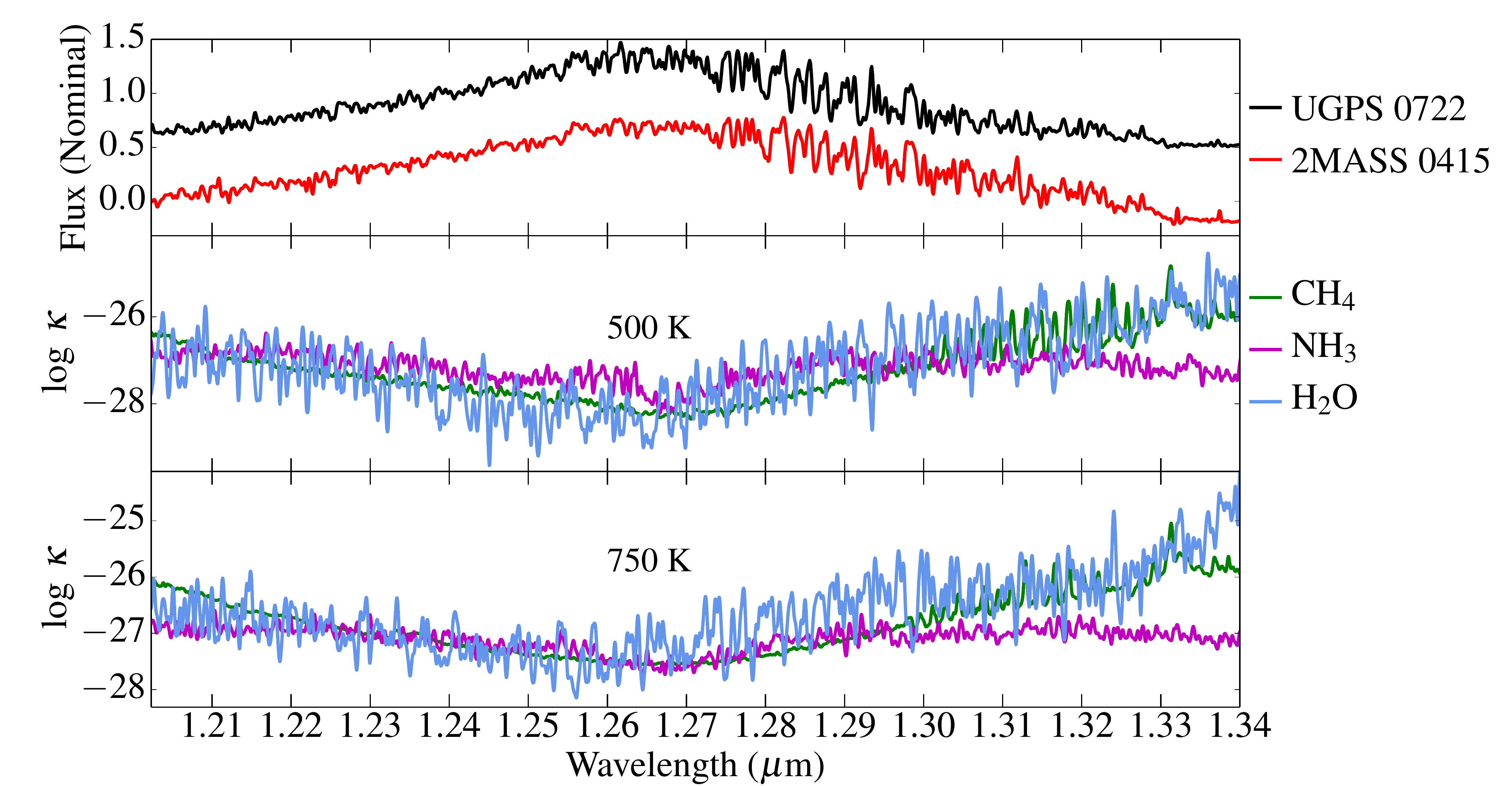} 
  \caption{The increasing dominance of water opacity on the red side of the $J$ band flux peak in late T dwarfs. The middle and lower graphs show the log of the scaled absorption cross-sections at 500~K and 750~K respectively for CH$_{4}$ (green), NH$_{3}$ (magenta) and H$_{2}$O (blue), smoothed to the same resolution as the T dwarf spectra.}
  \label{fig:jwater}
\end{minipage}
\end{figure*}

\begin{figure*}
\centering
\subfigure{
\hspace*{0.5cm}\includegraphics[scale=0.42]{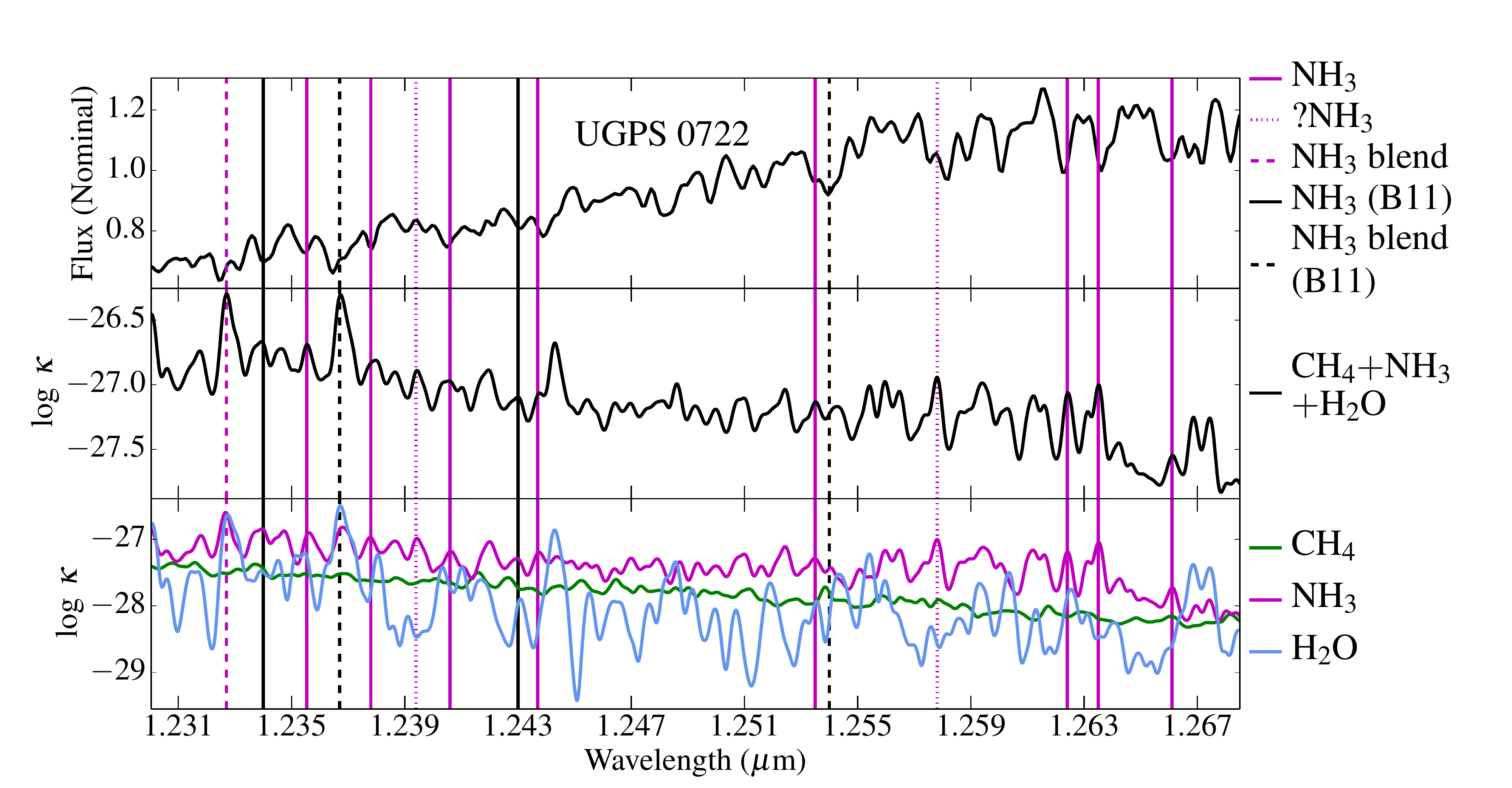}
                 }                                                           
\subfigure{
   \includegraphics[scale=0.4]{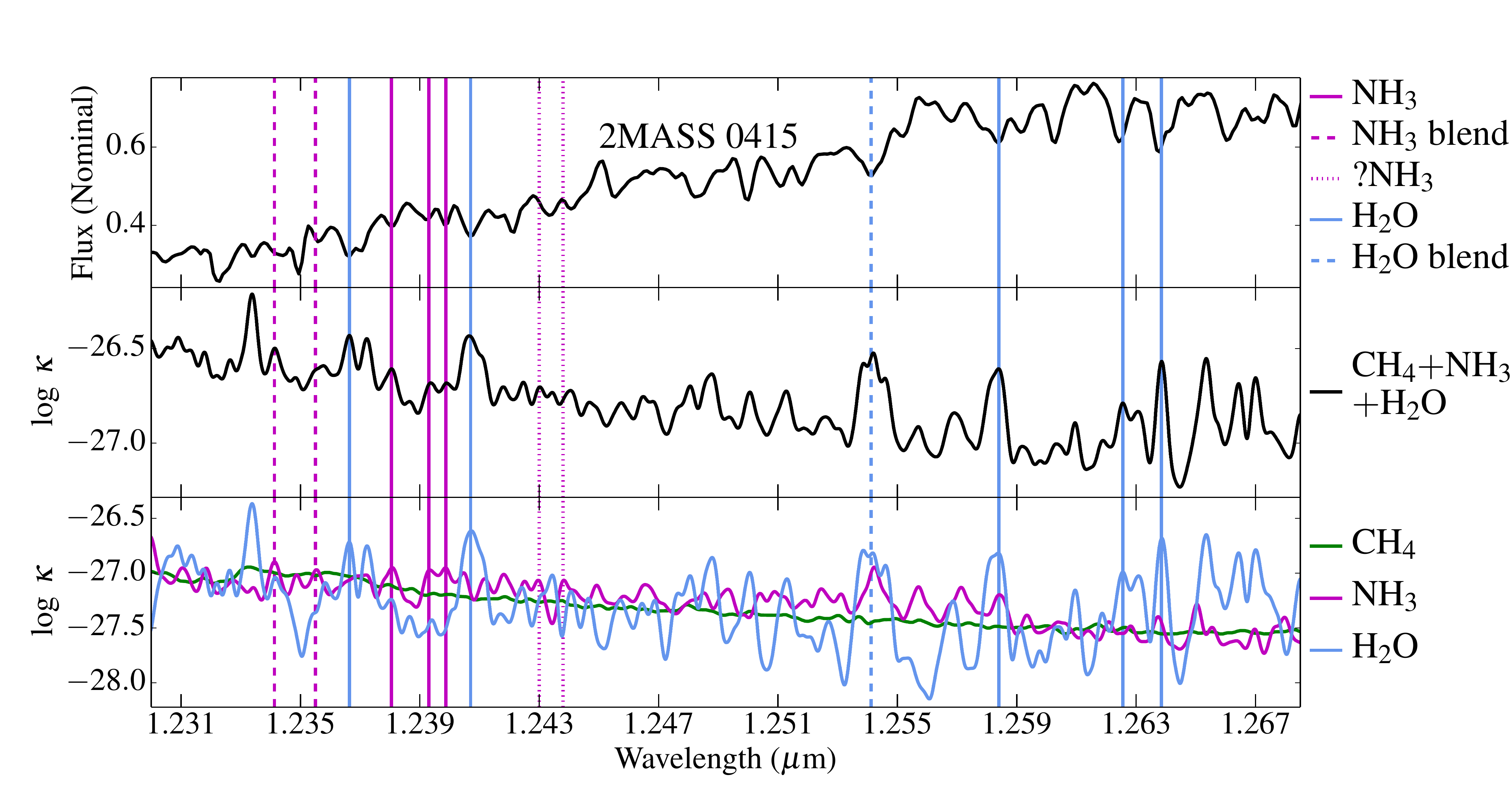}  
                 }
  \caption{Opacities on the short side of the $J$-band flux peak. \textbf{Top.} Solid black lines identify NH$_{3}$ features detected by B11 in the spectrum of UGPS~0722. Dashed black lines are NH$_{3}$ blends identified by B11. Magenta lines are NH$_{3}$ features identified in this work. Most of the features appear to be due to ammonia opacity. The lower graph shows the log of the scaled absorption cross-sections at 500~K for CH$_{4}$ (green), NH$_{3}$ (magenta) and H$_{2}$O (blue), smoothed to the same resolution as the T dwarf spectrum. The middle graph is the sum of the scaled CH$_{4}$, NH$_{3}$ and  H$_{2}$O opacities.  \textbf{Bottom.} The same region in the spectrum of 
  2MASS~0415. Scaled opacity cross-sections have been made at 750~K. In this case, the dominant opacity 
  source is H$_{2}$O.}
  \label{fig:figshortj}
\end{figure*}

\begin{figure*}
\centering
\subfigure{
   \includegraphics[scale=0.42, angle=0]{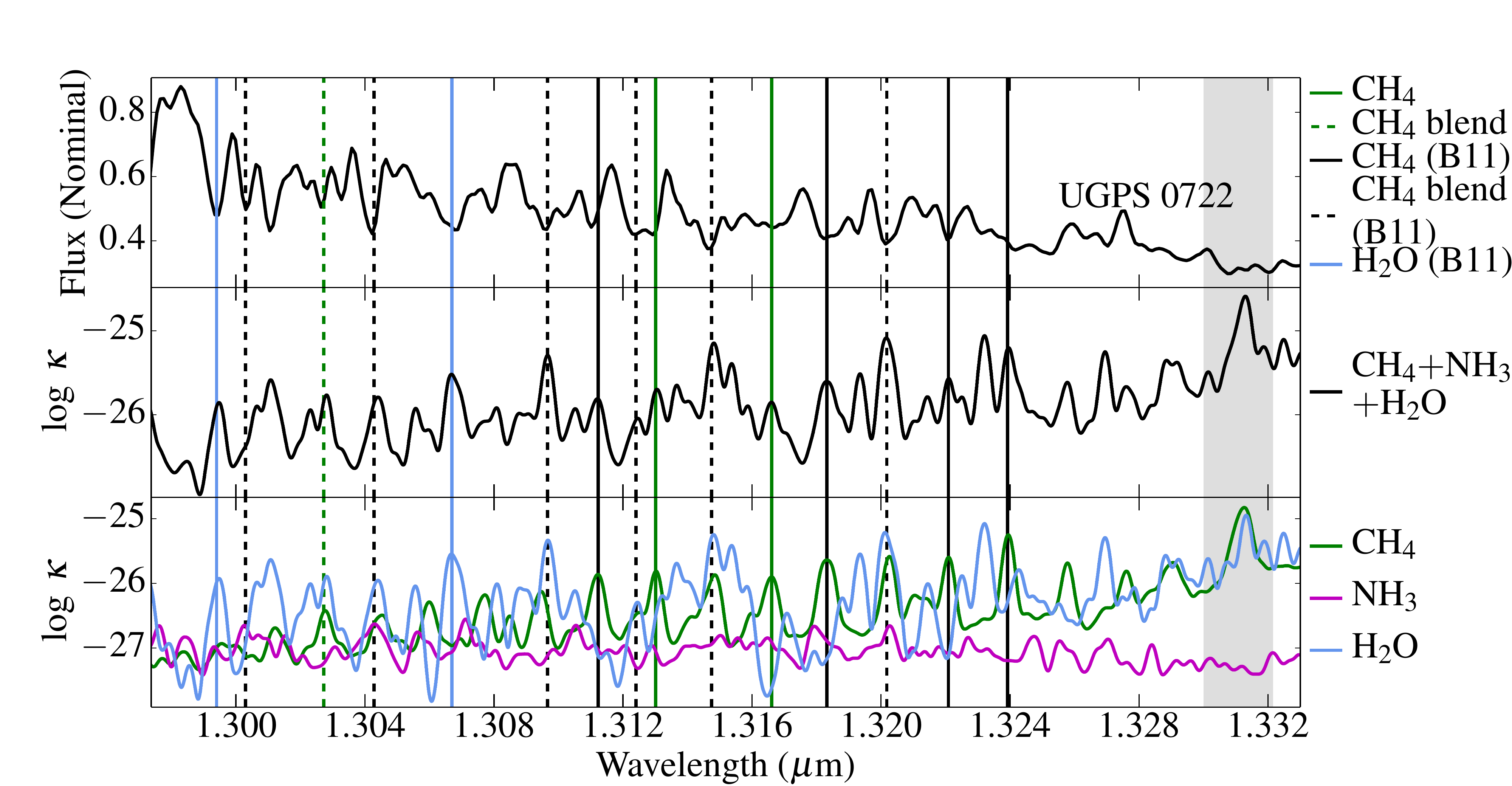} 
   \label{fig:subfig3}
                 }                                                           
\subfigure{
   \includegraphics[scale=0.42, angle=0]{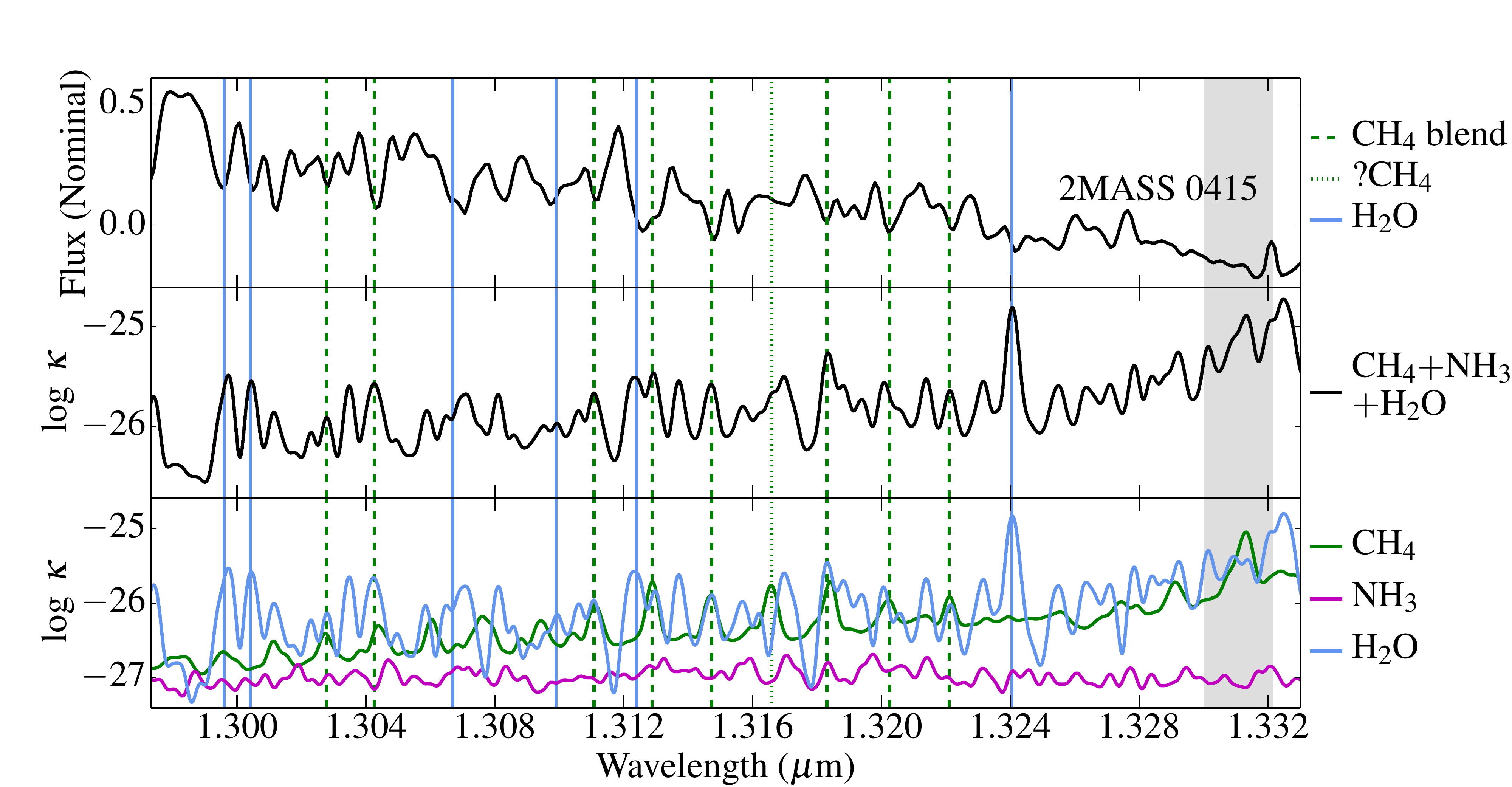} 
   \label{fig:subfig4}
                 }
  \caption{\textbf{Top.}  CH$_{4}$ absorption features on the long side of the $J$-band flux peak in the spectrum 
  of UGPS~0722 (see Table \ref{tab:2}). Solid black lines are CH$_{4}$ features identified in B11. Solid green lines are CH$_{4}$ features identified in this 
  work. Dashed black lines are mixed CH$_{4}$ features identified in B11. The dashed green line is a mixed CH$_{4}$ 
  feature identified in this work. Solid blue lines indicate H$_{2}$O features previously identified as mixed CH$_{4}$ features. 
  The lower graph shows the log of the absorption cross-sections at 500~K for 
  CH$_{4}$ (green), NH$_{3}$ (magenta) and H$_{2}$O (blue), scaled by molecular abundances and smoothed to 
  the same resolution as the T dwarf spectrum. The middle graph is the sum of the scaled CH$_{4}$, NH$_{3}$ and  H$_{2}$O 
  opacities at 500~K. Absorption features are produced by a blend of individual 
  transition lines. \textbf{Bottom.}  The same region in the spectrum of 2MASS~0415 showing 
  the corresponding absorption features (see Table \ref{tab:2} again). Absorption cross-sections calculated at 750~K. }
  \label{fig:0722j_0415j}
\end{figure*}



\subsection{The $J$-Band}
\label{sec:ch4j}
There has been very little work done on intermediate resolution spectroscopy of T dwarfs. Up till now, 
we are only aware of one paper that has been published in this area, that by  B11. By comparing their 
spectrum of UGPS~0722 with molecular line lists, B11 were able to identify absorption features due to 
H$_{2}$O and CH$_{4}$. They have also made the first confirmed detections of NH$_{3}$ opacity in 
the near-infrared spectrum of a T dwarf. While B11 used the BT2 and BYTe line lists we use in 
our analysis, they used an old CH$_{4}$ line list~\citep{nassar03}, supplemented by the HITRAN 2008 
database~\citep{rothman09}.
 
Figure \ref{fig:jlong} shows the $J$-band spectra of UGPS~0722 and 2MASS~0415 
and the corresponding CH$_{4}$ absorption cross-sections at 500~K and 750~K.
The figure shows CH$_{4}$ opacity reducing with increasing wavelength over the width of the short side of 
the $J$-band flux peak. This decrease in opacity is particularly smooth in the 750~K cross-section. 
On the long side of the $J$-band flux peak, the opacity shown by both cross-sections increases, producing 
regular patterns of peaks, separated by $\sim$0.002~$\mu$m, from $\sim$1.30 - 1.33~$\mu$m.

B11 identified a number of blended CH$_{4}$ features on the short side of the $J$-band flux peak in the 
T dwarf's spectrum at 1.2390~$\mu$m, 1.2406~$\mu$m, 1.2439~$\mu$m, 1.2540~$\mu$m, 1.2578~$\mu$m, 
1.2624~$\mu$m, 1.2635~$\mu$m, and 1.2661~$\mu$m. We have found absorption features at these 
wavelengths in the spectrum of UGPS~0722 (see Figure \ref{fig:figshortj}),
but only the feature at 1.2540~$\mu$m corresponds to a peak in the CH$_{4}$ opacity (and a much stronger peak 
in ammonia opacity). 
CH$_{4}$ opacity between $\sim$1.235-1.270~$\mu$m is generally flat. While the mean CH$_{4}$ opacity 
is greater than the H$_{2}$O opacity in this region, in several places it is surpassed by peaks in the H$_{2}$O opacity.
In fact, NH$_{3}$ is the main opacity source in this region and we find that the majority of these features 
can be attributed to ammonia opacity. (See Section \ref{sec:nh3j} for a fuller discussion of ammonia opacity in the $J$-band spectra of these T dwarfs). 
In contrast, water is the major opacity source in the same region in the spectrum of 2MASS 0415 (see Figure \ref{fig:jwater}).

B11 identified two possible CH$_{4}$/H$_{2}$O absorption features either side of the peak in the spectrum 
of UGPS~0722 at $\sim$1.30~$\mu$m (see Figure \ref{fig:0722j_0415j}). Neither cross-section shows a peak in CH$_{4}$ opacity at these 
wavelengths. We have determined that the feature at 1.2994~$\mu$m is due to water opacity, and that at 
1.3004~$\mu$m is produced by a combination of water and ammonia opacity. B11 were more confident in 
identifying two CH$_{4}+$H$_{2}$O absorption features at 1.3043~$\mu$m and 1.3067~$\mu$m. While there 
is a peak in the CH$_{4}$ opacity at 1.3043~$\mu$m, we find that water is again the major opacity source 
at this wavelength, with a smaller contribution from NH$_{3}$ opacity. There is no corresponding peak in the CH$_{4}$ opacity at 1.3067~$\mu$m and we conclude 
that this is another water feature. A  blended CH$_{4}+$H$_{2}$O absorption feature at 1.3097~$\mu$m is 
consistent with a peak in the CH$_{4}$ opacity, and a much stronger peak in water opacity. A CH$_{4}$ feature 
identified by B11 at 1.3112~$\mu$m is consistent with methane as the main opacity source. Opacities 
calculated using the 10to10 line list show that three CH$_{4}$ features at 1.3124~$\mu$m, 1.3148~$\mu$m, 
and 1.3202~$\mu$m are actually H$_{2}$O$+$CH$_{4}$ blends, while a CH$_{4}+$H$_{2}$O blend at 
1.3183~$\mu$m is a ``pure" methane feature. We can confirm the methane feature at 1.3221~$\mu$m. 
While CH$_{4}$ is the strongest opacity at 1.3284 $\mu$m, there is no peak in the CH$_{4}$ opacity corresponding 
to a previously identified CH$_{4}$ feature at this wavelength. 

While the 10to10 line list has enabled us to correct previous mis-identifications of spectral features, 
it has also allowed us to identify new CH$_{4}$ features at 1.3130~$\mu$m and 1.3166~$\mu$m. 
We have also identified a new H$_{2}$O$+$CH$_{4}$ feature at 1.3027~$\mu$m. See Figure \ref{fig:0722j_0415j}, and Table \ref{tab:2}. 

These are late T dwarfs, and we would expect CH$_{4}$ absorption features to be prominent in both 
objects' spectra. However, in places the methane absorption in the T9 is deeper and/or broader than 
in the T8. For example, the absorption features at $\sim$1.3183~$\mu$m, 1.3202~$\mu$m and 1.3239~$\mu$m.
That being said, we have identified fewer pure methane features in the $J$-band spectrum of 
UGPS 0722 than earlier studies, and find that a number of features previously identified as pure methane features are 
actually water features or water/methane blends.
Additional water features are seen in the spectrum of 2MASS 0415 at 1.2334~$\mu$m, 1.2500~$\mu$m, 
and 1.2568~$\mu$m in Figure \ref{fig:figshortj} and at 1.3012~$\mu$m, 1.3035~$\mu$m, 1.3081~$\mu$m, and 1.3255~$\mu$m in Figure \ref{fig:0722j_0415j}. 
We have not marked these features as we are focussing here on methane and ammonia features. Nonetheless, it is clear that water is an important 
opacity source in the $J$-band spectra of these T dwarfs. 
In fact, our analysis suggests that water rather than methane is the dominant opacity 
source in the red half of the $J$-band in 2MASS 0415. In the case of UGPS 0722, water is the principal absorber until 
$\sim$1.31 $\mu$m, when methane opacity starts to dominate (see again Figure \ref{fig:jwater}).

\begin{table*}
\small
\centering
\newcommand{\footstar}[1]{$^*$ \footnotetext{$^*$#1}}
\caption{CH$_{4}$ Absorption Features in the $J$ Band Spectra of Late T Dwarfs (see Figure \ref{fig:0722j_0415j})}. 
\begin{tabular}{c c c c c c c } 
\vspace{-1.1em}\\
& Source & $\lambda$ ($\mu$m) & Opacity Source (B11) & Opacity Source (500~K/750~K) & \head{2.5cm}{Absorption feature in UGPS~0722} & \head{2.5cm}{Absorption feature in 2MASS~0415}\\ [0.5ex] 
& B11          & 1.2994    &  CH$_{4}$(?)+H$_{2}$O    & H$_{2}$O/H$_{2}$O                                                            & Yes & Yes\\
& B11          & 1.3004    &  CH$_{4}$(?)+H$_{2}$O    & (NH$_{3}$+H$_{2}$O)/H$_{2}$O                                          & Yes & Yes\\
\rowcolor{bubblegum}
& This work & 1.3027   &              ---                      & (H$_{2}$O+CH$_{4}$)/(H$_{2}$O+CH$_{4}$)                         & Yes & Yes\\
\rowcolor{bubblegum}
& B11          & 1.3043   &  CH$_{4}$+H$_{2}$O        & (H$_{2}$O+CH$_{4}$+NH$_{3}$)/(H$_{2}$O+CH$_{4}$)         & Yes & Yes\\
& B11          & 1.3067   &  CH$_{4}$+H$_{2}$O        & H$_{2}$O/H$_{2}$O                                           & Yes & Yes\\
\rowcolor{bubblegum}
& B11          & 1.3097   &  CH$_{4}$+H$_{2}$O        & (H$_{2}$O+CH$_{4}$)/H$_{2}$O                                         & Yes & Yes\\
\rowcolor{bubblegum}
& B11          & 1.3112   &  CH$_{4}$                        & CH$_{4}$/(H$_{2}$O+CH$_{4}$)                                           & Yes & Yes\\
\rowcolor{bubblegum}
& B11          & 1.3124   &  CH$_{4}$                        & (H$_{2}$O+CH$_{4}$)/H$_{2}$O                                           & Yes & Yes\\
\rowcolor{bubblegum}
& This work & 1.3130   &       ---                            & CH$_{4}$/(CH$_{4}$+H$_{2}$O)                                                            & Yes & Yes\\
\rowcolor{bubblegum}
& B11          & 1.3148   &  CH$_{4}$                        & (H$_{2}$O+CH$_{4}$)/(H$_{2}$O+CH$_{4}$)                         & Yes & Yes\\
\rowcolor{bubblegum}
& This work & 1.3166   &       ---                            & CH$_{4}$/CH$_{4}$                                          & Yes & No(?)\\
\rowcolor{bubblegum}
& B11          & 1.3183   &  CH$_{4}$+H$_{2}$O        & CH$_{4}$/(H$_{2}$O+CH$_{4}$)                                           & Yes & Yes\\
\rowcolor{bubblegum}
& B11          & 1.3202   &  CH$_{4}$                        & (H$_{2}$O+CH$_{4}$)/(H$_{2}$O+CH$_{4}$)         & Yes & Yes\\
\rowcolor{bubblegum}
& B11          & 1.3221   &  CH$_{4}$                        & CH$_{4}$/(CH$_{4}$+H$_{2}$O)                                           & Yes & Yes\\
\rowcolor{bubblegum}
& B11          & 1.3239   &  CH$_{4}$+H$_{2}$O        & CH$_{4}$/H$_{2}$O                                           & Yes & Yes\\
\end{tabular}
\begin{tablenotes}
\item In tables \ref{tab:2}, \ref{tab:h}, and \ref{tab:ch4k}, R-, Q-, and P-branch methane features are highlighted in red, grey, and blue respectively. Non-shaded regions are features produced by other opacity sources, which had previously been identified as due to methane or methane blends. 
\end{tablenotes}
\label{tab:2}
\end{table*}

\begin{table*}
\small
\centering
\caption{R-branch ro-vibrational transition lines responsible for the CH$_{4}$ absorption features in the $J$ band spectra of UGPS 0722 and 2MASS~0415 (shaded). Intensities were calculated using Equation 7 in \citealt{hill13}.}
\begin{tabular}{ccccccccccccc}
\vspace{-1.1em}\\
$\lambda$~($\mu$m) & Intensity & $\Delta\Gamma$  & $\Delta$J  & $\Delta\nu_{2}$ & $\Delta$L2 & $\Delta\nu_{3}$ & $\Delta$L3 & $\Delta$M3\\
                                    &  (cm molecule$^{-1}$) &&&&&&&\\
1.3111927 & 1.94E-23 & A$_{1}$~$\rightarrow$~A$_{2}$ &  10~$\rightarrow$~11  & 0~$\rightarrow$~1  & 0~$\rightarrow$~1  & 0~$\rightarrow$~2     & 0~$\rightarrow$~2     & 0~$\rightarrow$~1\\ 
\rowcolor{bisque}
\cellcolor{white} & 1.38E-23 & A$_{1}$~$\rightarrow$~A$_{2}$ &  10~$\rightarrow$~11  & 0~$\rightarrow$~1  & 0~$\rightarrow$~1  & 0~$\rightarrow$~2     & 0~$\rightarrow$~2     & 0~$\rightarrow$~1\\ 
\hline
1.3128914  & 3.59E-23 & A$_{1}$~$\rightarrow$~A$_{2}$    & 9~$\rightarrow$~10   & 0~$\rightarrow$~1  & 0~$\rightarrow$~1  & 0~$\rightarrow$~2     & 0~$\rightarrow$~2     & 0~$\rightarrow$~1\\   
\rowcolor{bisque}
\cellcolor{white} & 2.32E-23 & A$_{1}$~$\rightarrow$~A$_{2}$    & 9~$\rightarrow$~10   & 0~$\rightarrow$~1  & 0~$\rightarrow$~1  & 0~$\rightarrow$~2     & 0~$\rightarrow$~2     & 0~$\rightarrow$~1\\   
\hline
1.3166163 & 6.50E-23 &A$_{2}$~$\rightarrow$~A$_{1}$     & 7~$\rightarrow$~8  & 0~$\rightarrow$~1  & 0~$\rightarrow$~1  & 0~$\rightarrow$~2     & 0~$\rightarrow$~2     & 0~$\rightarrow$~1\\ 
\rowcolor{bisque}
 \cellcolor{white} & 3.54E-23 & A$_{2}$~$\rightarrow$~A$_{1}$   & 7~$\rightarrow$~8   & 0~$\rightarrow$~1  & 0~$\rightarrow$~1  & 0~$\rightarrow$~2     & 0~$\rightarrow$~2     & 0~$\rightarrow$~1\\   
\hline
1.3184149 & 7.76E-23 & F$_{1}$~$\rightarrow$~F$_{2}$    & 6~$\rightarrow$~7  & 0~$\rightarrow$~1  & 0~$\rightarrow$~1  & 0~$\rightarrow$~2     & 0~$\rightarrow$~2     & 0~$\rightarrow$~1\\  
\rowcolor{bisque}
\cellcolor{white} & 3.94E-23 & F$_{1}$~$\rightarrow$~F$_{2}$     & 6~$\rightarrow$~7  & 0~$\rightarrow$~1  & 0~$\rightarrow$~1  & 0~$\rightarrow$~2     & 0~$\rightarrow$~2     & 0~$\rightarrow$~1\\ 
1.3184160 & 7.66E-23 & A$_{1}$~$\rightarrow$~A$_{2}$    & 6~$\rightarrow$~7  & 0~$\rightarrow$~1  & 0~$\rightarrow$~1  & 0~$\rightarrow$~2     & 0~$\rightarrow$~2     & 0~$\rightarrow$~1\\ 
\rowcolor{bisque}
\cellcolor{white} & 3.89E-23 & A$_{1}$~$\rightarrow$~A$_{2}$     & 6~$\rightarrow$~7  & 0~$\rightarrow$~1  & 0~$\rightarrow$~1  & 0~$\rightarrow$~2     & 0~$\rightarrow$~2     & 0~$\rightarrow$~1\\ 
1.3184211 & 7.48E-23 &F$_{2}$~$\rightarrow$~F$_{1}$    & 6~$\rightarrow$~7  & 0~$\rightarrow$~1  & 0~$\rightarrow$~1  & 0~$\rightarrow$~2     & 0~$\rightarrow$~2     & 0~$\rightarrow$~1\\  
\rowcolor{bisque}
\cellcolor{white} & 3.80E-23 & F$_{2}$~$\rightarrow$~F$_{1}$     & 6~$\rightarrow$~7  & 0~$\rightarrow$~1  & 0~$\rightarrow$~1  & 0~$\rightarrow$~2     & 0~$\rightarrow$~2     & 0~$\rightarrow$~1\\ 
1.3184622 & 7.26E-23 &E~$\rightarrow$~E          & 6~$\rightarrow$~7 & 0~$\rightarrow$~1  & 0~$\rightarrow$~1  & 0~$\rightarrow$~2     & 0~$\rightarrow$~2     & 0~$\rightarrow$~1\\ 
\rowcolor{bisque}
\cellcolor{white} & 3.69E-23 & E~$\rightarrow$~E     & 6~$\rightarrow$~7  & 0~$\rightarrow$~1  & 0~$\rightarrow$~1  & 0~$\rightarrow$~2     & 0~$\rightarrow$~2     & 0~$\rightarrow$~1\\ 
1.3184644 & 6.57E-23 &F$_{2}$~$\rightarrow$~F$_{1}$      & 6~$\rightarrow$~7  & 0~$\rightarrow$~1  & 0~$\rightarrow$~1  & 0~$\rightarrow$~2     & 0~$\rightarrow$~2     & 0~$\rightarrow$~1\\
\rowcolor{bisque}
\cellcolor{white} & 3.33E-23 &  F$_{2}$~$\rightarrow$~F$_{1}$     & 6~$\rightarrow$~7  & 0~$\rightarrow$~1  & 0~$\rightarrow$~1  & 0~$\rightarrow$~2     & 0~$\rightarrow$~2     & 0~$\rightarrow$~1\\ 
1.3184738 & 6.14E-23 &A$_{2}$~$\rightarrow$~A$_{1}$     & 6~$\rightarrow$~7  & 0~$\rightarrow$~1  & 0~$\rightarrow$~1  & 0~$\rightarrow$~2     & 0~$\rightarrow$~2     & 0~$\rightarrow$~1\\ 
\rowcolor{bisque}
\cellcolor{white}  & 3.12E-23 &A$_{2}$~$\rightarrow$~A$_{1}$     & 6~$\rightarrow$~7  & 0~$\rightarrow$~1  & 0~$\rightarrow$~1  & 0~$\rightarrow$~2     & 0~$\rightarrow$~2     & 0~$\rightarrow$~1\\ 
\hline
1.3220974 & 1.54E-22 & A$_{1}$~$\rightarrow$~A$_{2}$    & 4~$\rightarrow$~5   & 0~$\rightarrow$~1  & 0~$\rightarrow$~1  & 0~$\rightarrow$~2     & 0~$\rightarrow$~2     & 0~$\rightarrow$~1\\   
\rowcolor{bisque}
\cellcolor{white} & 7.00E-23 &  A$_{1}$~$\rightarrow$~A$_{2}$     & 4~$\rightarrow$~5  & 0~$\rightarrow$~1  & 0~$\rightarrow$~1  & 0~$\rightarrow$~2     & 0~$\rightarrow$~2     & 0~$\rightarrow$~1\\  
1.3220991 & 1.17E-22 &F$_{1}$~$\rightarrow$~F$_{2}$    & 4~$\rightarrow$~5  & 0~$\rightarrow$~1     & 0~$\rightarrow$~1  & 0~$\rightarrow$~2     & 0~$\rightarrow$~2     & 0~$\rightarrow$~1\\ 
\rowcolor{bisque}
\cellcolor{white} & 5.31E-23 & F$_{1}$~$\rightarrow$~F$_{2}$     & 4~$\rightarrow$~5  & 0~$\rightarrow$~1  & 0~$\rightarrow$~1  & 0~$\rightarrow$~2     & 0~$\rightarrow$~2     & 0~$\rightarrow$~1\\    
1.3221013 & 1.01E-22 &F$_{2}$~$\rightarrow$~F$_{1}$    & 4~$\rightarrow$~5 & 0~$\rightarrow$~1     & 0~$\rightarrow$~1 & 0~$\rightarrow$~2     & 0~$\rightarrow$~2     & 0~$\rightarrow$~1\\    
\rowcolor{bisque}
\cellcolor{white} & 4.59E-23 &F$_{2}$~$\rightarrow$~F$_{1}$     & 4~$\rightarrow$~5  & 0~$\rightarrow$~1  & 0~$\rightarrow$~1  & 0~$\rightarrow$~2     & 0~$\rightarrow$~2     & 0~$\rightarrow$~1\\ 
1.3221177 & 1.32E-22 &E~$\rightarrow$~E     & 4~$\rightarrow$~5 & 0~$\rightarrow$~1     & 0~$\rightarrow$~1     & 0~$\rightarrow$~2     & 0~$\rightarrow$~2     & 0~$\rightarrow$~1\\   
\rowcolor{bisque}
\cellcolor{white} & 6.00E-23 &E~$\rightarrow$~E     & 4~$\rightarrow$~5  & 0~$\rightarrow$~1  & 0~$\rightarrow$~1  & 0~$\rightarrow$~2     & 0~$\rightarrow$~2     & 0~$\rightarrow$~1\\ 
\hline
1.3239472 & 1.47E-22 &F$_{1}$~$\rightarrow$~F$_{2}$    & 3~$\rightarrow$~4  & 0~$\rightarrow$~1  & 0~$\rightarrow$~1  & 0~$\rightarrow$~2     & 0~$\rightarrow$~2     & 0~$\rightarrow$~1\\   
\rowcolor{bisque}
\cellcolor{white} & 6.40E-23 &F$_{1}$~$\rightarrow$~F$_{2}$     & 3~$\rightarrow$~4  & 0~$\rightarrow$~1  & 0~$\rightarrow$~1  & 0~$\rightarrow$~2     & 0~$\rightarrow$~2     & 0~$\rightarrow$~1\\
1.3239478 & 1.88E-22 &F$_{2}$~$\rightarrow$~F$_{1}$    & 3~$\rightarrow$~4  & 0~$\rightarrow$~1  & 0~$\rightarrow$~1  & 0~$\rightarrow$~2     & 0~$\rightarrow$~2     & 0~$\rightarrow$~1\\    
\rowcolor{bisque}
\cellcolor{white} & 8.20E-23 &F$_{2}$~$\rightarrow$~F$_{1}$     & 3~$\rightarrow$~4  & 0~$\rightarrow$~1  & 0~$\rightarrow$~1  & 0~$\rightarrow$~2     & 0~$\rightarrow$~2     & 0~$\rightarrow$~1\\  
1.3239481 & 2.02E-22 &A$_{2}$~$\rightarrow$~A$_{1}$    & 3~$\rightarrow$~4 & 0~$\rightarrow$~1  & 0~$\rightarrow$~1  & 0~$\rightarrow$~2     & 0~$\rightarrow$~2     & 0~$\rightarrow$~1\\    
\rowcolor{bisque}
\cellcolor{white} & 8.80E-23 &A$_{2}$~$\rightarrow$~A$_{1}$     & 3~$\rightarrow$~4  & 0~$\rightarrow$~1  & 0~$\rightarrow$~1  & 0~$\rightarrow$~2     & 0~$\rightarrow$~2     & 0~$\rightarrow$~1\\ 
\end{tabular}
\begin{tablenotes}
\item   $\Delta\Gamma$, $\Delta$J, and $\Delta\nu$ describe the changes in symmetry, rotational, and vibrational energy level between initial and final states. 
\end{tablenotes}
\label{tab:ch4_j_rbranch}
\end{table*}

\begin{figure*}
\centering
\begin{minipage}{\linewidth}
   \includegraphics[scale=0.4]{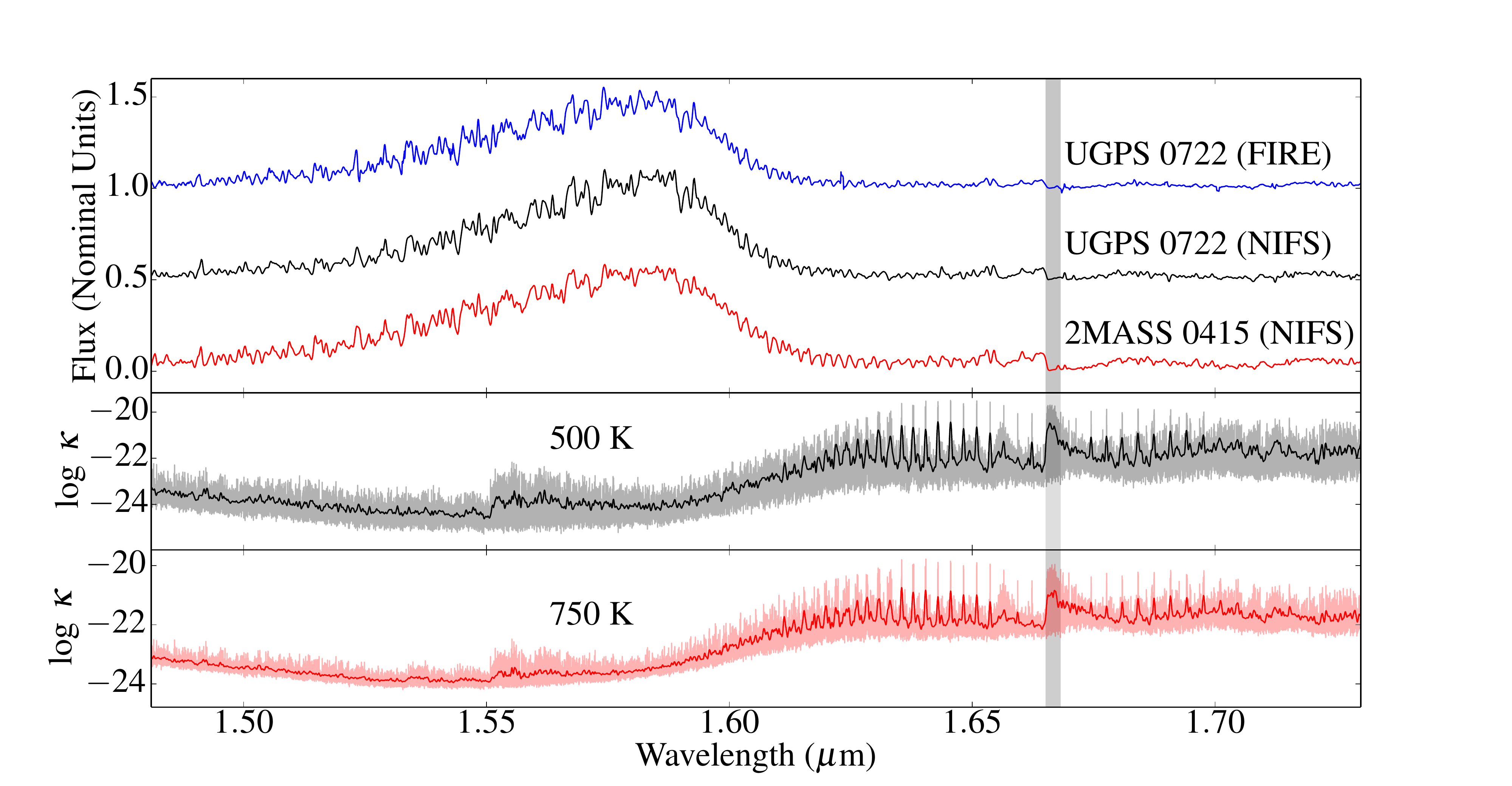} 
  \caption{CH$_{4}$ absorption in the $H$-band spectra of 2MASS~0415 (red), UGPS~0722 (black), and the Magellan/FIRE spectrum of UGPS 0722 (blue).  
  The shaded region indicates the Q-branch starting at $\sim$1.665~$\mu$m. The centre and lower graphs 
  contains the cross-sections as described in Figure \ref{fig:jlong}. (T dwarf spectra have been offset 
  to aid identification of spectral features.)}
  \label{fig:hlong}
\end{minipage}
\end{figure*}

\begin{figure*}
\centering 
\subfigure{
  \hspace*{0.3cm}\includegraphics[scale=0.40, angle=0]{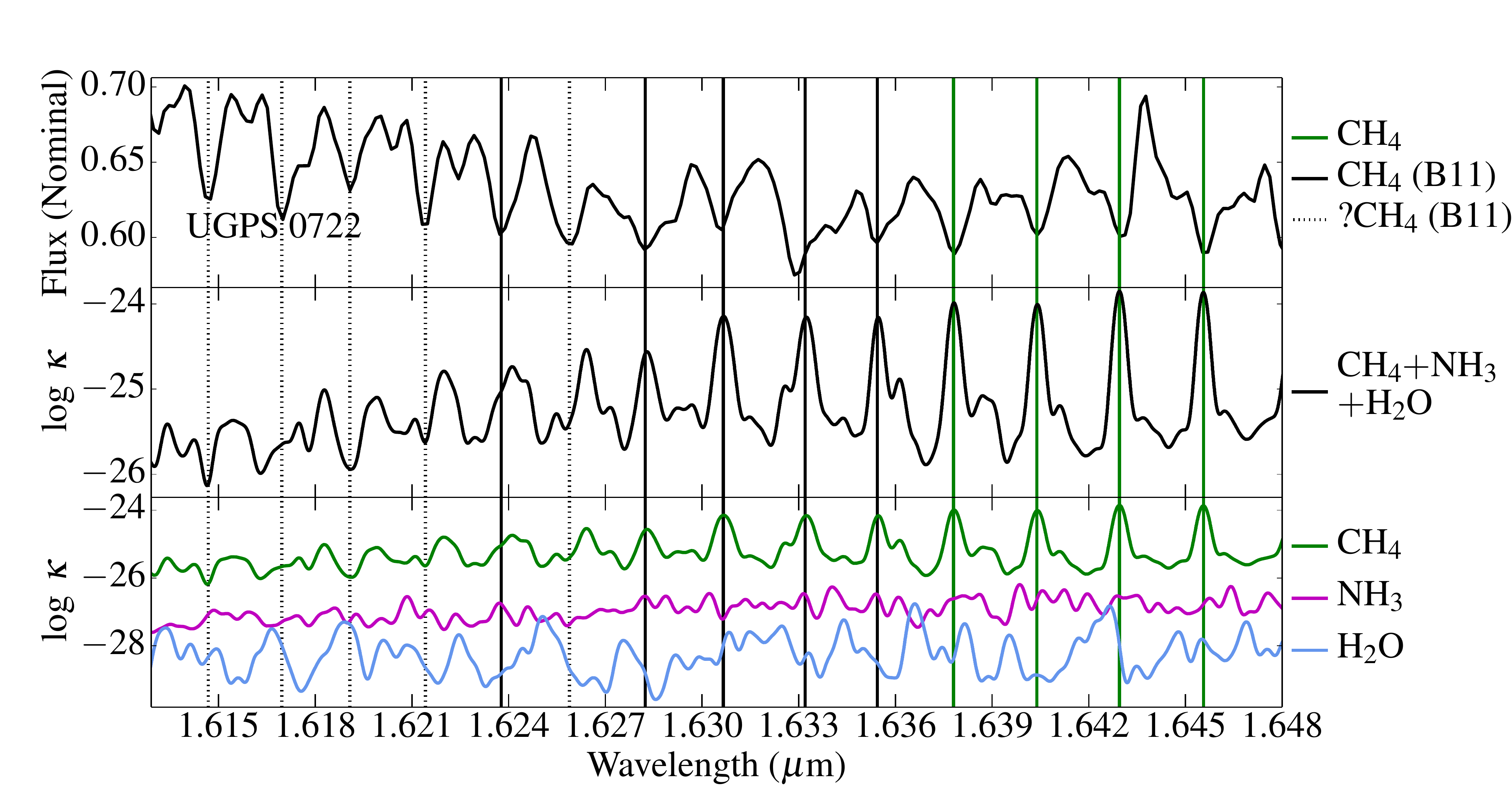} 
                 }                                                           
\subfigure{
  \hspace*{-0.6cm}\includegraphics[scale=0.375, angle=0]{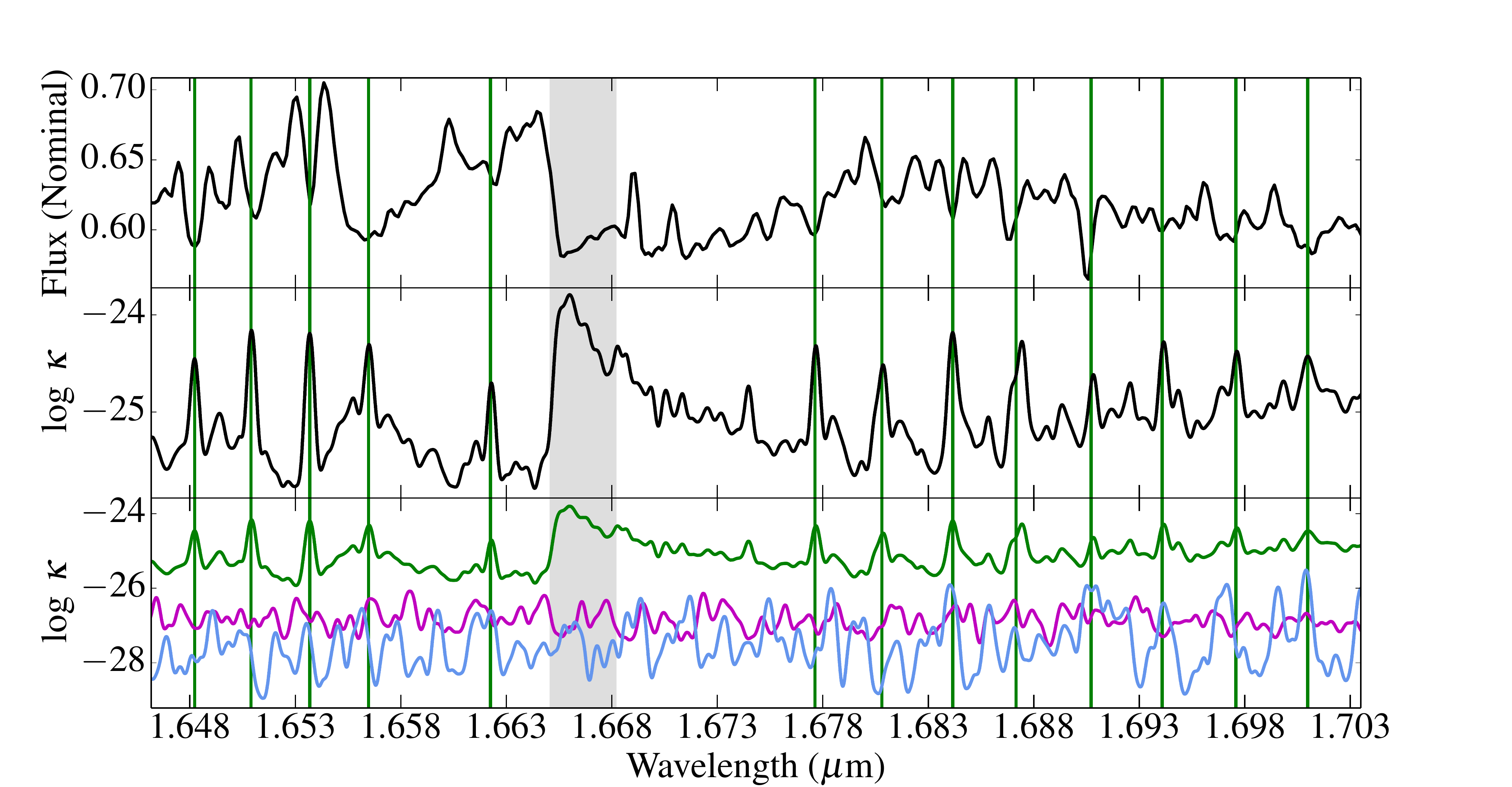}
   \label{fig:subfig10}
                 }
  \caption{CH$_{4}$ absorption features in the $H$-band spectrum of UGPS~0722 (see Table \ref{tab:h}). 
  A large number of new methane features are detected (green lines) and several previously identified 
  features are recovered (solid black lines). The black dotted lines are the locations of features which 
  B11 identified as CH$_{4}$ absorption features but which do not correspond to peaks in the CH$_{4}$ opacity. The 
  proximity of the features to peaks in the CH$_{4}$ opacity suggest that they are methane features. (See the text 
  for more discussion of these features). The shaded region is the Q-branch centred at $\sim$1.665 $\mu$m. 
  Other features are as those described in Figure \ref{fig:0722j_0415j}.}
  \label{fig:0722h}
\end{figure*}

\begin{figure*}
\centering 
\subfigure{
  \hspace*{0.3cm}\includegraphics[scale=0.40, angle=0]{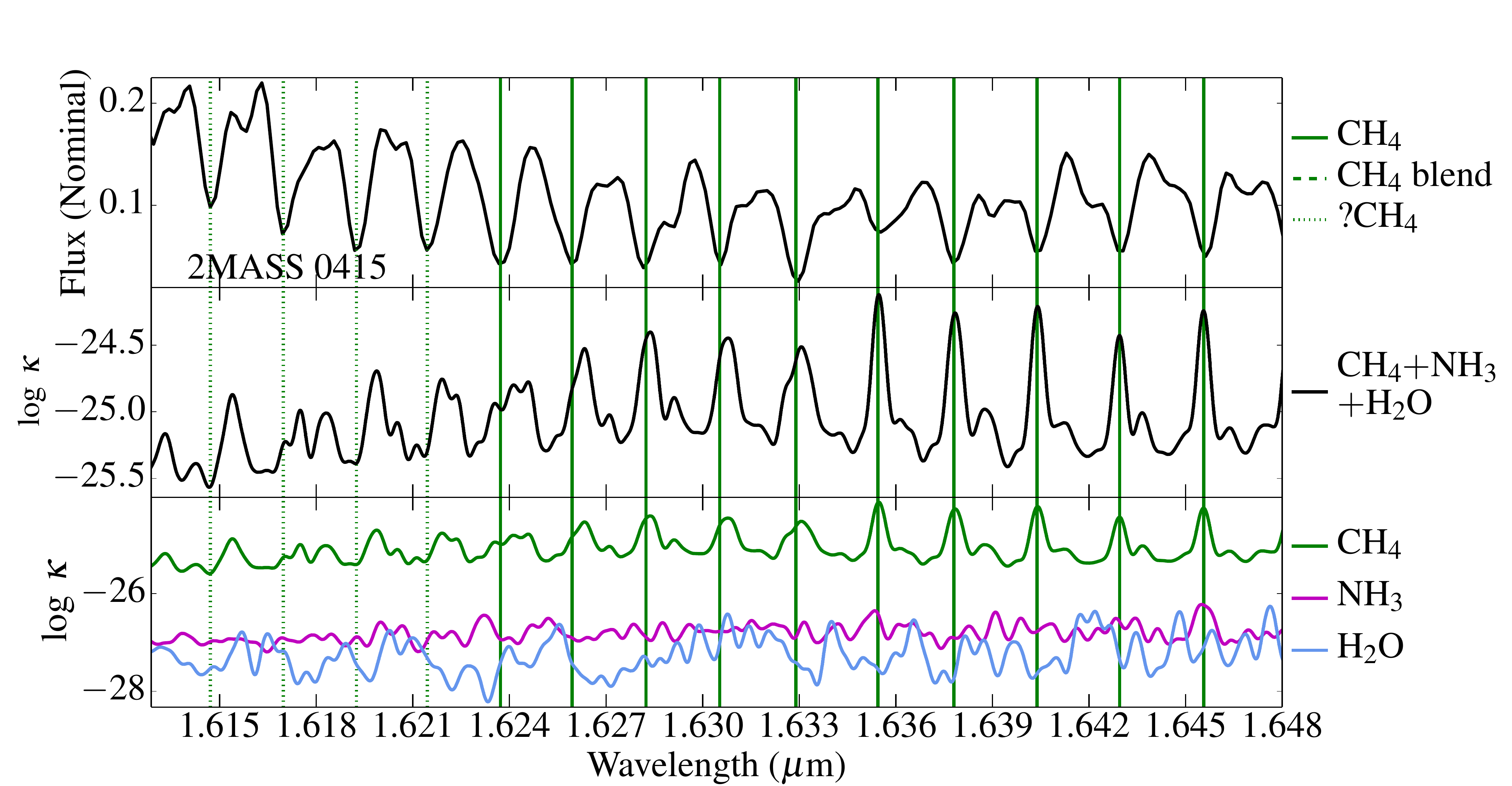} 
                }                                                           
\subfigure{
  \hspace*{-0.6cm}\includegraphics[scale=0.375, angle=0]{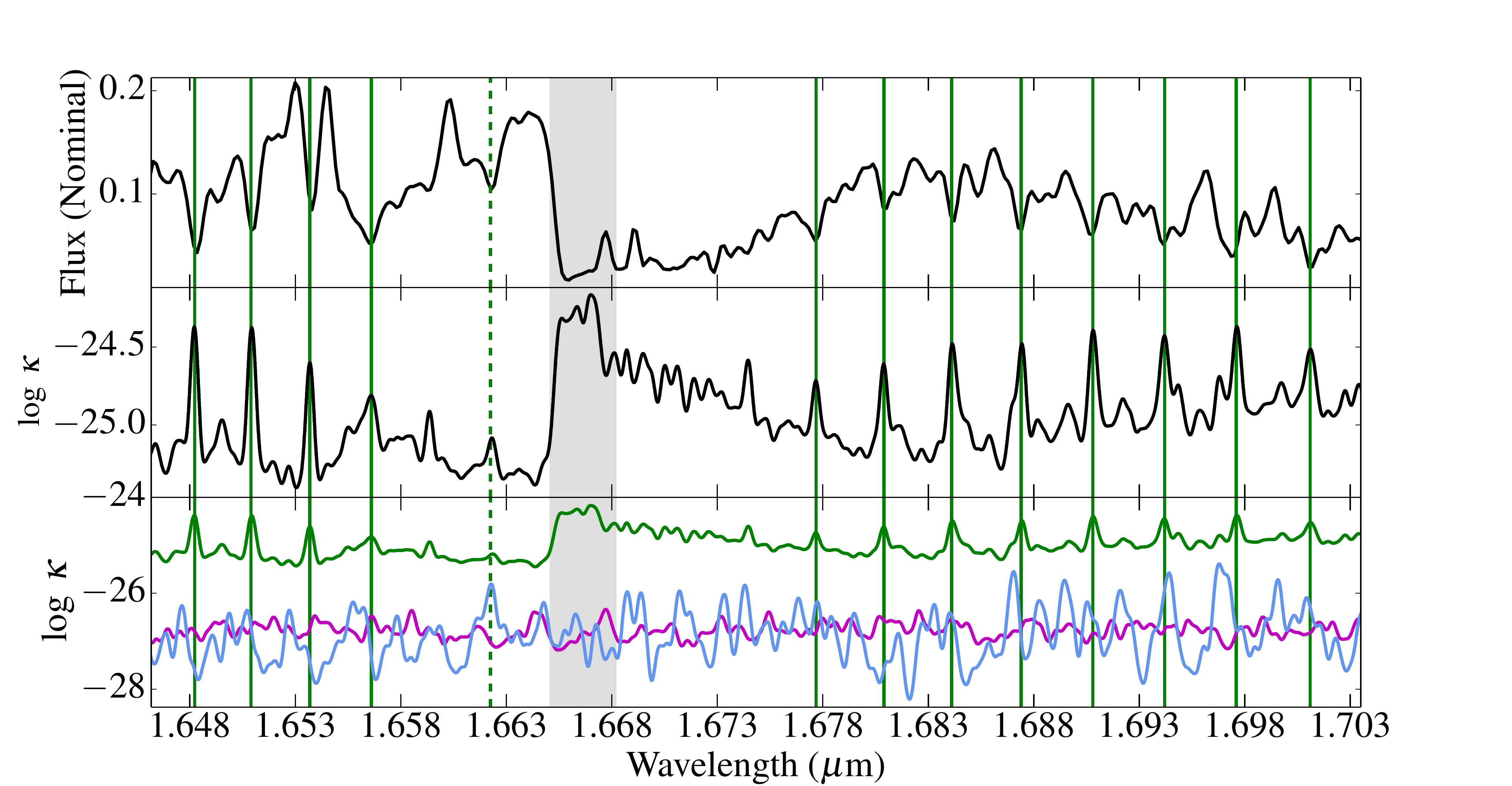}
   \label{fig:subfig10}
                 }
  \caption{CH$_{4}$ absorption features in 
  the $H$-band spectrum of 2MASS~0415 (see Table \ref{tab:h}).}
  \label{fig:0415h}
\end{figure*}

We have found that $J$-band absorption features in both T dwarfs' spectra are produced by R-branch 
line transitions belonging to the $\nu_{2}$ + 2$\nu_{3}$ vibrational band.
Table \ref{tab:ch4_j_rbranch} shows the changes in symmetries and ro-vibrational quantum numbers 
of the strongest line transitions for the CH$_{4}$ absorption features we have detected in the $J$-band 
spectra of UGPS~0722 and 2MASS~0415. Tables A1-A5 and B1-B6 showing the changes in symmetries and ro-vibrational quantum 
numbers of the strongest line transitions responsible, respectively, for the CH$_{4}$ and NH$_{3}$ 
absorption features we have detected in the near-infrared spectra of each T dwarf can be found in the 
electronic version of this paper.



\subsection{The $H$-Band}
\label{sec:ch4h}
The CH$_{4}$ absorption cross-sections at 500~K and 750~K show CH$_{4}$ opacity 
reaching a maximum on the long side of the $H$-band flux peak, producing regular patterns 
of peaks, separated by $\sim$0.002~$\mu$m, from $\sim$1.61-1.70~$\mu$m. 
The spectrum has a strong Q-branch composed of many transitions (see Figure \ref{fig:hlong}). 
A comparison of the observed and modelled spectral energy 
distributions (SEDs) of three T7.5-8 dwarfs~(\citealt{saumon06};~\citealt{saumon07}), ascribed 
a divergence in the model SEDs at 1.6-1.7~$\mu$m to an incomplete CH$_{4}$ line list. 
We expect that this discrepancy will be resolved with the inclusion of the 10to10 line list in
model spectra.

B11 identified CH$_{4}$ absorption features at 1.6145~$\mu$m, 1.6168~$\mu$m, 1.6191~$\mu$m, 
1.6213~$\mu$m, and 1.6258~$\mu$m. While CH$_{4}$ is the dominant opacity source in 
this wavelength range, the absorption cross-sections at 500~K show no peaks in the 
CH$_{4}$ opacity corresponding to these wavelengths. The same is true for the CH$_{4}$ 
absorption cross-sections made at 750~K, apart from the feature at 1.6258~$\mu$m (see Figures \ref{fig:0722h} and \ref{fig:0415h}). 
However, the absorption features in the T dwarfs' spectra in this wavelength region are a clear extension of 
the pattern in absorption features at longer wavelengths in the data, and are only very slightly off-set 
from peaks in the CH$_{4}$ opacity. We suspected that these were methane features, and the 
failure of the opacity cross-sections to correspond with these features was most likely due to uncertainties in the 
10to10 line list in this region. To examine this, the ExoMol group compiled a hybrid version of the 10to10 line list 
where some of the line positions, including lines in this region, were replaced with 
experimental values. This line list was then used to generate absorption intensities in the wavelength 
range 1.615-1.710~$\mu$m at T=500~K. The hybrid spectrum (not shown) produces peaks in the opacity which 
are in better agreement with the data, particularly at 1.6213~$\mu$m, and 1.6258~$\mu$m, indicating 
that the 10to10 features in this region are affected by deficiencies in ExoMol's theoretical model.

B11 did not identify CH$_{4}$ absorption features longward of 1.6354~$\mu$m. We have identified 
numerous CH$_{4}$ absorption features either side of the CH$_{4}$ Q-branch starting at 
$\sim$1.6650~$\mu$m. See Figure \ref{fig:0722h} and Table \ref{tab:h}. 
In this region of the $H$-band, the CH$_{4}$ absorption cross-sections are $\sim$2 magnitudes 
stronger than other opacity sources, and all the features we have identified are ``pure" CH$_{4}$ absorption features.

While the $J$-band absorption features we observed are from the R-branch, $H$-band absorption 
features in both T dwarfs are produced by ro-vibrational transitions in the R-, P-, and Q-branches. 
We did not identify isolated Q-branch absorption features. Instead, we note that the cluster of transition 
lines forming the Q-branch correspond to broad absorption troughs in the T dwarfs' spectra between 
$\sim$1.6651-1.6682~$\mu$m.  These and the identified absorption features in 2MASS~0415 
and UGPS~0722 belong to the 2$\nu_{3}$ vibrational band. 
We consider that the pattern of strong peaks in the methane opacity in the $H$-band is most likely 
due to the 2$\nu_{3}$ band being an allowed asymmetric stretching band with a non-zero dipole moment, in contrast, 
for example, to the 2$\nu_{1}$ stretching band, which is also an overtone.
While the features at 1.6976~$\mu$m 
and 1.7010~$\mu$m are largely due to P-branch transitions, each feature does contain an R-branch 
transition line. The R-branch lines have a similar intensity to the P-branch transition lines and 
belong to the $\nu_{2}$+$\nu_{3}$+$\nu_{4}$ vibrational band  (1.6976~$\mu$m) and $\nu_{1}$+$\nu_{3}$ 
vibrational band (1.7010~$\mu$m). We consider it likely that these two lines are outlying members 
of the next set of R-branch transition lines.

\begin{table*}
\small
\centering
\newcommand{\footstar}[1]{$^*$ \footnotetext{$^*$#1}}
\caption{CH$_{4}$ Absorption Features in the $H$ Band Spectra of Late T Dwarfs (see Figures \ref{fig:0722h} and \ref{fig:0415h})}  
\begin{tabular}{c c c c c c c } 
\vspace{-1.1em}\\
& Source & $\lambda$ ($\mu$m) & Opacity Source (B11) & Opacity Source (500~K/750~K) & \head{2.5cm}{Absorption feature in UGPS~0722} & \head{2.5cm}{Absorption feature in 2MASS~0415}\\ [0.5ex] 
\rowcolor{bubblegum}
& B11          & 1.6145 &  CH$_{4}$                       & CH$_{4}$(?)/CH$_{4}$(?)      & Yes    &  Yes\\
\rowcolor{bubblegum}
& B11          & 1.6168 &  CH$_{4}$                       & CH$_{4}$(?)/CH$_{4}$(?)      & Yes    &  Yes\\
\rowcolor{bubblegum}
& B11          & 1.6191 &  CH$_{4}$                       & CH$_{4}$(?)/CH$_{4}$(?)      & Yes    &  Yes\\
\rowcolor{bubblegum}
& B11          & 1.6213 &  CH$_{4}$                       & CH$_{4}$(?)/CH$_{4}$(?)      & Yes   &   Yes\\
\rowcolor{bubblegum}
& B11          & 1.6236 &  CH$_{4}$                       & CH$_{4}$/CH$_{4}$          &  Yes   &   Yes\\
\rowcolor{bubblegum}
& B11          & 1.6258 &  CH$_{4}$                       & CH$_{4}$(?)/CH$_{4}$                 & Yes    & Yes\\
\rowcolor{bubblegum}
& B11          & 1.6282 &  CH$_{4}$                       & CH$_{4}$/CH$_{4}$                & Yes & Yes\\
\rowcolor{bubblegum}
& B11          & 1.6307 &  CH$_{4}$                       & CH$_{4}$/CH$_{4}$                & Yes & Yes\\
\rowcolor{bubblegum}
& B11          & 1.6332 &  CH$_{4}$                       & CH$_{4}$/CH$_{4}$                & Yes & Yes\\
\rowcolor{bubblegum}
& B11          & 1.6354 &  CH$_{4}$                       & CH$_{4}$/CH$_{4}$                & Yes & Yes\\
\rowcolor{bubblegum}
& This work & 1.6378 &  ---                               & CH$_{4}$/CH$_{4}$                 & Yes & Yes\\
\rowcolor{bubblegum}
& This work & 1.6404 &  ---                               & CH$_{4}$/CH$_{4}$                 & Yes & Yes\\
\rowcolor{bubblegum}
& This work & 1.6430 &  ---                               & CH$_{4}$/CH$_{4}$                 & Yes & Yes\\
\rowcolor{bubblegum}
& This work & 1.6456 &  ---                               & CH$_{4}$/CH$_{4}$                 & Yes & Yes\\
\rowcolor{bubblegum}
& This work & 1.6482 &  ---                               & CH$_{4}$/CH$_{4}$                 & Yes & Yes\\
\rowcolor{bubblegum}
& This work & 1.6509 &  ---                               & CH$_{4}$/CH$_{4}$                 & Yes & Yes\\
\rowcolor{bubblegum}
& This work & 1.6537 &  ---                               & CH$_{4}$/CH$_{4}$                 & Yes & Yes\\
\rowcolor{bubblegum}
& This work & 1.6565 &  ---                               & CH$_{4}$/CH$_{4}$                 & Yes & Yes\\
\rowcolor{bubblegum}
& This work & 1.6623 &  ---                               & CH$_{4}$/(CH$_{4}$+H$_{2}$O)                 & Yes & Yes\\
\rowcolor{gray}
& & & & & &\\
\rowcolor{gray}
& This work & $\sim$(1.6651$\rightarrow$1.6682) &  ---                               & CH$_{4}$/CH$_{4}$                 & Yes & Yes\\
\rowcolor{gray}
& & & & & &\\
\rowcolor{beaublue}
& This work & 1.6776 &  ---                               & CH$_{4}$/CH$_{4}$                 & Yes & Yes\\
\rowcolor{beaublue}
& This work & 1.6808 &  ---                               & CH$_{4}$/CH$_{4}$                 & Yes & Yes\\
\rowcolor{beaublue}
& This work & 1.6840 &  ---                               & CH$_{4}$/CH$_{4}$                 & Yes & Yes\\
\rowcolor{beaublue}
& This work & 1.6874 &  ---                               & CH$_{4}$/CH$_{4}$                 & Yes & Yes\\
\rowcolor{beaublue}
& This work & 1.6907 &  ---                               & CH$_{4}$/CH$_{4}$                 & Yes & Yes\\
\rowcolor{beaublue}
& This work & 1.6941 &  ---                               & CH$_{4}$/CH$_{4}$                 & Yes & Yes\\
\rowcolor{beaublue}
& This work & 1.6976 &  ---                               & CH$_{4}$/CH$_{4}$                 & Yes & Yes\\
\rowcolor{beaublue}
& This work & 1.7010 &  ---                               & CH$_{4}$/CH$_{4}$                & Yes & Yes\\
\end{tabular}
\label{tab:h}
\end{table*}

\begin{figure*}
\centering
\begin{minipage}{\linewidth}
   \includegraphics[scale=0.4]{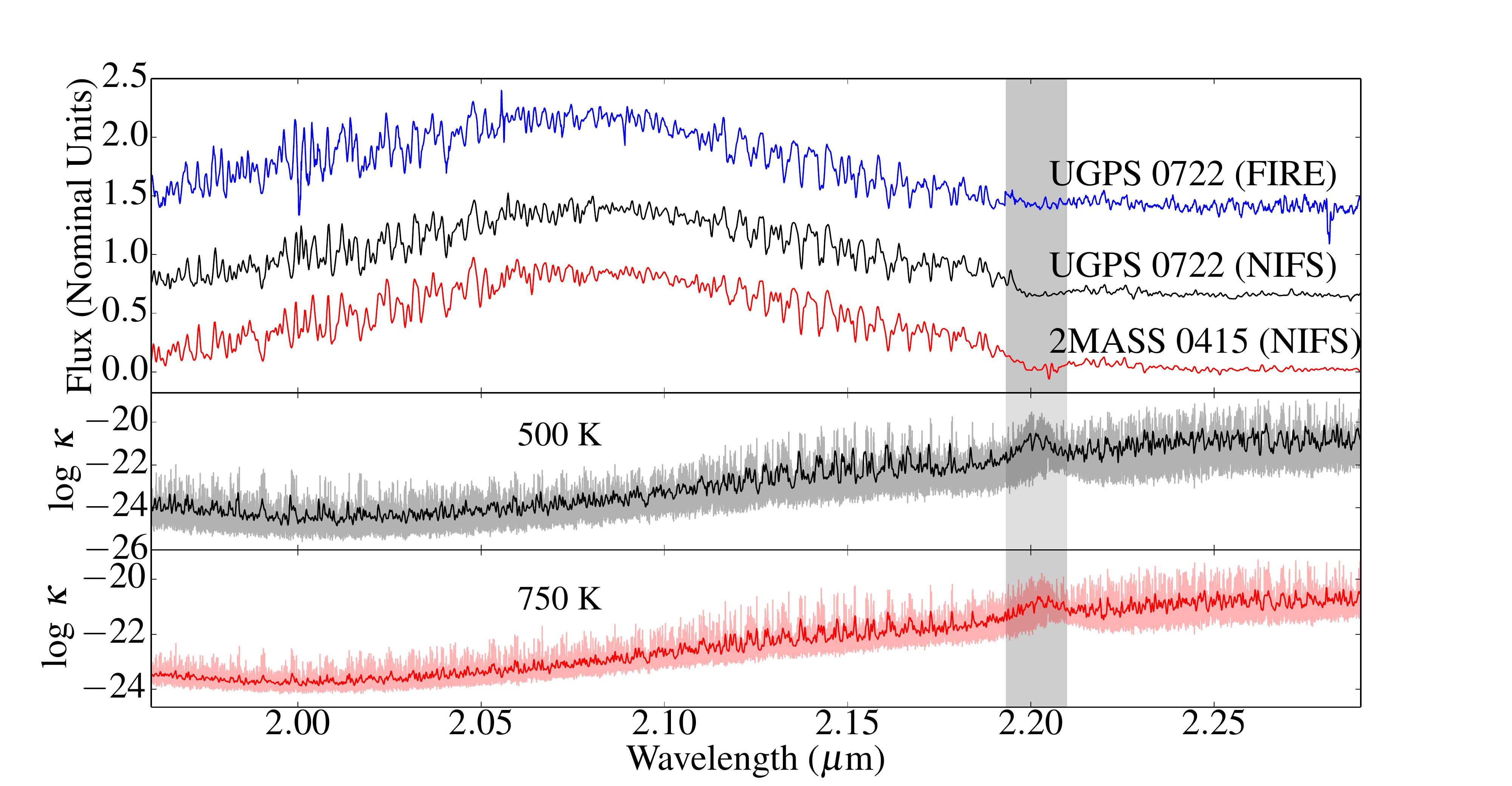} 
  \caption{CH$_{4}$ absorption in the $K$-band spectra of 2MASS~0415 (red), UGPS~0722 (black), and the Magellan/FIRE spectrum of UGPS 0722 (blue).  
  The shaded region indicates the Q-branch transitions centred at $\sim$2.20~$\mu$m. 
  The lower graphs contains the cross-sections as described in Figure \ref{fig:jlong}. 
  (T dwarf spectra have been offset to aid identification of spectral features.)}
  \label{fig:klong}
\end{minipage}
\end{figure*}

\begin{figure*}
\centering  
\subfigure{
    \hspace*{0.8cm}\includegraphics[scale=0.365, angle=0]{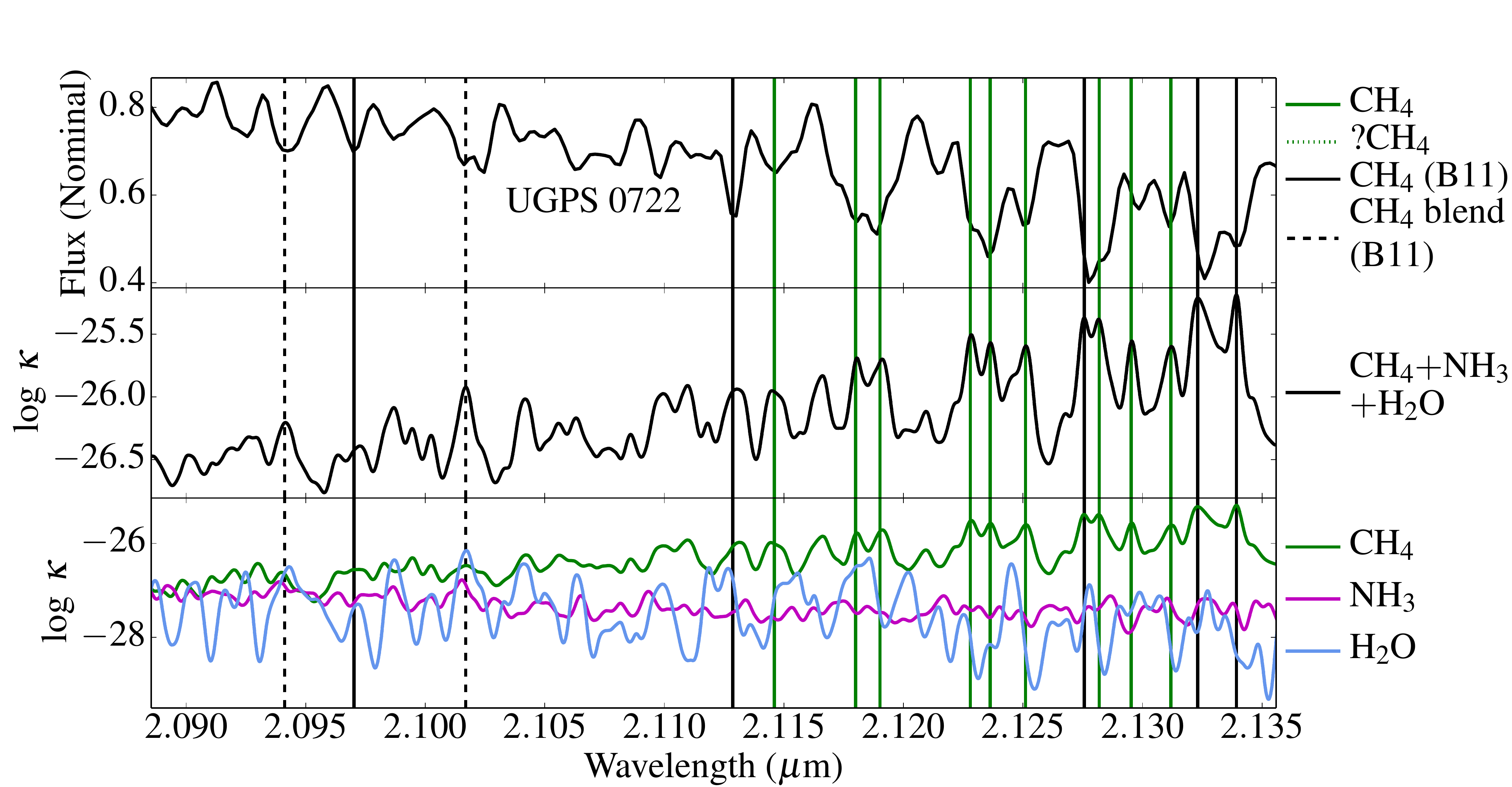} 
   \label{fig:subfig11}
                 }                                                           
\subfigure{
   \includegraphics[scale=0.34, angle=0]{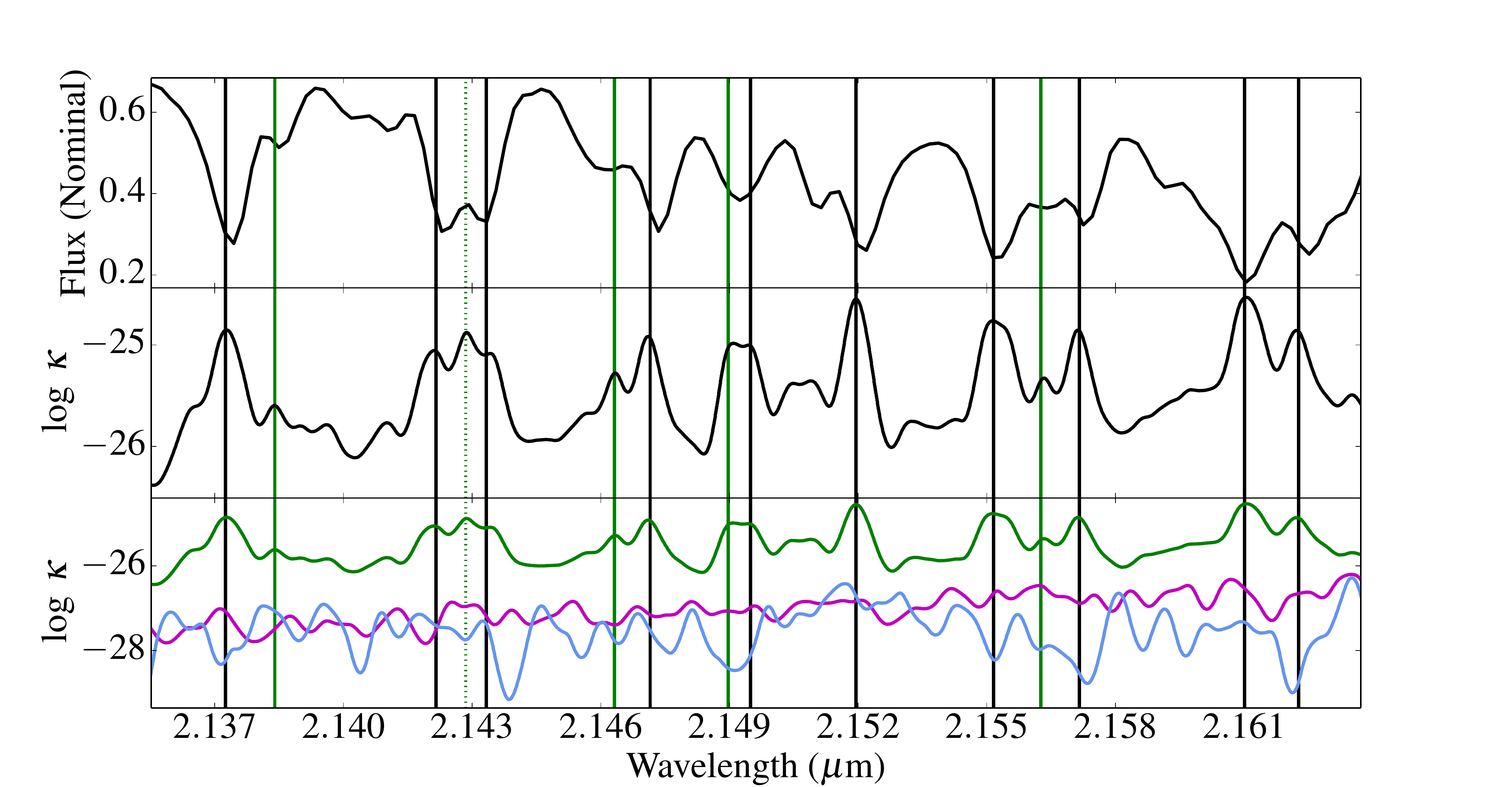} 
   \label{fig:subfig12}
                 }
                 \subfigure{
   \includegraphics[scale=0.34, angle=0]{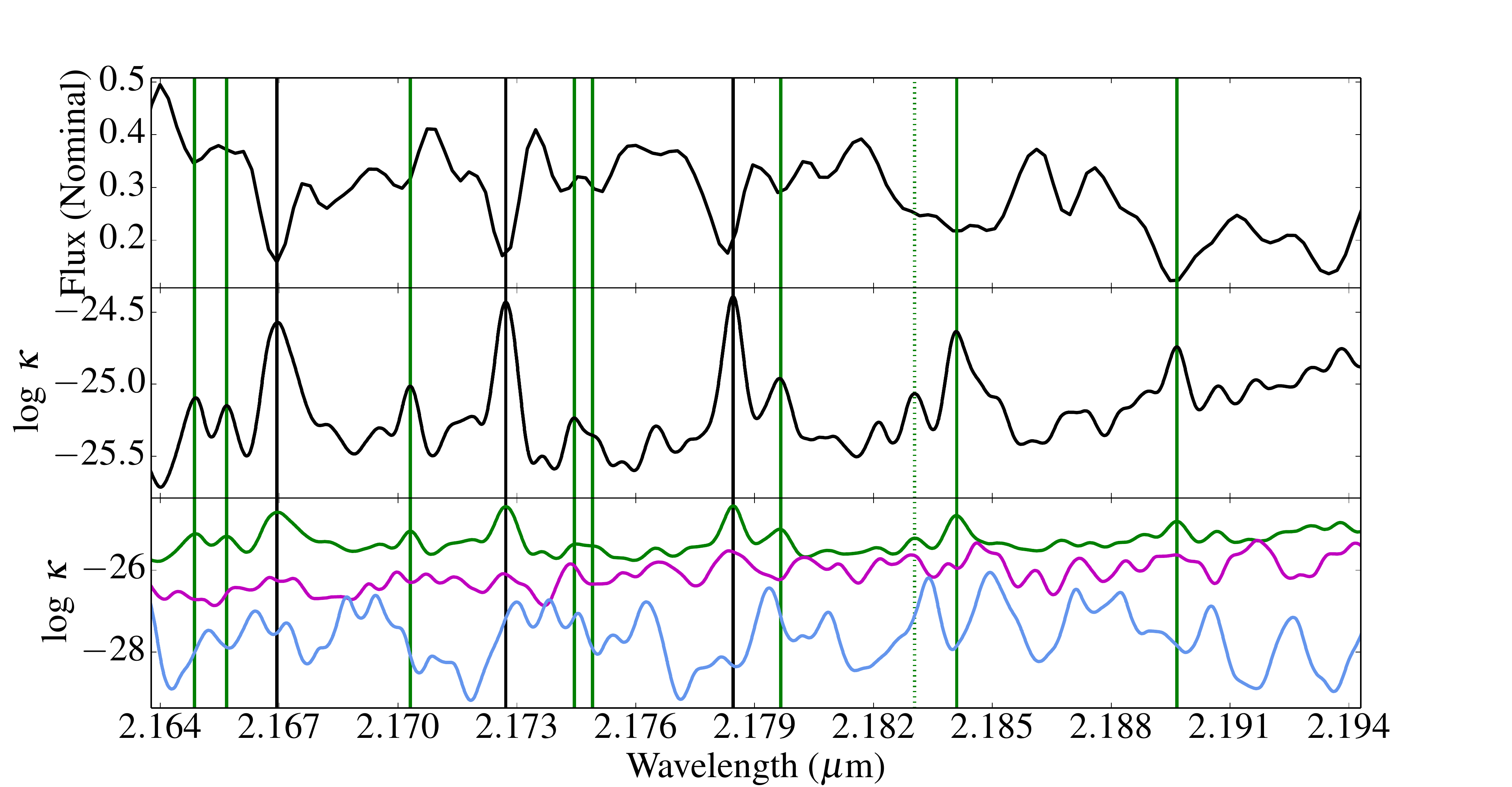} 
   \label{fig:subfig13}
                                 }
  \caption{CH$_{4}$ absorption features in the $K$-band spectrum of UGPS~0722 (see Table \ref{tab:ch4k}).
  Solid black lines are features previously detected by B11. Solid green lines are features identified 
  in this work. Black dashed lines are features identified by B11 as CH$_{4}$ features which we 
  find to be due to CH$_{4}$ and one or more other molecular species.}
  \label{fig:ch4_0722k}
\end{figure*}

\begin{figure*}
\centering  
\subfigure{
   \hspace*{0.8cm}\includegraphics[scale=0.365, angle=0]{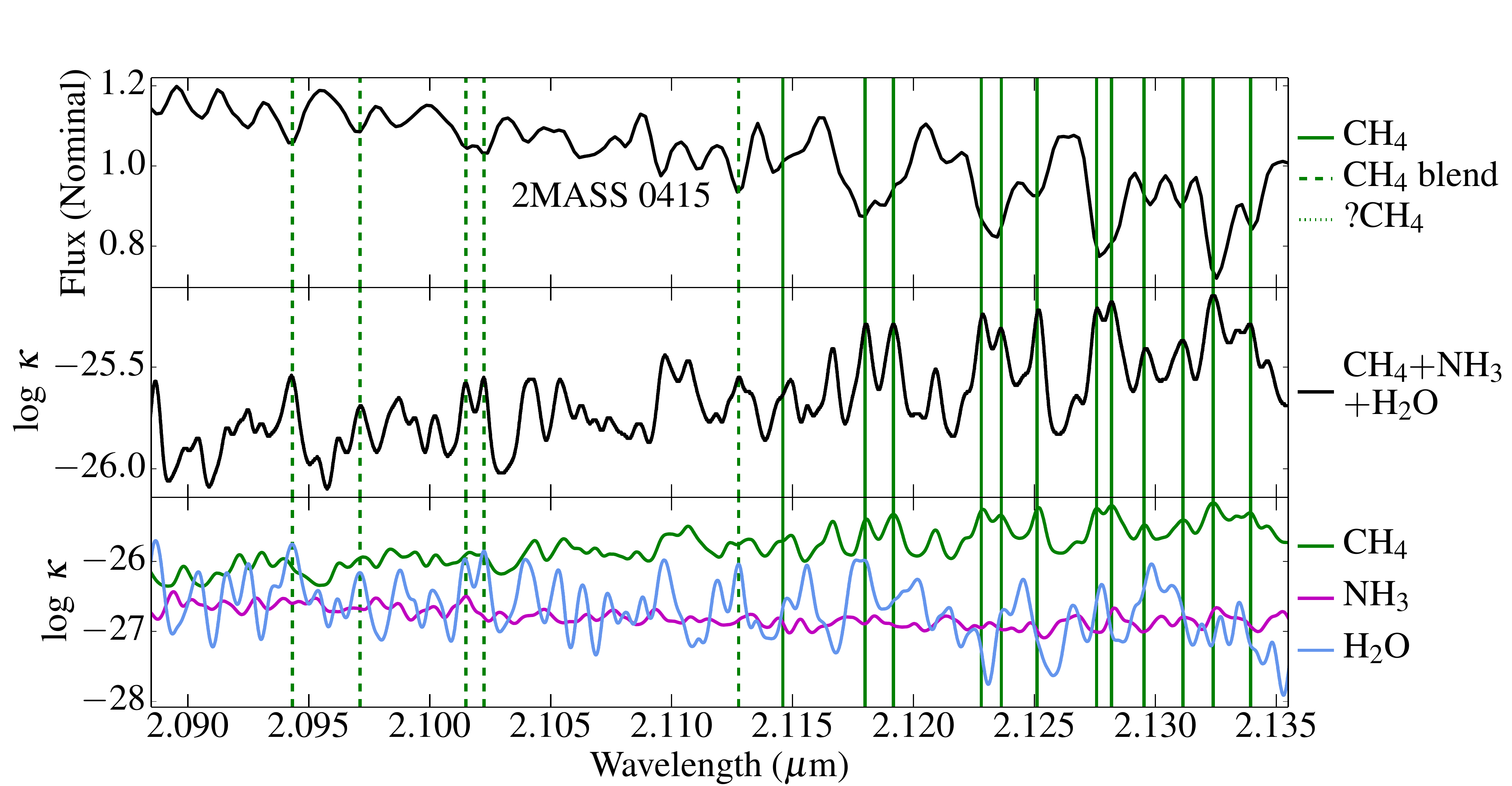} 
   \label{fig:subfig11}
                 }                                                           
\subfigure{
   \includegraphics[scale=0.34, angle=0]{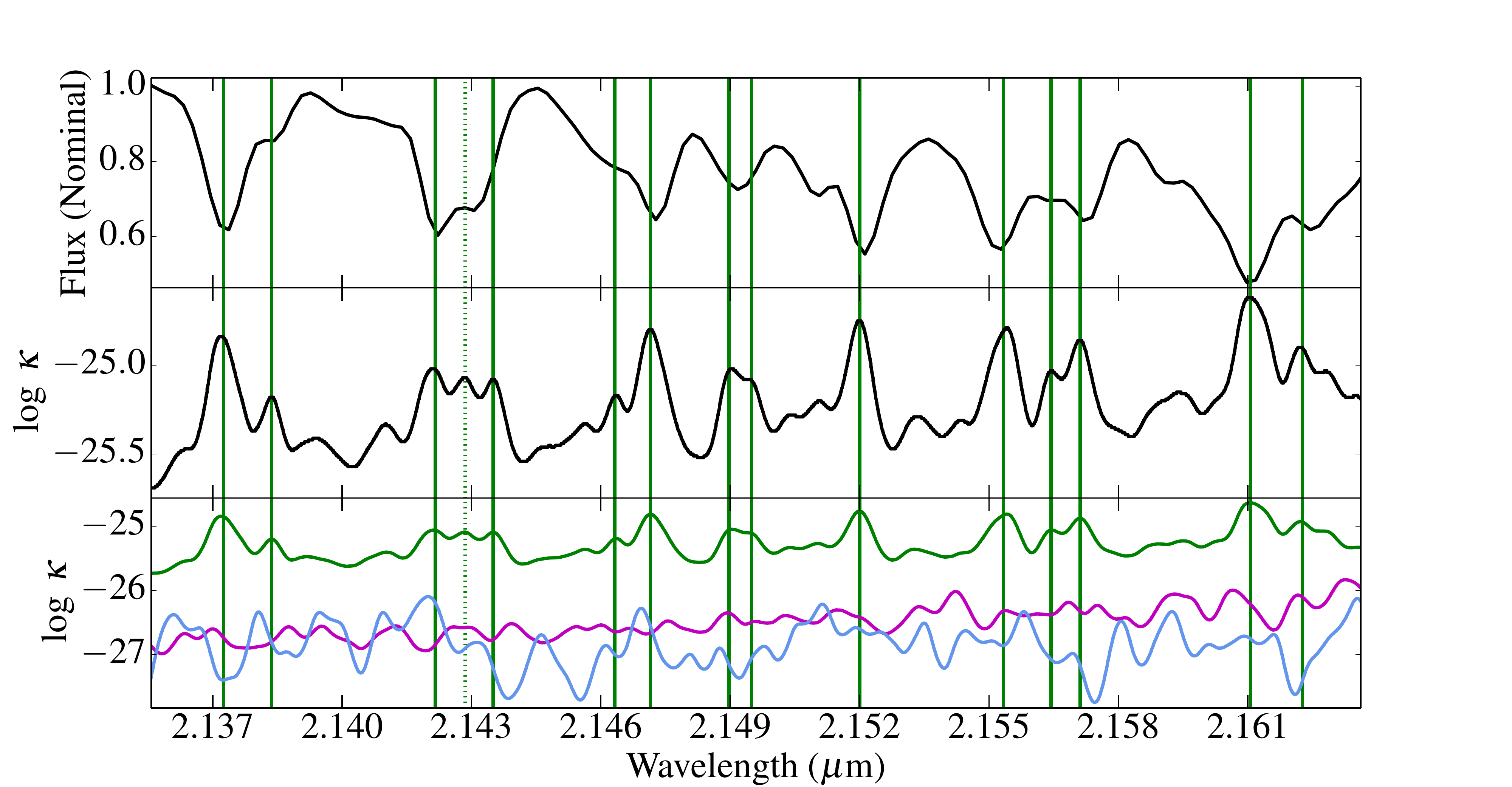} 
   \label{fig:subfig12}
                 }
                 \subfigure{
   \includegraphics[scale=0.34, angle=0]{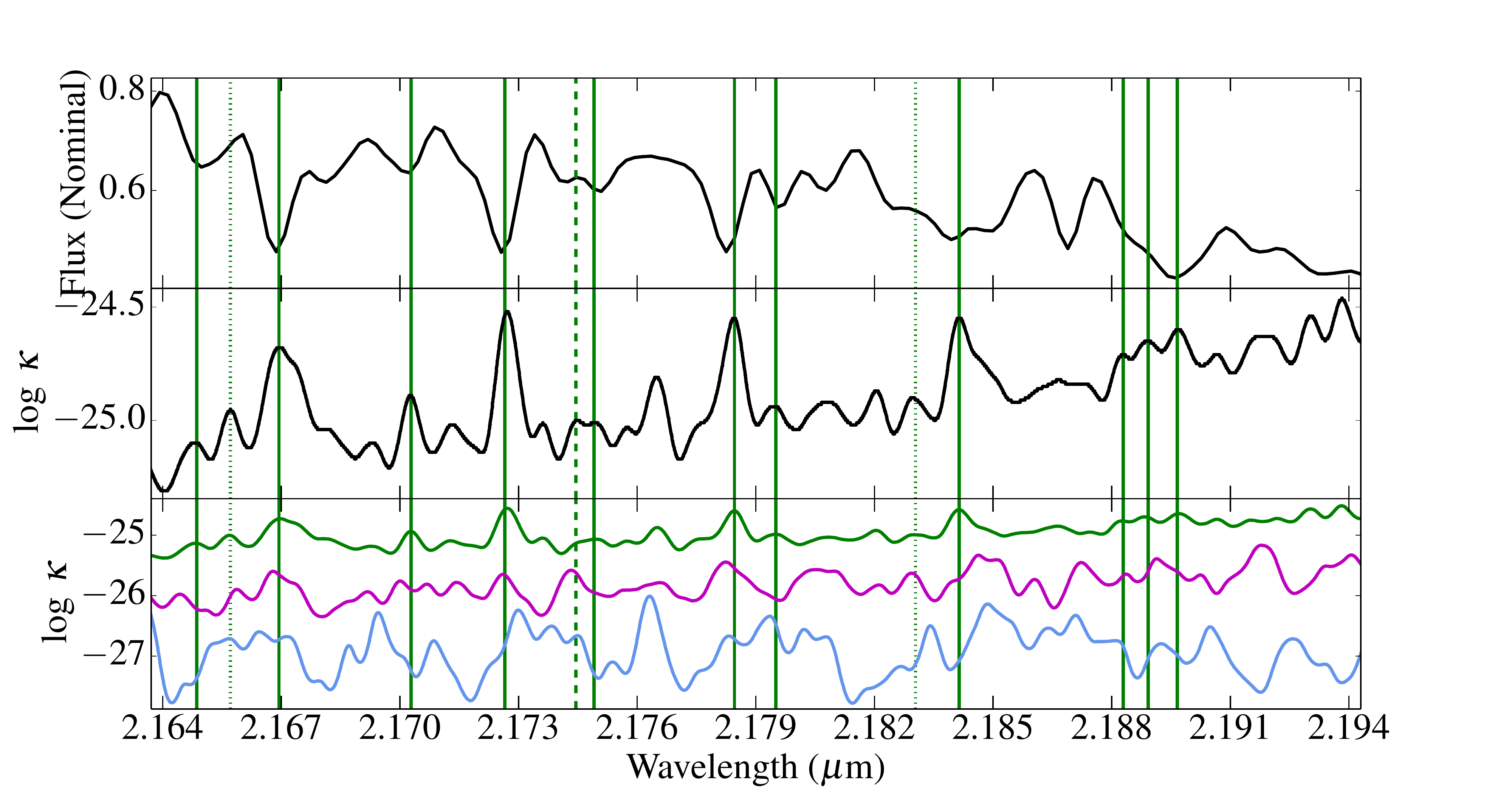} 
   \label{fig:subfig13}
                                 }
  \caption{CH$_{4}$ features in the $K$-band spectrum of 2MASS 0415 (see Table \ref{tab:ch4k}).}
  \label{fig:ch4_0415k}
\end{figure*}

\begin{table*}
\begin{threeparttable}
\small
\centering
\newcommand{\footstar}[1]{$^*$ \footnotetext{$^*$#1}}
\caption{CH$_{4}$ Absorption Features in the $K$ Band Spectra of Late T Dwarfs (see Figures \ref{fig:ch4_0722k} and \ref{fig:ch4_0415k})}  
\begin{tabular}{c c c c c c c } 
\vspace{-1.1em}\\
& Source & $\lambda$ ($\mu$m) & Opacity Source (B11) & Opacity Source (500~K/750~K) & \head{2.5cm}{Absorption feature in UGPS~0722} & \head{2.5cm}{Absorption feature in 2MASS~0415}\\ [0.5ex] 
\rowcolor{bubblegum}
& B11                               & 2.0943                    &  CH$_{4}$                                 & (CH$_{4}$+H$_{2}$O+NH$_{3}$)/(H$_{2}$O+CH$_{4}$)       & Yes & Yes\\
\rowcolor{bubblegum}
& B11                               & 2.0971                    &  CH$_{4}$                                 & CH$_{4}$/(CH$_{4}$+H$_{2}$O)                                         & Yes & Yes\\
\rowcolor{bubblegum}
& B11                               & 2.1017\tnote{\textdagger}           & CH$_{4}$          & (CH$_{4}$+H$_{2}$O+NH$_{3}$)/(H$_{2}$O+CH$_{4}$)         & Yes &  Yes\\
\rowcolor{bubblegum}
& B11                               & 2.1129                    & CH$_{4}$                                 & CH$_{4}$/(CH$_{4}$+H$_{2}$O)                                         & Yes & Yes\\
\rowcolor{bubblegum}
& This work                      & 2.1146                    &  ---                                         & CH$_{4}$/CH$_{4}$                                                           & Yes & Yes\\
\rowcolor{bubblegum}
& This work                      & 2.1180                    & ---                                          & CH$_{4}$/CH$_{4}$                                                           & Yes & Yes\\
\rowcolor{bubblegum}
& This work                      & 2.1190                    & ---                                          & CH$_{4}$/CH$_{4}$                                                           & Yes & Yes\\
\rowcolor{bubblegum}
& This work                      & 2.1228                    & ---                                          & CH$_{4}$/CH$_{4}$                                                           & Yes & Yes\\
\rowcolor{bubblegum}
& This work                      & 2.1236                    &  ---                                         & CH$_{4}$/CH$_{4}$                                                           & Yes & Yes\\
\rowcolor{bubblegum}
& This work                      & 2.1251                    &  ---                                         & CH$_{4}$/CH$_{4}$                                                           & Yes & Yes\\
\rowcolor{bubblegum}
& B11                               & 2.1276                    &  CH$_{4}$                                 & CH$_{4}$/CH$_{4}$                                                           & Yes & Yes\\
\rowcolor{bubblegum}
& This work                      & 2.1282                    &  ---                                          & CH$_{4}$/CH$_{4}$                                                           & Yes & Yes\\
\rowcolor{bubblegum}
& This work                      & 2.1295                    & ---                                           & CH$_{4}$/CH$_{4}$                                                           & Yes & Yes\\
\rowcolor{bubblegum}
& This work                      & 2.1312                    &  ---                                          & CH$_{4}$/CH$_{4}$                                                           & Yes & Yes\\
\rowcolor{bubblegum}
& B11                               & 2.1323                    &  CH$_{4}$                                 & CH$_{4}$/CH$_{4}$                                                           & Yes & Yes\\
\rowcolor{bubblegum}
& B11                               & 2.1339                    &  CH$_{4}$                                 & CH$_{4}$/CH$_{4}$                                                           & Yes & Yes\\
\rowcolor{bubblegum}
& B11                               & 2.1373                    &  CH$_{4}$                                 & CH$_{4}$/CH$_{4}$                                                           & Yes & Yes\\
\rowcolor{bubblegum}
& This work                     & 2.1384                    &  ---                                              & CH$_{4}$/CH$_{4}$                                                           & Yes & Yes(?)\\
\rowcolor{bubblegum}
& B11                               & 2.1421                    &  CH$_{4}$                                 & CH$_{4}$/CH$_{4}$                                                           & Yes & Yes\\
\rowcolor{bubblegum}
& This work                      & 2.1429                    &   ---                                         & CH$_{4}$/CH$_{4}$                                                           & Yes(?) & Yes(?)\\
\rowcolor{bubblegum}
& B11                               & 2.1433                    &  CH$_{4}$                                 & CH$_{4}$/CH$_{4}$                                                            & Yes & Yes\\
\rowcolor{bubblegum}
& This work                      & 2.1463                    & ---                                           & CH$_{4}$/CH$_{4}$                                                            & Yes & Yes\\
\rowcolor{bubblegum}
& B11                               & 2.1472                    &  CH$_{4}$                                  & CH$_{4}$/CH$_{4}$                                                            & Yes & Yes\\
\rowcolor{bubblegum}
& This work                      & 2.1490                    &  ---                                          & CH$_{4}$/CH$_{4}$                                                            & Yes & Yes\\
\rowcolor{bubblegum}
& B11                               & 2.1495                    &  CH$_{4}$                                 & CH$_{4}$/CH$_{4}$                                                             & Yes & Yes\\
\rowcolor{bubblegum}
& B11                               & 2.1520                    &  CH$_{4}$                                 & CH$_{4}$/CH$_{4}$                                                            & Yes & Yes\\
\rowcolor{bubblegum}
& B11                               & 2.1552                    &  CH$_{4}$                                 & CH$_{4}$/CH$_{4}$                                                             & Yes & Yes\\
\rowcolor{bubblegum}  
& This work                      & 2.1564                    &     ---                                       & CH$_{4}$/CH$_{4}$                                                             & Yes & Yes\\
\rowcolor{bubblegum}
& B11                               & 2.1572                    &  CH$_{4}$                                 & CH$_{4}$/CH$_{4}$                                                              & Yes & Yes\\
\rowcolor{bubblegum}
& B11                               & 2.1610                    &  CH$_{4}$                                 & CH$_{4}$/CH$_{4}$                                                              & Yes & Yes\\
\rowcolor{bubblegum}
& B11                               & 2.1623                    &  CH$_{4}$                                 & CH$_{4}$/CH$_{4}$                                                              & Yes & Yes\\
\rowcolor{bubblegum}
& This work                      & 2.1649                    & ---                                           & CH$_{4}$/CH$_{4}$                                                              & Yes & Yes\\
\rowcolor{bubblegum}
& This work                      & 2.1656                    &  ---                                          & CH$_{4}$/CH$_{4}$                                                              & Yes & Yes(?)\\
\rowcolor{bubblegum}
& B11                               & 2.1669                    &  CH$_{4}$                                  & CH$_{4}$/CH$_{4}$                                                              & Yes & Yes\\
\rowcolor{bubblegum}
& This work                     & 2.1703                    &  ---                                            & CH$_{4}$/CH$_{4}$                                                              & Yes & Yes\\
\rowcolor{bubblegum}
& B11                               & 2.1727                    &  CH$_{4}$                                  & CH$_{4}$/CH$_{4}$                                                              & Yes & Yes\\
\rowcolor{bubblegum}
& This work                      & 2.1745                    & ---                                            & CH$_{4}$/(CH$_{4}$+NH$_{3}$)                                          & Yes & Yes\\
\rowcolor{bubblegum}
& This work                     & 2.1749                     &  ---                                           & CH$_{4}$/CH$_{4}$                                                             & Yes & Yes\\
\rowcolor{bubblegum}
& B11                               & 2.1784                    &  CH$_{4}$                                   & CH$_{4}$/CH$_{4}$                                                             & Yes & Yes\\
\rowcolor{bubblegum}
& This work                      & 2.1797                    &  ---                                            & CH$_{4}$/CH$_{4}$                                                            & Yes & Yes\\
\rowcolor{bubblegum}
& This work                      & 2.1830                    &  ---                                            & CH$_{4}$/CH$_{4}$                                                            & Yes(?) & No\\
\rowcolor{bubblegum}
& This work                      & 2.1841                    &  ---                                            & CH$_{4}$/CH$_{4}$                                                            & Yes & Yes\\
\rowcolor{bubblegum}
& This work                      & 2.1897\tnote{\textdaggerdbl}                      &  ---        & CH$_{4}$/CH$_{4}$                                                            & Yes & Yes\\

\end{tabular}
\begin{tablenotes}
\item[\textdagger]In 2MASS 0415, two adjacent peaks in H$_{2}$O/CH$_{4}$ opacity coincide with absorption features at 2.1015~$\mu$m and 2.1023~$\mu$m.
\item[\textdaggerdbl]In 2MASS 0415, this feature corresponds to three adjacent peaks in CH$_{4}$ opacity at 2.1883~$\mu$m, 2.1889~$\mu$m and 2.1897~$\mu$m.
\end{tablenotes}
\label{tab:ch4k}\end{threeparttable}
\end{table*}



\subsection{The $K$-Band}
\label{sec:ch4k}

The CH$_{4}$ absorption cross-sections show CH$_{4}$ opacity increasing with increasing wavelength 
over the long side of the $K$-band flux peak, reaching a maximum at $\sim$2.20~$\mu$m 
(see Figure \ref{fig:klong}). The absorption cross-sections clearly show the 
Q-branch transitions centred at $\sim$2.20~$\mu$m. 

B11 detected a number of absorption features between $\sim$2.14~$\mu$m and $\sim$2.18~$\mu$m. 
Using the high quality of the Gemini/NIFS spectra and the greater accuracy of the 10to10 line list 
we have identified a number of new absorption features within this range (see Figure \ref{fig:ch4_0722k}  
and Table \ref{tab:ch4k}). Note that the features centred at $\sim$2.1278~$\mu$m and at $\sim$2.1429~$\mu$m 
are actually due to multiple peaks in the methane opacity which at this resolution (R$\sim$5000) are marginally resolved. 

The peak in the T dwarfs' $K$-band flux at $\sim$2.06~$\mu$m coincides with the start of increasing CH$_{4}$ 
opacity (see Figure \ref{fig:klong}). 
The continuum opacity of H$_{2}$ CIA is very high throughout the $K$-band [S12] and 
this will tend to veil narrow molecular features.
H$_{2}$ CIA is produced by ro-vibrational transitions in H$_{2}$ molecules, 
forming a series of bands corresponding to $\Delta\nu$=0, 1, 2, 3, etc. [S12]. The $\Delta\nu$=1 band 
is centred at $\sim$0.42~$\mu$m$^{-1}$ ($\sim$2.4~$\mu$m).  
The roles played by these mechanisms 
in the $K$-band spectra of late T dwarfs will be clearer once the 10to10 line list is included in model 
atmospheres. 

The shape of the $K$-band continuum is also affected by metallicity, but model fits to the SED
and spectrum of UGPS~0722, combined with its small space motion, indicate that it is a young object and 
therefore low metallicity is unlikely (\citealt{leggett12}; \citealt{morley12}, hereafter M12).

In T dwarfs, the $K$-band flux weakens with increasing spectral type, i.e. decreasing temperature. 
H$_{2}$ CIA is dependent on pressure (surface gravity) rather than temperature. On the other hand, 
methane absorption becomes stronger as the temperature drops. It is possible that methane absorption 
is sufficiently strong on its own to completely suppress the $K$-band in this region. For example, the 
Q-branch peak transition at $\sim$2.20~$\mu$m is an order of magnitude stronger than most of the 
transitions responsible for the absorption features detected by B11 and us. The Q-branch peak transition 
at $\sim$2.32~$\mu$m (not shown) is an order of magnitude stronger still.

All the identified absorption features in UGPS~0722 and 2MASS~0415 arise from the $\nu_{2}$ + $\nu_{3}$ 
vibrational band.  As with the $J$-band, these are all R-branch transitions.

\begin{table*}
\begin{threeparttable}
\centering
\small
\caption{Absorption features and spectroscopic signatures of NH$_{3}$ in the $J$ band spectra of late T dwarfs (see Figures \ref{fig:nh3_0722j} and \ref{fig:nh3_0415j})} 
\begin{tabular}{c c c c c c c} 
\vspace{-1.1em}\\
 & Source & $\lambda$ ($\mu$m) & Opacity Source (500~K/750 K) & \head{2.5cm}{NH$_{3}$ feature in synthetic spectrum (500~K/750~K)} & \head{2.5cm}{Feature in UGPS~0722} & \head{2.5cm}{Feature in 2MASS~0415}\\ [0.5ex] 
\rowcolor{beaublue}
& This work & 1.2249 & NH$_{3}$/(H$_{2}$O+NH$_{3}$)                & Yes/No & Yes & Yes\\ 
\rowcolor{beaublue}
& This work & 1.2265 & NH$_{3}$/(NH$_{3}$+CH$_{4}$+H$_{2}$O)                & Yes/No & Yes(?) & Yes(?)\\ 
\rowcolor{beaublue}
& This work & 1.2284\tnote{\textsection}  & NH$_{3}$/---                           & Yes/No & Yes(?) & Yes\\
\rowcolor{beaublue}
& This work & 1.2300 & (H$_{2}$O+NH$_{3}$)/NH$_{3}$                & Yes/Yes & Yes(?) & No\\
\rowcolor{beaublue}
& This work & 1.2326\tnote{\textsection}   & (NH$_{3}$+H$_{2}$O)/H$_{2}$O                      & No(?)/No & Yes & Yes\\
\rowcolor{beaublue}
& B11 & 1.2340 & NH$_{3}$/(NH$_{3}$+H$_{2}$O)                      & No(?)/No & Yes & Yes(?)\\
\rowcolor{beaublue}
& This work & 1.2355 & NH$_{3}$/(NH$_{3}$+CH$_{4}$)                      & Yes/Yes & Yes & Yes(?)\\
\rowcolor{beaublue}
& B11 & 1.2367 & (H$_{2}$O+NH$_{3}$)/H$_{2}$O                      & No/No & Yes & Yes\\
\rowcolor{beaublue}
& This work & 1.2378\tnote{\textdagger} & NH$_{3}$/NH$_{3}$                           & Yes/No & Yes & Yes\\
\rowcolor{beaublue}
& This work & 1.2394\tnote{\textdagger\textdagger} & NH$_{3}$/NH$_{3}$                           & Yes/Yes & No & Yes\\
\rowcolor{beaublue}
& This work & 1.2406 &   NH$_{3}$/H$_{2}$O                                 & Yes/No & Yes & Yes\\
\rowcolor{beaublue}
& B11 & 1.2430\tnote{\textsection}  &   NH$_{3}$/---                                & No(?)/No & Yes & No\\
\rowcolor{beaublue}
& B11 & 1.2438\tnote{\textsection}  & NH$_{3}$/---                                 & No(?)/No & Yes & No\\
\rowcolor{beaublue}
& This work & 1.2494 & NH$_{3}$/NH$_{3}$                           & Yes/Yes(?) & No & Yes(?)\\
\rowcolor{beaublue}
& This work & 1.2535\tnote{\textsection}  &   NH$_{3}$/---                                 & No(?)/No & Yes & No\\
\rowcolor{beaublue}
& B11 & 1.2540 &   (NH$_{3}$+CH$_{4}$)/(H$_{2}$O+NH$_{3}$)                                 & Yes/No & Yes & Yes\\
\rowcolor{beaublue}
& This work & 1.2578\tnote{\textsection} &   NH$_{3}$/---                                & Yes(?)/No & No & Yes\\
\rowcolor{beaublue}
& This work & 1.2624 &   NH$_{3}$/H$_{2}$O                                & No/No & Yes & Yes\\
\rowcolor{beaublue}
& This work & 1.2635\tnote{\textdaggerdbl}  &   NH$_{3}$/H$_{2}$O                                & No/No & Yes & Yes\\
\rowcolor{beaublue}
& B11 & 1.2661\tnote{\textsection}  &   (NH$_{3}$+CH$_{4}$)/---                                & No/No & Yes & Yes\\

\end{tabular}
\begin{tablenotes}
\item[\textsection]There is no obvious opacity source corresponding to an absorption feature at this wavelength in the spectrum of 2MASS 0415.
\item[\textdagger]In 2MASS 0415, the peak in NH$_{3}$ opacity is at 1.2381~$\mu$m.
\item[\textdagger\textdagger]In 2MASS 0415, this feature corresponds to two adjacent peaks in NH$_{3}$ opacity at 1.2394~$\mu$m, and 1.2399~$\mu$m.
\item[\textdaggerdbl]In 2MASS 0415, the peak in NH$_{3}$ opacity is at 1.2638~$\mu$m.
\item In tables \ref{tab:5}, \ref{tab:6}, and \ref{tab:7}, ammonia features produced by R-, Q-, and P-branch line transitions are highlighted in red, grey, and blue respectively. 
\end{tablenotes}
\label{tab:5}
\end{threeparttable}
\end{table*}



\begin{table*}
\begin{threeparttable}
\centering
\small
\caption{Absorption features and spectroscopic signatures of NH$_{3}$ in the $H$ band spectra of late T dwarfs (see Figures \ref{fig:nh3_0722h} and \ref{fig:nh3_0415h})}
\begin{tabular}{c c c c c c c} 
\vspace{-1.1em}\\
& Source & $\lambda$ ($\mu$m) & Opacity Source (500~K/750 K) & \head{2.5cm}{NH$_{3}$ feature in synthetic spectrum (500~K/750~K)} & \head{2.5cm}{Feature in UGPS~0722} & \head{2.5cm}{Feature in 2MASS~0415}\\ [0.5ex] 
\rowcolor{bubblegum}
& B11             & 1.4996           & (H$_{2}$O+NH$_{3}$)/(H$_{2}$O+NH$_{3}$)         & No/No               & Yes & Yes\\     
\rowcolor{bubblegum}
& B11             & 1.5020           & (NH$_{3}$+H$_{2}$O)/H$_{2}$O                           & No/No               & Yes & Yes\\     
\rowcolor{gray}
& B11             & 1.5086           & NH$_{3}$/H$_{2}$O                                             & No/No               & Yes & Yes\\     
\rowcolor{gray}
& B11             & 1.5140           & (NH$_{3}$+H$_{2}$O)/(NH$_{3}$+H$_{2}$O)         & No/No               & Yes & Yes\\     
\rowcolor{gray}
& B11             & 1.5152           & NH$_{3}$/(H$_{2}$O+NH${_3}$)                           & Yes/No              & Yes & Yes\\    
\rowcolor{gray}
& B11             & 1.5179\tnote{\textsection}           & NH$_{3}$/---                              & Yes/No              & Yes & Yes\\    
\rowcolor{gray}
& B11             & 1.5201           & NH$_{3}$/(NH$_{3}$+H$_{2}$O)                                & Yes/Yes             & Yes & Yes\\    
\rowcolor{beaublue}
& This work    & 1.5224           & NH$_{3}$/(NH$_{3}$+H$_{2}$O)            & Yes/No               & Yes & Yes   \\    
\rowcolor{beaublue}
& This work    & 1.5240           & NH$_{3}$/(NH$_{3}$+H$_{2}$O)                          & Yes/No(?)                & Yes & Yes(?)\\    
\rowcolor{beaublue}
& B11             & 1.5260           & NH$_{3}$/(NH$_{3}$+H$_{2}$O)                          & Yes(?)/No                 & No(?) & Yes\\     
\rowcolor{beaublue}
& B11             & 1.5270           & NH$_{3}$/(H$_{2}$O+NH$_{3}$)                          & Yes(?)/No                 & Yes & Yes\\     
\rowcolor{beaublue}
& B11             & 1.5282           & NH$_{3}$/(H$_{2}$O+NH$_{3}$)                          & Yes/No                  & Yes & Yes \\
\rowcolor{beaublue}
& This work    & 1.5297           & NH$_{3}$/(H$_{2}$O+NH$_{3}$)                          & Yes(?)/No                & Yes & Yes\\    
\rowcolor{beaublue}\     
& B11             & 1.5305           & (NH$_{3}$+H$_{2}$O)/(H$_{2}$O+NH$_{3}$)(?)                          & Yes(?)/No                 & Yes & Yes\\     
\rowcolor{beaublue}
& B11             & 1.5327           & NH$_{3}$/(NH$_{3}$+H$_{2}$O)                          & Yes/No                & Yes & Yes\\    
\rowcolor{beaublue}
& B11             & 1.5352           & (NH$_{3}$+H$_{2}$O)/(H$_{2}$O+NH$_{3}$)        & Yes(?)/No                 & Yes & Yes\\     
\rowcolor{beaublue}
& This work    & 1.5367           & NH$_{3}$/(NH$_{3}$+H$_{2}$O)                          & Yes/No                & Yes & Yes\\    
\rowcolor{beaublue}
& This work    & 1.5382\tnote{\textdagger}            & NH$_{3}$/NH$_{3}$                               & Yes/Yes(?)              & No & No\\  
\rowcolor{beaublue}     
& This work    & 1.5395\tnote{\textsection}            & NH$_{3}$/---                          & Yes/No                & Yes & Yes(?)\\   
\rowcolor{beaublue}
& B11             & 1.5408           & NH$_{3}$/(NH$_{3}$+H$_{2}$O)             & Yes(?)/No                 & Yes & Yes\\     
\rowcolor{beaublue}
& This work    & 1.5427 \tnote{\textdagger\textdagger}	& NH$_{3}$/(H$_{2}$O+NH$_{3}$)             & Yes/No                & Yes & Yes  \\   
\rowcolor{beaublue} 
& B11             & 1.5440           & NH$_{3}$/H$_{2}$O                                             & Yes(?)/No                 & Yes & Yes\\    
\rowcolor{beaublue}   
& B11             & 1.5480\tnote{\textdaggerdbl}          & NH$_{3}$/(H$_{2}$O+NH$_{3}$)    & Yes(?)/No     & Yes & Yes\\    
\rowcolor{beaublue}
& B11             & 1.5504\tnote{\textsection}      & NH$_{3}$/---            & Yes/No                 & Yes & Yes\\    
\rowcolor{beaublue}
& This work    & 1.5533           & NH$_{3}$/NH$_{3}$                                            & Yes/Yes                & No & No\\   
\rowcolor{beaublue}
& This work    & 1.5545           & NH$_{3}$/NH$_{3}$                                           &  Yes/No                  & Yes & Yes  \\    
\rowcolor{beaublue}
& B11             & 1.5609\tnote{\textsection} & (NH$_{3}$+CH$_{4}$+H$_{2}$O)/ \textemdash{}   & No/No                   & Yes & Yes\\    
\rowcolor{beaublue}
& B11             & 1.5660           & NH$_{3}$/(NH$_{3}$+H$_{2}$O)                          & Yes/No                   & Yes & Yes\\    
\rowcolor{beaublue}
& B11             & 1.5687           & NH$_{3}$/(NH$_{3}$+H$_{2}$O)                          & No/No                   & Yes & Yes\\     
\rowcolor{beaublue}
& B11             & 1.5735           & NH$_{3}$/NH$_{3}$                         & Yes/No                   & Yes & Yes\\     
\rowcolor{beaublue}
& B11             & 1.5805\tnote{\textdaggerdbl\textdaggerdbl}           & (NH$_{3}$+H$_{2}$O)/(NH$_{3}$+H$_{2}$O)                                            & No/No                   & Yes & Yes\\    
\rowcolor{beaublue}
& B11             & 1.5897\tnote{\textsection}           & (NH$_{3}$+CH$_{4}$)/---          & No/No                   & No(?) & No(?)\\    
\rowcolor{beaublue}
& B11             & 1.5905\tnote{\textsection}           & (NH$_{3}$+CH$_{4}$+H$_{2}$O)/---          & No/No                   & No(?) & No(?)\\    
\end{tabular}
\begin{tablenotes}
\item[\textsection]The absorption feature in 2MASS 0415 does not correspond to a peak in any of the opacity sources considered here.
\item[\textdagger]In 2MASS 0415, the peak in NH$_{3}$ opacity is at 1.5386~$\mu$m. 
\item[\textdagger\textdagger]In 2MASS 0415, the peak in NH$_{3}$ opacity is at 1.5425~$\mu$m.
\item[\textdaggerdbl]In 2MASS 0415, the absorption feature due to H$_{2}$O$+$NH$_{3}$ opacity is centred at $\sim$1.5477~$\mu$m.
\item[\textdaggerdbl\textdaggerdbl]In 2MASS 0415, the peak in NH$_{3}$ opacity is at 1.5803~$\mu$m.
\end{tablenotes}
\label{tab:6}
\end{threeparttable}
\end{table*}

\begin{figure*}
\centering  
\subfigure{
  \hspace*{0.3cm}\includegraphics[scale=0.425, angle=0]{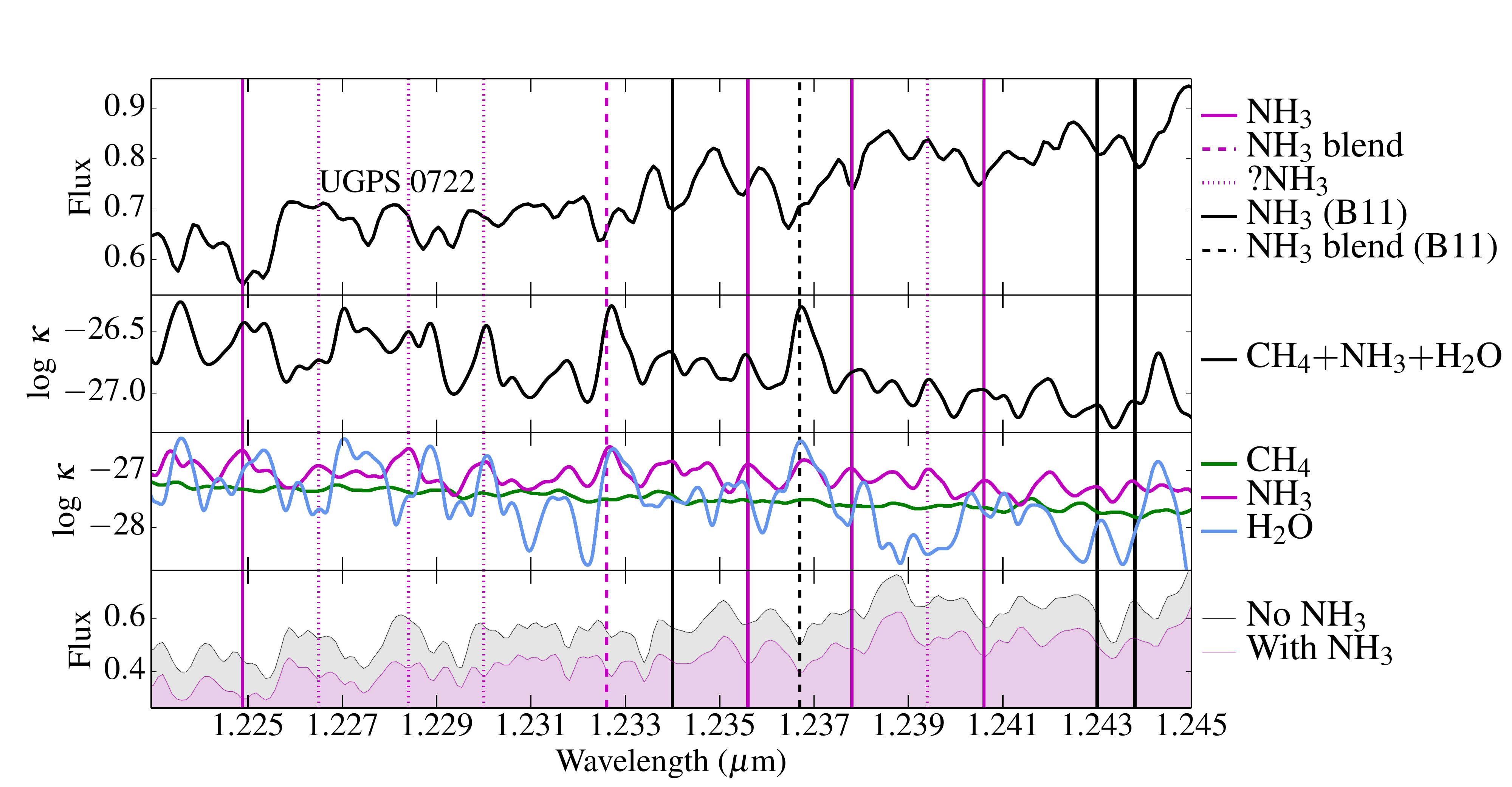} 
   \label{fig:subfig9}
                 }                                                           
\subfigure{
  \hspace*{-1.6cm}\includegraphics[scale=0.37, angle=0]{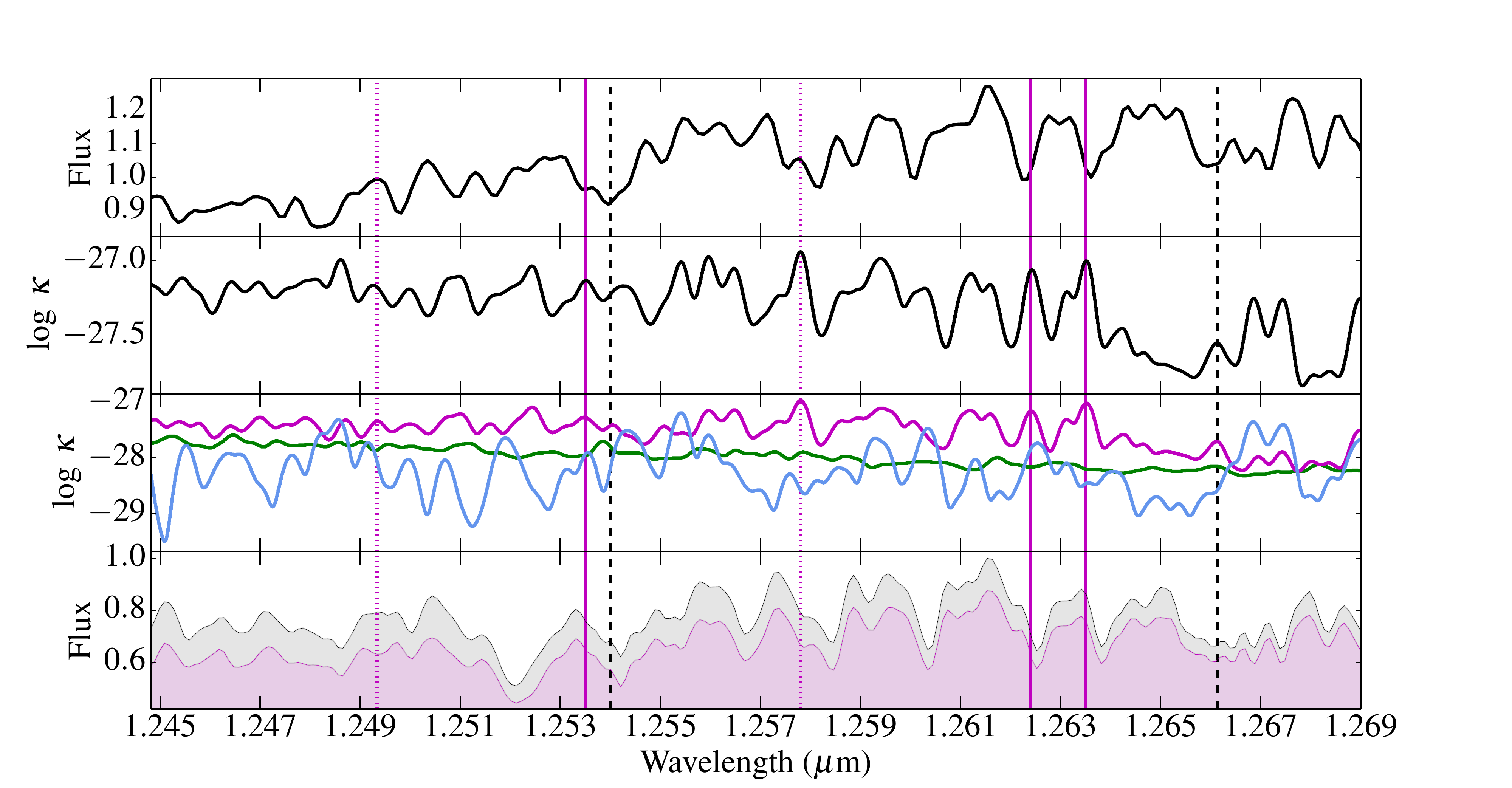} 
   \label{fig:subfig10}
                 }
  \caption{NH$_{3}$ absorption in the $J$-band spectrum of UGPS~0722 (see Table \ref{tab:5}). Absorption 
  cross-sections, scaled for molecular abundances, are calculated at 500~K for  CH$_{4}$ (green), 
  H$_{2}$O (blue) and NH$_{3}$ (magenta).  The lowest graph contains S12 synthetic spectra with 
  NH$_{3}$ opacity (magenta shade) and without NH$_{3}$ opacity (grey shade), also calculated at 500~K. 
  Solid magenta lines are NH$_{3}$ features identified in this work. The dashed magenta line is an NH$_{3}$/H$_{2}$O 
  feature identified in this work. Dotted magenta lines are features which are predicted by the scaled opacity 
  cross-section and the S12 models but which are either missing or ambiguous. Solid and dashed black lines 
  indicate features identified by B11.}
  \label{fig:nh3_0722j}
\end{figure*}

\begin{figure*}
\centering 
\subfigure{
  \hspace*{0.3cm}\includegraphics[scale=0.40, angle=0]{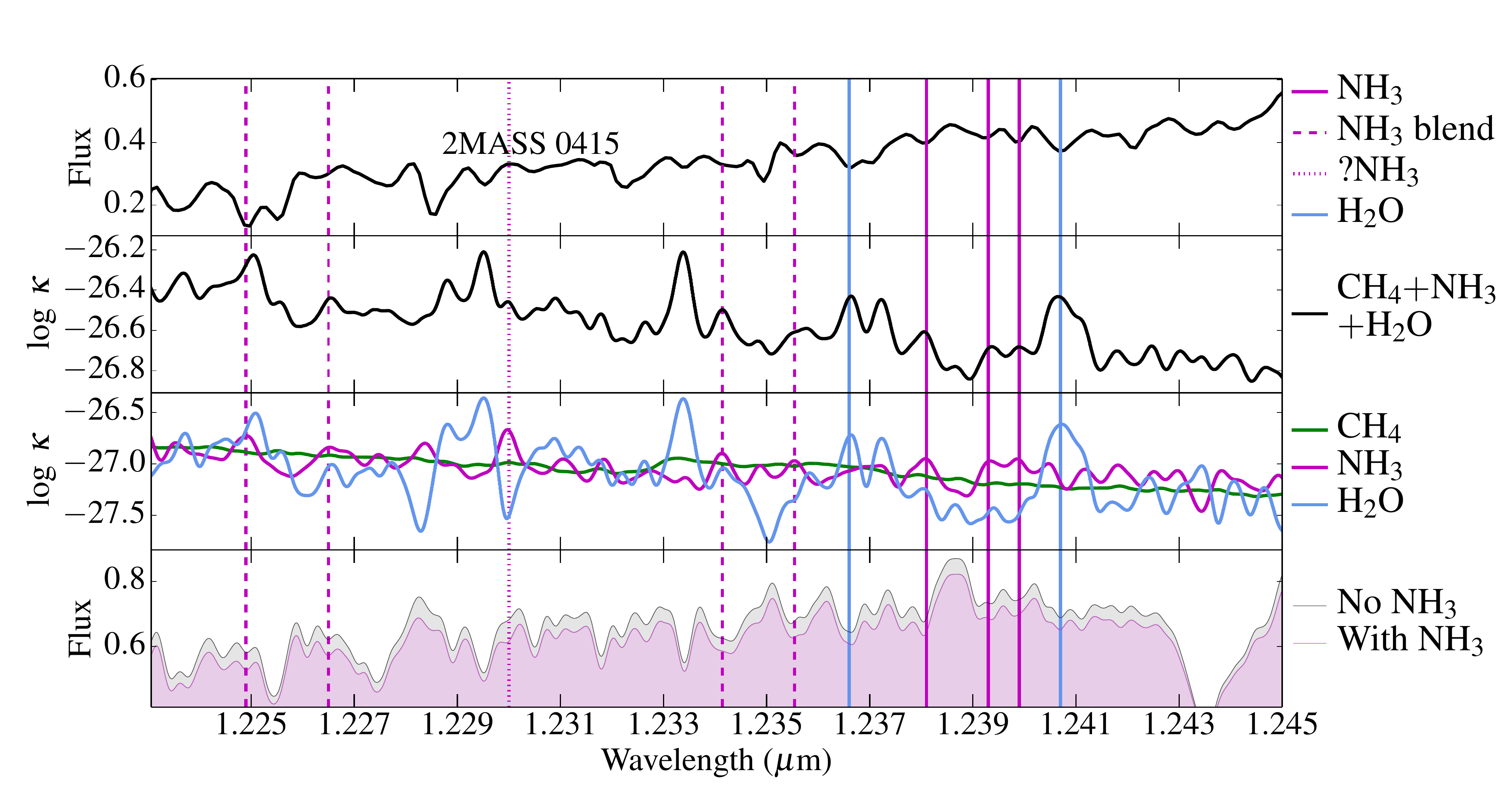} 
   \label{fig:subfig9}
                 }                                                           
\subfigure{
  \hspace*{-0.6cm}\includegraphics[scale=0.375, angle=0]{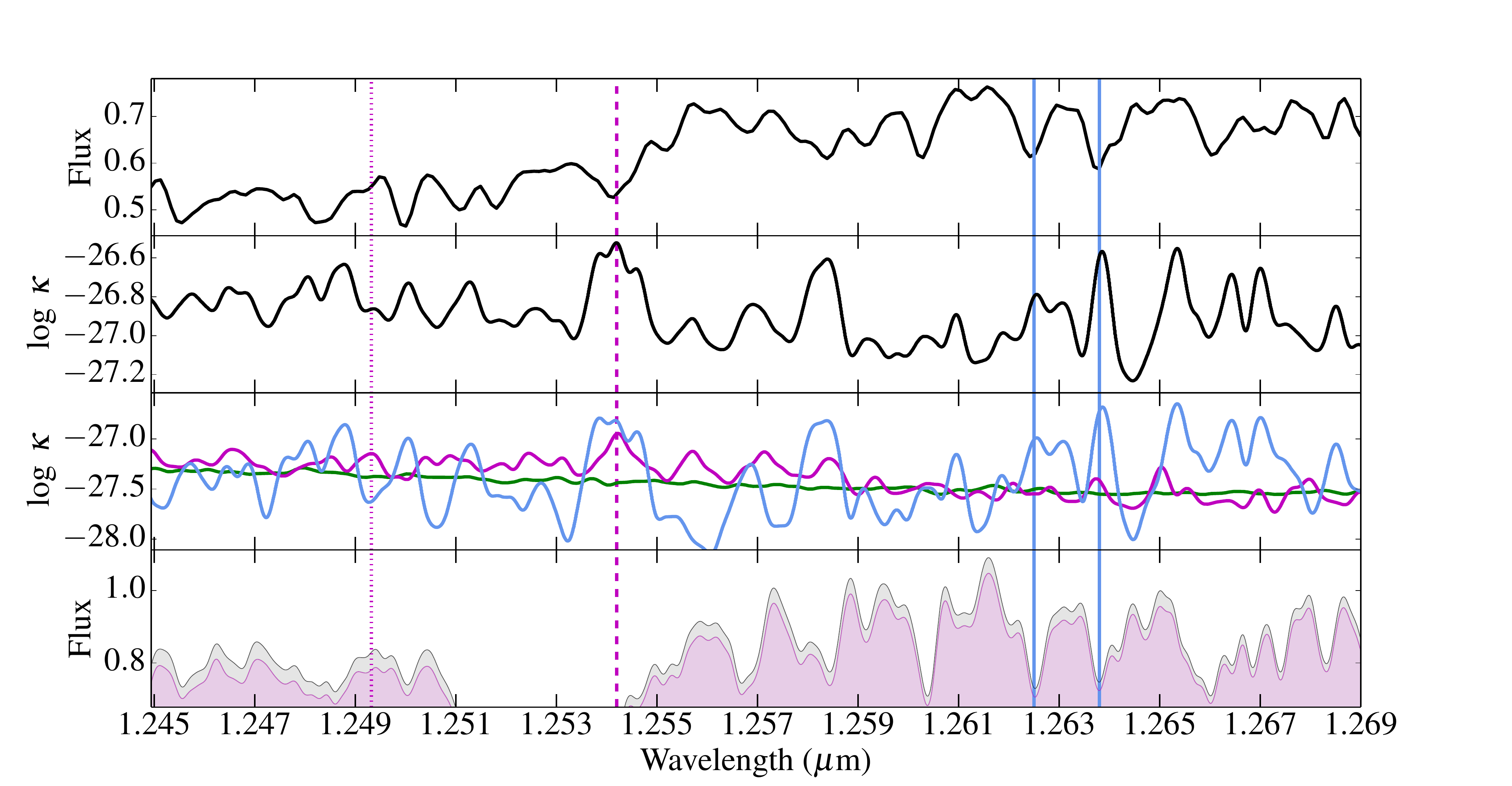} 
   \label{fig:subfig10}
                 }
  \caption{NH$_{3}$ absorption features in the $J$-band spectrum of 2MASS~0415 (see Table \ref{tab:5}). Features are as described in 
  the $J$-band spectrum of UGPS~0722. The solid blue lines are pure H$_{2}$O features which were NH$_{3}+$H$_{2}$O blends in the spectrum 
  of UGPS 0722. The deep absorption features centred at $\sim$1.243 $\mu$m and $\sim$1.253 $\mu$m in the synthetic  
  spectra, are the blue and red components of the K {\scriptsize I} doublet.}
  \label{fig:nh3_0415j}
\end{figure*}

\section{Ammonia}
\label{sec:ammonia}
In the following discussion, the terms NH$_{3}$ and ammonia refer to the main isotopologue of ammonia, 
$^{14}$NH$_{3}$. The conversion of N$_{2}$ into NH$_{3}$ occurs at T$_{eff}$ $\sim$800~K [S12]. Bands 
of NH$_{3}$ opacity are first seen in the mid-infrared spectra of T2 dwarfs~\citep{cushing06}, while the 
first unequivocal detection of NH$_{3}$ opacity in brown dwarfs was made in the mid-infrared spectra of 
the T dwarf binary $\epsilon$ Indi Bab (T1+T6)~(\citealt{roellig04};~\citealt{mainzer07}). Features in low resolution 
near-infrared spectra of late T dwarfs and the recently discovered Y dwarfs have been attributed to weak 
bands of NH$_{3}$ opacity~(\citealt{delorme08}; C11), but the first confirmed identification of 
ammonia features in the near-infrared spectrum of a late T dwarf was made by B11 using medium resolution 
spectra. This discovery has been significant in justifying the Y spectral class.  Indeed, as we observed in 
Section \ref{sec:ch4j}, B11 may have underestimated the number of ammonia features in the $J$-band of 
UGPS~0722. By comparing synthetic spectra derived from full model atmosphere calculations with and 
without NH$_{3}$ opacity with the Magellan/FIRE spectrum of UGPS~0722 in B11, S12 were able to identify 
a number of new NH$_{3}$ absorption features. We have applied the technique described in S12 to the 
Gemini/NIFS spectra of 2MASS~0415 and UGPS~0722. The synthetic spectra were produced with solar 
metallicity and log~$g=$4.25. Non-equilibrium chemistry was assumed, with eddy diffusion constant 
$K_{zz}$=10$^4$ cm$^2$ s$^{-1}$.  Separate models were produced at 750~K and 500~K, the respective 
effective temperatures of the T8 and T9 objects. The models included the most recent calculations of all 
major opacity sources, except the 10to10 line list. With that caveat in mind, with the possible exception of 
regions where methane is the dominant opacity source, the model spectra should appear similar to the 
T dwarf spectra. It is also useful to have spectra from two dwarfs of successive spectral types since 
absorption features missing in the spectrum of one T dwarf, may be present in the spectrum of the other. 

By comparing the T dwarf spectra with the S12 model spectra and the scaled cross-sections for 
CH$_{4}$, H$_{2}$O, and NH$_{3}$, we have been able to identify a number of new ammonia features 
in the near-infrared spectra of our T dwarfs 
(see Figures \ref{fig:nh3_0722j} $-$ \ref{fig:nh3_0415k}, and Tables \ref{tab:5} $-$ \ref{tab:7}).
 
In order to be confident in identifying an absorption feature, 
the shape of the S12 model spectra with and without NH$_{3}$ opacity should appear 
different, not simply in amplitude. In places, peaks in the NH$_{3}$ opacity do not coincide exactly with 
absorption troughs in the science spectrum. In these cases, we identify an ammonia feature at the 
wavelength of the peak in the NH$_{3}$ opacity when the science spectrum at this wavelength appears 
similar to the model spectrum with NH$_{3}$ opacity, e.g., the feature at 1.5240~$\mu$m in the spectrum of 
UGPS 0722 (see Figure \ref{fig:nh3_0722h}).
In a number of cases, features are predicted by the model spectra at 500~K and/or 750~K 
but are not found in the respective data. These results are also shown in Tables \ref{tab:5}, \ref{tab:6}, and \ref{tab:7}.


\subsection{The $Z$-Band}
\label{sec:nh3z}
In the $Z$-band, the S12 models predict NH$_{3}$ opacity only between 1.005~$\mu$m and 
1.074~$\mu$m. Across most of this wavelength range the S12 model spectra differ 
only in amplitude. Only at 1.0256~$\mu$m do we find a slight variation in structure 
between the two models, corresponding to a peak in NH$_{3}$ opacity and weak 
absorption features in both T dwarfs. However, the detection is weak, and we cannot 
say with confidence that we have found any evidence of NH$_{3}$ absorption features in the $Z$-band.



\subsection{The $J$-Band}
\label{sec:nh3j}
The scaled opacity cross-sections at 500~K show peaks in the NH$_{3}$ opacity 
between $\sim$1.210~$\mu$m and $\sim$1.276~$\mu$m. We compared these 
cross-sections with the S12 synthetic spectra at 500~K and identified a number of 
features between $\sim$1.2250$-$1.2690~$\mu$m, (see Figures \ref{fig:nh3_0722j}, \ref{fig:nh3_0415j} and Table \ref{tab:5}).

B11 found isolated NH$_{3}$ features at 1.2340~$\mu$m and 1.2430~$\mu$m in the FIRE spectrum of UGPS 0722. 
There are peaks in the NH$_{3}$ absorption cross-sections at these wavelengths, 
corresponding to absorption features in our spectrum of UGPS 0722. There are no well-defined
differences in the synthetic spectra with and without NH$_{3}$ opacity at these wavelengths. It may be that 
there are ammonia features at 1.2340~$\mu$m and 1.2430~$\mu$m which the model spectra are unable to identify. At the 
moment, these features are unconfirmed. 
B11 identified an NH$_{3}$+H$_{2}$O feature at 1.2367~$\mu$m. We find that this feature is an H$_{2}$O+NH$_{3}$ blend.
B11 identified a feature at 1.2438~$\mu$m as an H$_{2}$O+CH$_{4}$ blend.
There is a peak in the NH$_{3}$ opacity at this wavelength, corresponding to an absorption feature in our spectrum of UGPS 0722.
However, the S12 synthetic spectra show no absorption feature at this location. 
We note that a synthetic spectrum naturally accounts for a range of temperatures in the line formation due to 
optical depth effects in and out of lines. This will affect lines that are particularly temperature sensitive and result in synthetic spectra 
that differ somewhat from opacity plots, even over a narrow range of wavelengths.
While we believe this feature is most likely an NH$_{3}$ feature, it remains unconfirmed. 
In general, higher resolution spectroscopy will greatly assist the identification of lines and absorbers.

While the peak in the NH$_{3}$ opacity at 1.2540~$\mu$m does not appear strong enough to account for the depth of the corresponding 
absorption feature in the spectrum of UGPS 0722, there is a change in slope at this wavelength between the synthetic spectra with and without NH$_{3}$ opacity, 
suggesting that ammonia opacity does make a contribution to this feature.

The ro-vibrational transitions responsible for the NH$_{3}$ absorption features in the 
$J$-band spectra of the two T dwarfs are P-branch transitions, with the exception of a 
relatively weak Q-branch transition line in the feature at 1.2284~$\mu$m, and a relatively 
strong R-branch transition line in the feature at 1.2494~$\mu$m. Contrast this with the 
ro-vibrational transition lines responsible for the CH$_{4}$ absorption features in the 
$J$-band, which are all R-branch transition lines. Here, the shorter wavelength P-branch 
transitions arise from the $\nu_{1}$+3$\nu_{4}$ vibrational band, while the longer 
wavelength P-branch transition lines, and the single Q-branch transition line, are from 
the $\nu_{1}$+$\nu_{3}$+$\nu_{4}$ vibrational band. The single R-branch transition 
line is from the $\nu_{2}$+2$\nu_{3}$ vibrational band. 

\begin{figure*}
\centering  
\subfigure{
\hspace*{0.5cm}\includegraphics[scale=0.355, angle=0]{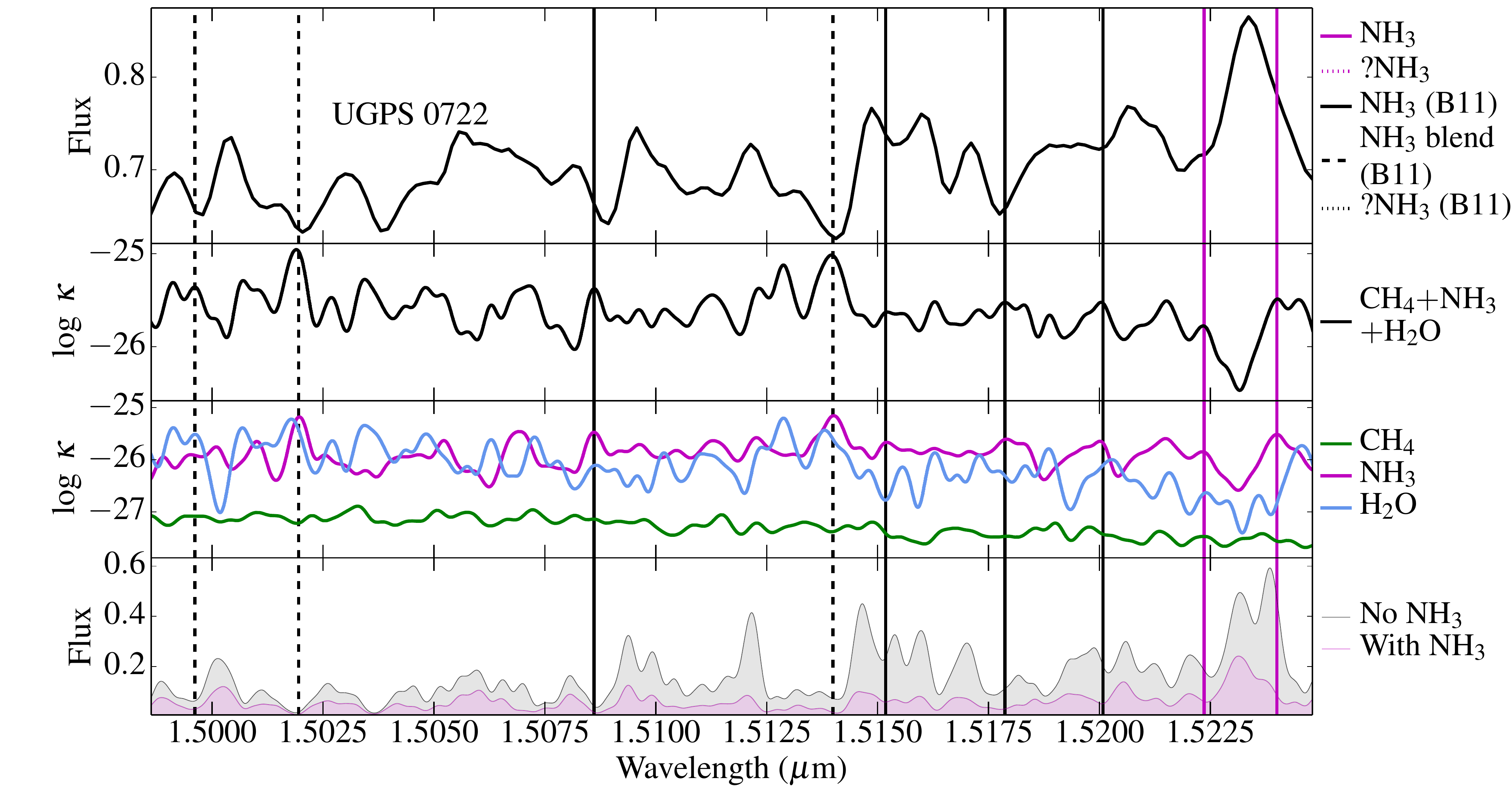} 
   \label{fig:subfig11}
                 }                                                           
\subfigure{
   \includegraphics[scale=0.34, angle=0]{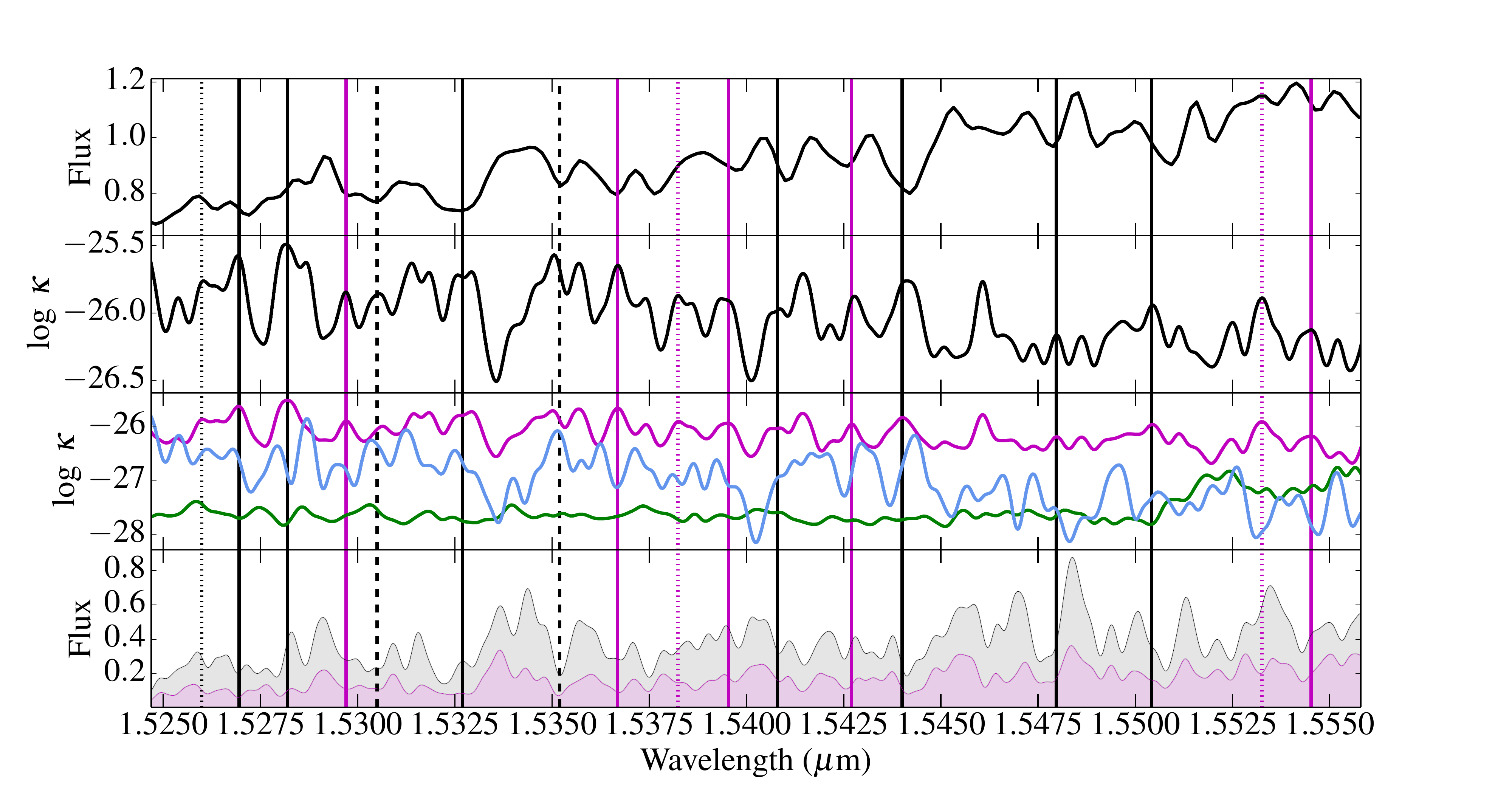} 
   \label{fig:subfig12}
                 }
                 \subfigure{
   \includegraphics[scale=0.34, angle=0]{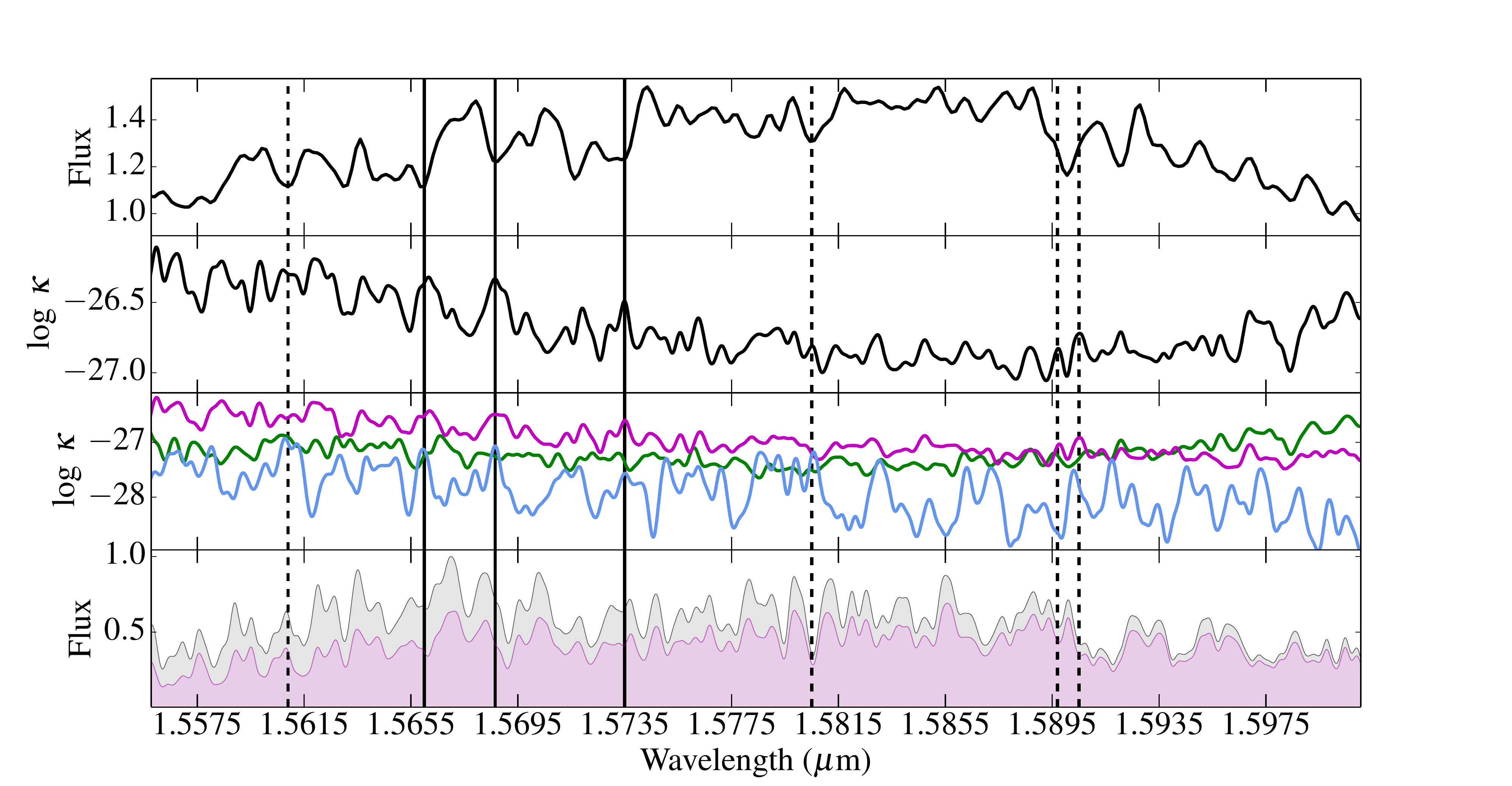} 
   \label{fig:subfig13}
                                 }
  \caption{NH$_{3}$ absorption features in the $H$-band spectrum of UGPS~0722 (see Table \ref{tab:6}). Scaled absorption 
  cross-sections are calculated at 500~K for CH$_{4}$ (green), 
  H$_{2}$O (blue) and NH$_{3}$ (magenta). Features are as 
  described in Figure \ref{fig:nh3_0722j}. Dashed lines are features produced by a combination 
  of molecular species. The dotted lines are NH$_{3}$ features predicted by the S12 models, but missing 
  or ambiguous in the data.}
  \label{fig:nh3_0722h}
\end{figure*}

\begin{figure*}
\centering  
\subfigure{
 \hspace*{0.5cm}\includegraphics[scale=0.355, angle=0]{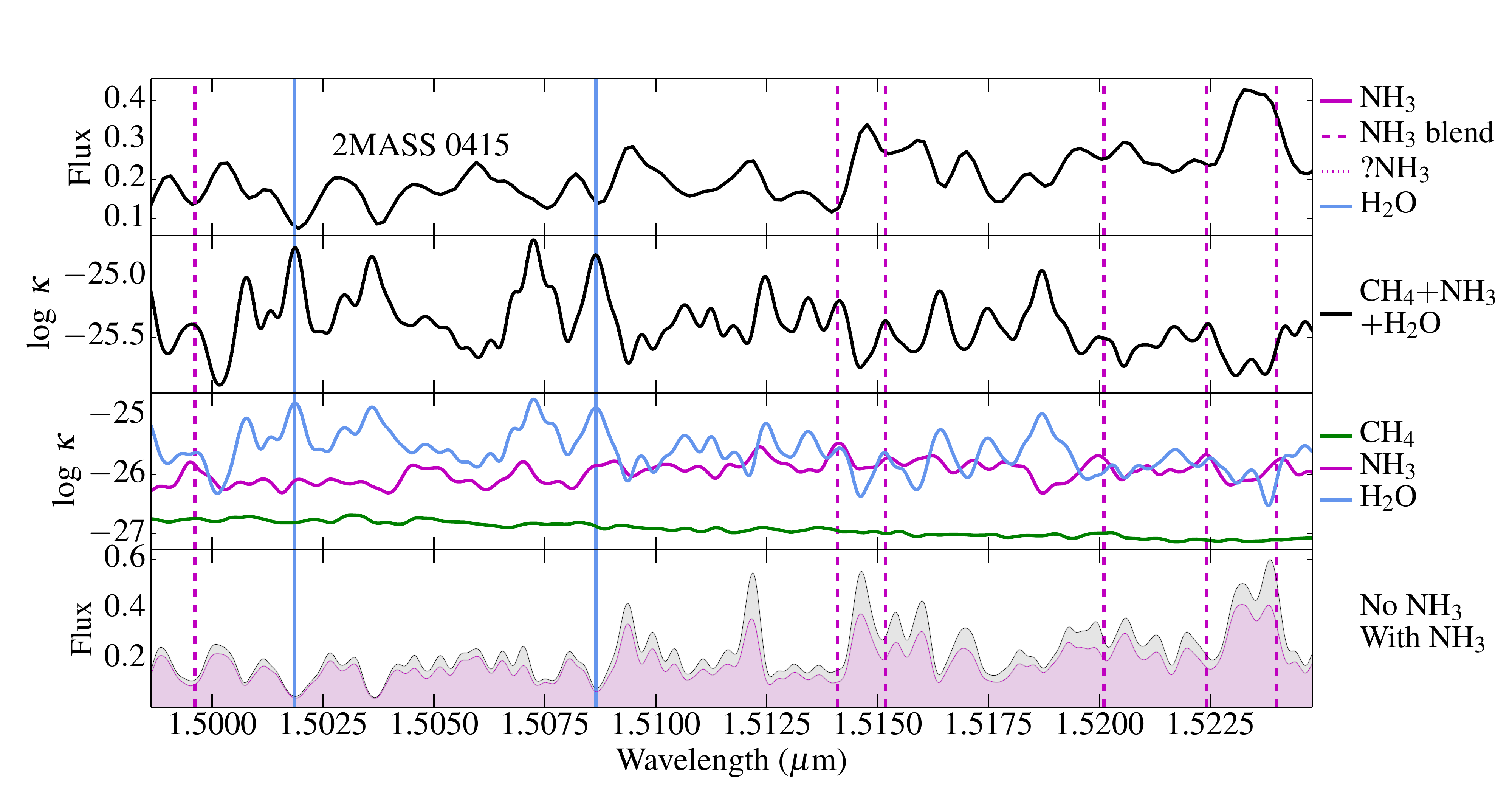} 
   \label{fig:subfig11}
                 }                                                           
\subfigure{
   \includegraphics[scale=0.34, angle=0]{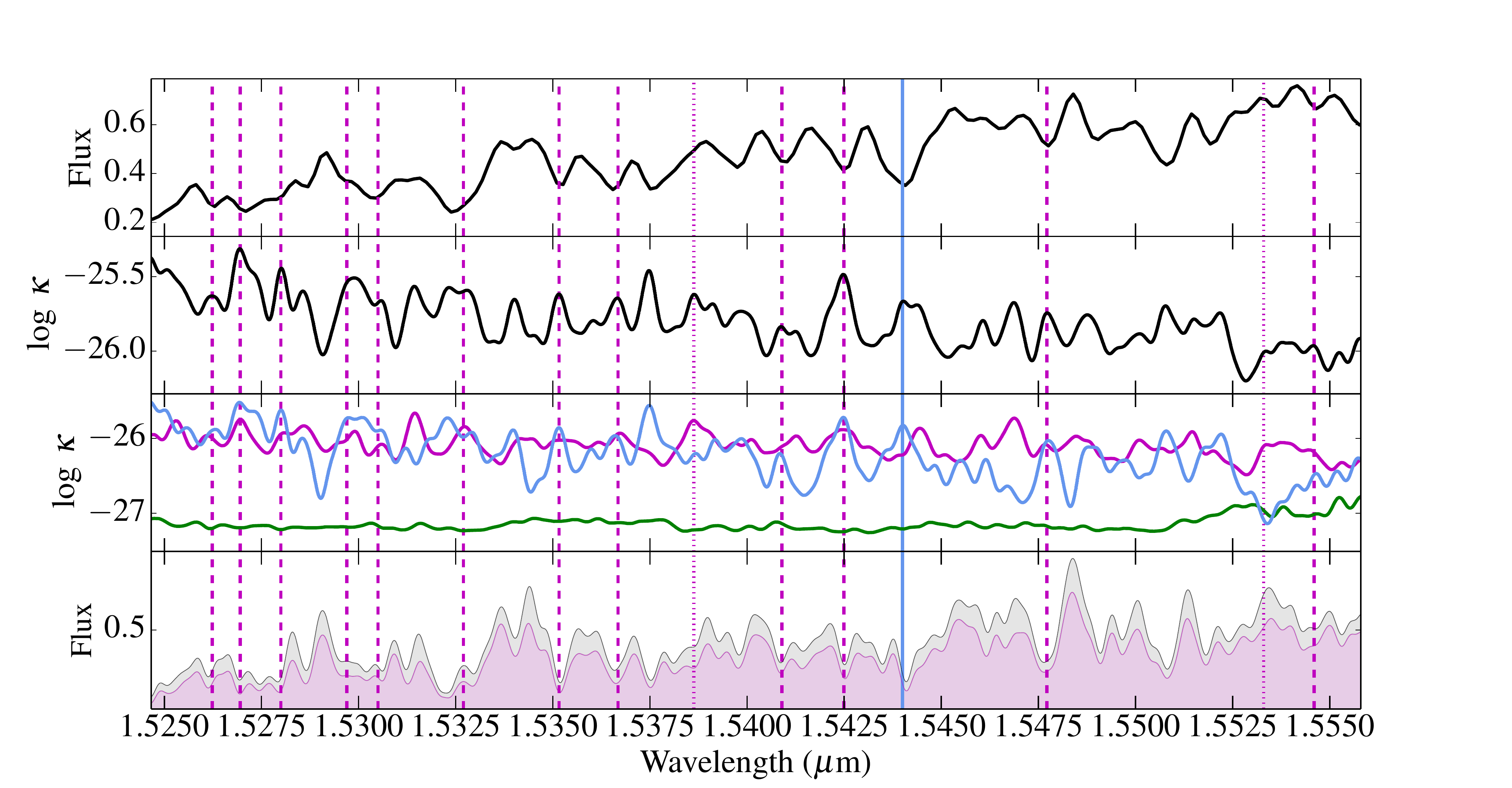} 
   \label{fig:subfig12}
                 }
                 \subfigure{
   \includegraphics[scale=0.34, angle=0]{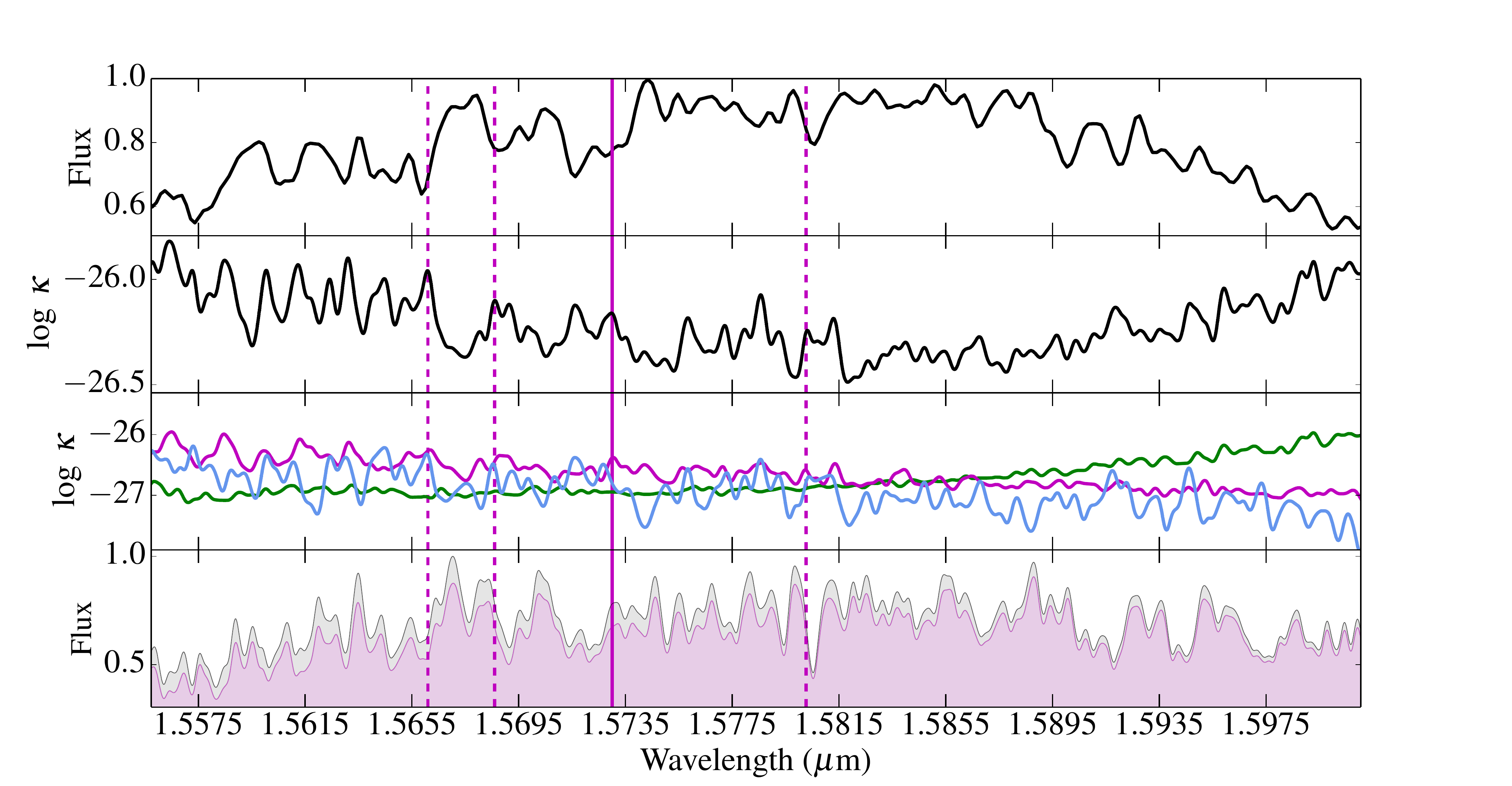} 
   \label{fig:subfig13}
                                 }
  \caption{Opacity sources responsible for the corresponding absorption features in the $H$-band spectrum of 2MASS~0415 (see Table \ref{tab:6}). Features are as described in Figures \ref{fig:nh3_0722j} and \ref{fig:nh3_0722h}. Scaled absorption cross-sections are calculated at 750~K for CH$_{4}$ (green), H$_{2}$O (blue) and NH$_{3}$ (magenta).}
  \label{fig:nh3_0415h}
\end{figure*}



\subsection{The $H$-Band}
\label{sec:nh3h}
Figure 5 in C11 shows that NH$_{3}$ is the dominant source of opacity across 
the blue wing of the $H$-band (1.50-1.59~$\mu$m) in T dwarfs with T$_{eff}$ $\leq$ 600~K. 
We also find that this is the case. In contrast, CH$_{4}$ is by two orders of magnitude the 
dominant opacity source on the red slope of the $H$-band peak flux. At 750~K, the NH$_{3}$ 
and H$_{2}$O scaled absorption cross-sections are of the same order of magnitude. 
Therefore, some features that are produced by NH$_{3}$ only in the spectrum of UGPS~0722, 
are produced by a combination of ammonia and water opacities in the spectrum of 2MASS~0415. 
These results are summarised in Figures \ref{fig:nh3_0722h}, \ref{fig:nh3_0415h} and Table \ref{tab:6}.

B11 observed a number of ``pure" NH$_{3}$ absorption features. 
Of these, the feature at 1.5140~$\mu$m corresponds to a peak in both NH$_{3}$ and H$_{2}$O 
opacity. The strengths of the opacities are broadly similar. There is no significant difference in 
the model spectra computed with and without NH$_3$ opacity, other than a small change in 
amplitude. Therefore, we conclude that this feature is most likely an ammonia/water blend.
The feature at 1.5152~$\mu$m corresponds to a peak in NH$_{3}$ opacity and absorption 
features in UGPS~0722 and 2MASS~0415. There appears to be a difference in opacity between 
the model spectrum computed at 500~K but not that computed at 750~K, other than in amplitude. 
We confirm that this NH$_{3}$ absorption feature is due entirely to NH$_{3}$ in the spectrum of 
UGPS~0722, but is an NH$_{3}$+H$_{2}$O feature in 2MASS~0415.
The feature at 1.5282~$\mu$m corresponds to a peak in NH$_{3}$ opacity and to a clear 
difference between the S12 model spectra computed at 500~K. The model spectra show that 
without NH$_{3}$ opacity a rather strong peak would be expected at 1.5282~$\mu$m while only a 
small one is present. Indeed, the spectrum 
of UGPS~0722 does look similar to the model spectrum with NH$_{3}$ opacity. The spectrum of 2MASS 0415 
at this wavelength looks similar to UGPS 0722, suggesting that NH$_{3}$ opacity is again the dominant opacity 
source. However, the synthetic spectra computed at 750~K show two relatively strong peaks, which differ only in
amplitude between the two spectra. The scaled molecular opacities at 750~K suggest that the feature in 2MASS 
0415's spectrum is produced by a blend of H$_{2}$O+NH$_{3}$ opacity and we have listed this as such in Table 
\ref{tab:6}. 

\begin{table*}
\begin{threeparttable}
\centering
\small
\caption{Absorption features and spectroscopic signatures of NH$_{3}$ in the $K$ band spectra of late T dwarfs (see Figures \ref{fig:nh3_0722k} and \ref{fig:nh3_0415k})} 
\begin{tabular}{c c c c c c c} 
\vspace{-1.1em}\\
 & Source & $\lambda$ ($\mu$m) & Opacity Source (500~K/750 K) & \head{2.5cm}{NH$_{3}$ feature in synthetic spectrum (500~K/750~K)} & \head{2.5cm}{Feature in UGPS~0722} & \head{2.5cm}{Feature in 2MASS~0415}\\ [0.5ex] 
\rowcolor{bubblegum}
& S12             & 1.9667 & NH$_{3}$/(NH$_{3}$+H$_{2}$O)        & Yes/Yes & No(?) &  No(?) \\   
\rowcolor{gray}
& S12             & 1.9698 & NH$_{3}$/NH$_{3}$         & Yes/Yes & No & No\\      
\rowcolor{gray}
& S12             & 1.9737 & NH$_{3}$/(H$_{2}$O+NH$_{3}$)         & Yes/No & Yes & Yes(?)\\    
\rowcolor{gray}
& S12       & 1.9784 & (NH$_{3}$+H$_{2}$O)/(NH$_{3}$+H$_{2}$O)                             & Yes/Yes & Yes & Yes(?)\\    
\rowcolor{gray}
& S12             & 1.9833 & (NH$_{3}$+H$_{2}$O)/(H$_{2}$O+NH$_{3}$)         & Yes/Yes(?) & Yes(?) & No\\        
\rowcolor{gray}
& This work       & 1.9858 & NH$_{3}$/(H$_{2}$O+NH$_{3}$)         & Yes/Yes & Yes & Yes\\   
\rowcolor{gray}
& This work       & 1.9894\tnote{\textsection} & NH$_{3}$/---                       & Yes/Yes(?) & No & Yes(?)\\  
\rowcolor{gray}
& B11             & 1.9900\tnote{\textdagger}  & (H$_{2}$O+NH$_{3}$)/H$_{2}$O         & No/No & Yes & Yes\\     
\rowcolor{beaublue}
& S12             & 1.9937 & NH$_{3}$/(H$_{2}$O+NH$_{3}$)         & Yes/Yes & Yes & Yes(?)\\      
\rowcolor{beaublue}
& S12             & 1.9972 & NH$_{3}$/NH$_{3}$        & Yes/Yes & Yes & Yes\\    
\rowcolor{beaublue}
& This work       & 2.0012 & NH$_{3}$/(H$_{2}$O+NH$_{3}$)         & Yes/No & No & Yes\\    
\rowcolor{beaublue}
& S12             & 2.0052 & NH$_{3}$/NH$_{3}$                               & Yes/Yes & Yes & Yes\\    
\rowcolor{beaublue}
& S12             & 2.0092 & NH$_{3}$/(NH$_{3}$+H$_{2}$O)                              & Yes/Yes & Yes & Yes\\   
\rowcolor{beaublue}
& This work             & 2.0097 & (NH$_{3}$+H$_{2}$O)/H$_{2}$O                              & Yes/No & Yes & Yes\\   
\rowcolor{beaublue}
& S12             & 2.0132 & NH$_{3}$/(NH$_{3}$+H$_{2}$O)                             & Yes/Yes & Yes & Yes\\   
\rowcolor{beaublue}
& S12             & 2.0177 & NH$_{3}$/(NH$_{3}$+H$_{2}$O)         & Yes/Yes & No(?) & Yes\\   
\rowcolor{beaublue}
& S12             & 2.0211 & NH$_{3}$/NH$_{3}$        & Yes/Yes & Yes & Yes\\    
\rowcolor{beaublue}
& This work             & 2.0221 & NH$_{3}$/NH$_{3}$                           & Yes/Yes & Yes & Yes\\    
\rowcolor{beaublue}
& S12             & 2.0256 & NH$_{3}$/(NH$_{3}$+H$_{2}$O)         & Yes/Yes & Yes & Yes\\    
\rowcolor{beaublue}
& This work       & 2.0265 & NH$_{3}$/(NH$_{3}$+H$_{2}$O)         & Yes/No & Yes & Yes\\    
\rowcolor{beaublue}
& S12             & 2.0294 & (NH$_{3}$+H$_{2}$O)/(H$_{2}$O+NH$_{3}$)         & Yes/No & Yes & Yes\\    
\rowcolor{beaublue}
& S12             & 2.0351 & NH$_{3}$/NH$_{3}$                    & Yes/Yes & No & No\\   
\rowcolor{beaublue}
& S12             & 2.0375 & NH$_{3}$/NH$_{3}$                    & Yes/No(?) & Yes & Yes\\      
\rowcolor{beaublue}
& S12             & 2.0387 & NH$_{3}$/(NH$_{3}$+H$_{2}$O)                    & Yes/Yes & Yes & No\\   
\rowcolor{beaublue}
& This work       & 2.0415 & NH$_{3}$/(H$_{2}$O+NH$_{3}$)         & Yes/No(?) & Yes & No(?)\\    
\rowcolor{beaublue}
& S12             & 2.0425\tnote{\textsection}  & NH$_{3}$/---                    & Yes/Yes & Yes & Yes\\    
\rowcolor{beaublue}
& S12             & 2.0454 & NH$_{3}$/NH$_{3}$                    & Yes/Yes & Yes & Yes\\    
\rowcolor{beaublue}
& This work       & 2.0498 & NH$_{3}$/(NH$_{3}$+H$_{2}$O)         & Yes/Yes(?) & Yes & Yes\\    
\end{tabular}
\begin{tablenotes}
\item[\textsection]The absorption feature in 2MASS 0415 does not correspond to a peak in any of the opacity sources considered here.
\item[\textdagger]In 2MASS 0415, the peak in H$_{2}$O opacity is at 1.9898~$\mu$m. 
\end{tablenotes}
\label{tab:7}
\end{threeparttable}
\end{table*}

\begin{figure*}
\centering  
\subfigure{
   \hspace*{0.5cm}\includegraphics[scale=0.355, angle=0]{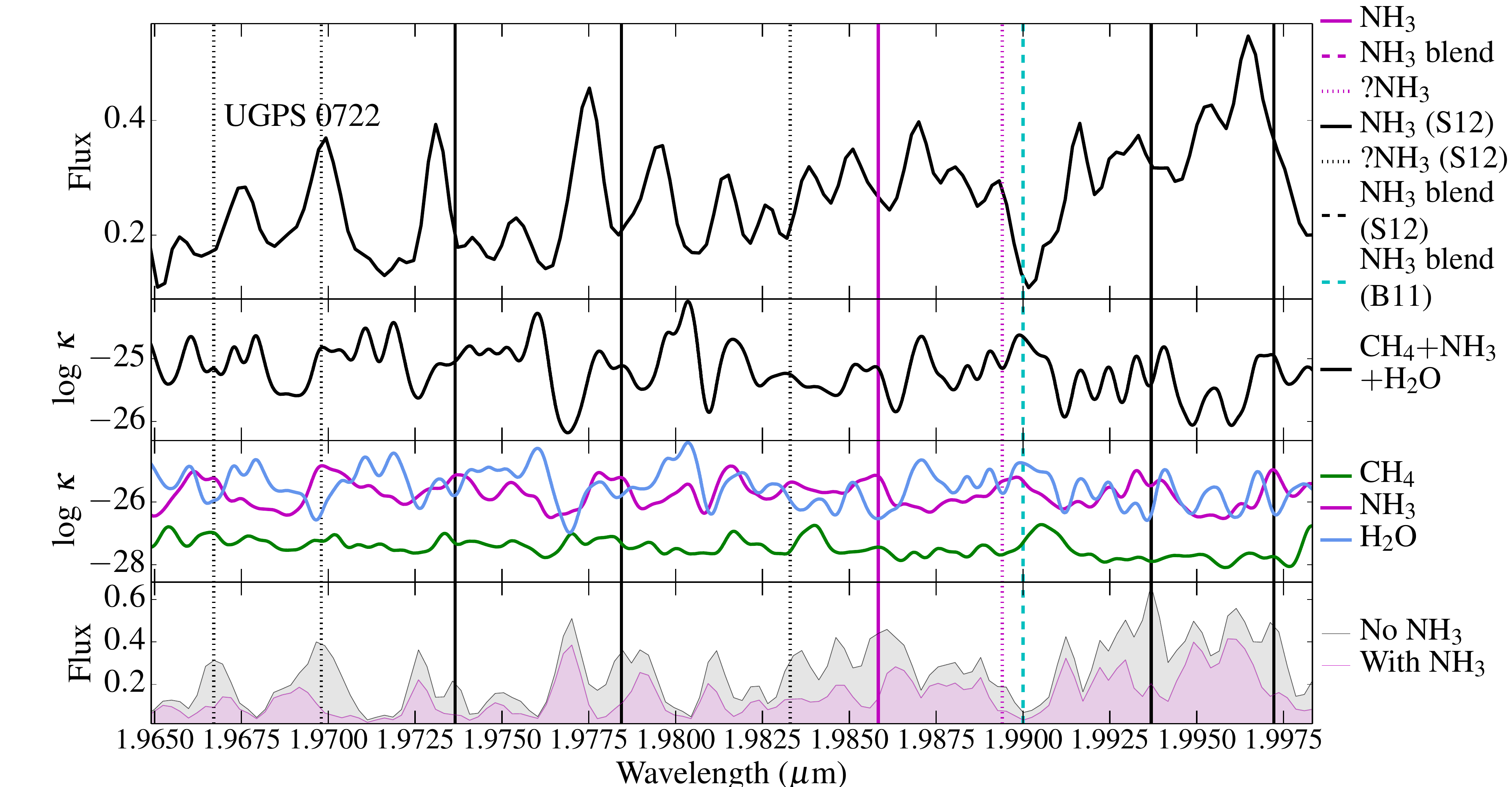} 
   \label{fig:subfig11}
                 }                                                           
\subfigure{
   \includegraphics[scale=0.34, angle=0]{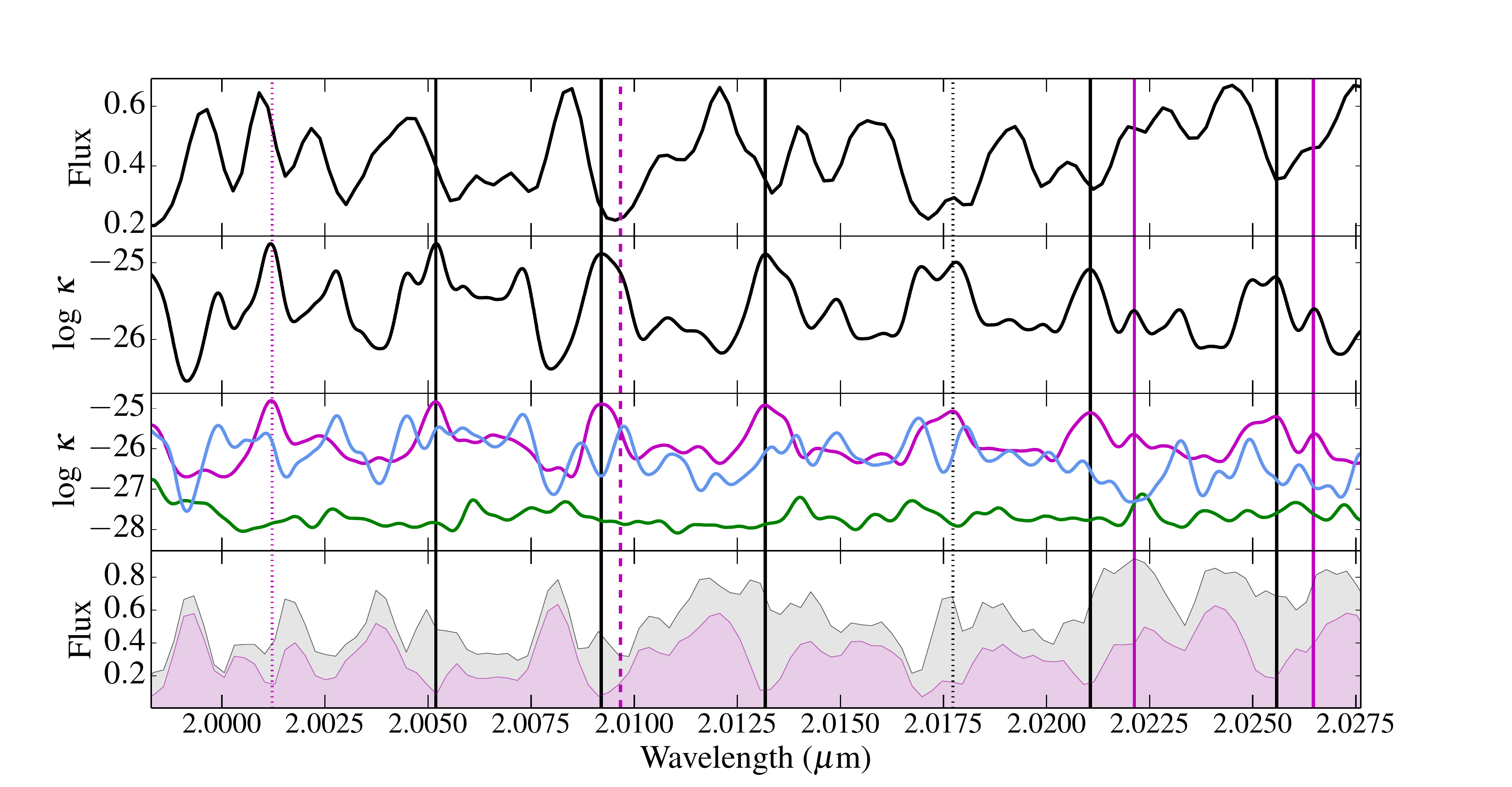} 
   \label{fig:subfig12}
                 }
                 \subfigure{
   \includegraphics[scale=0.34, angle=0]{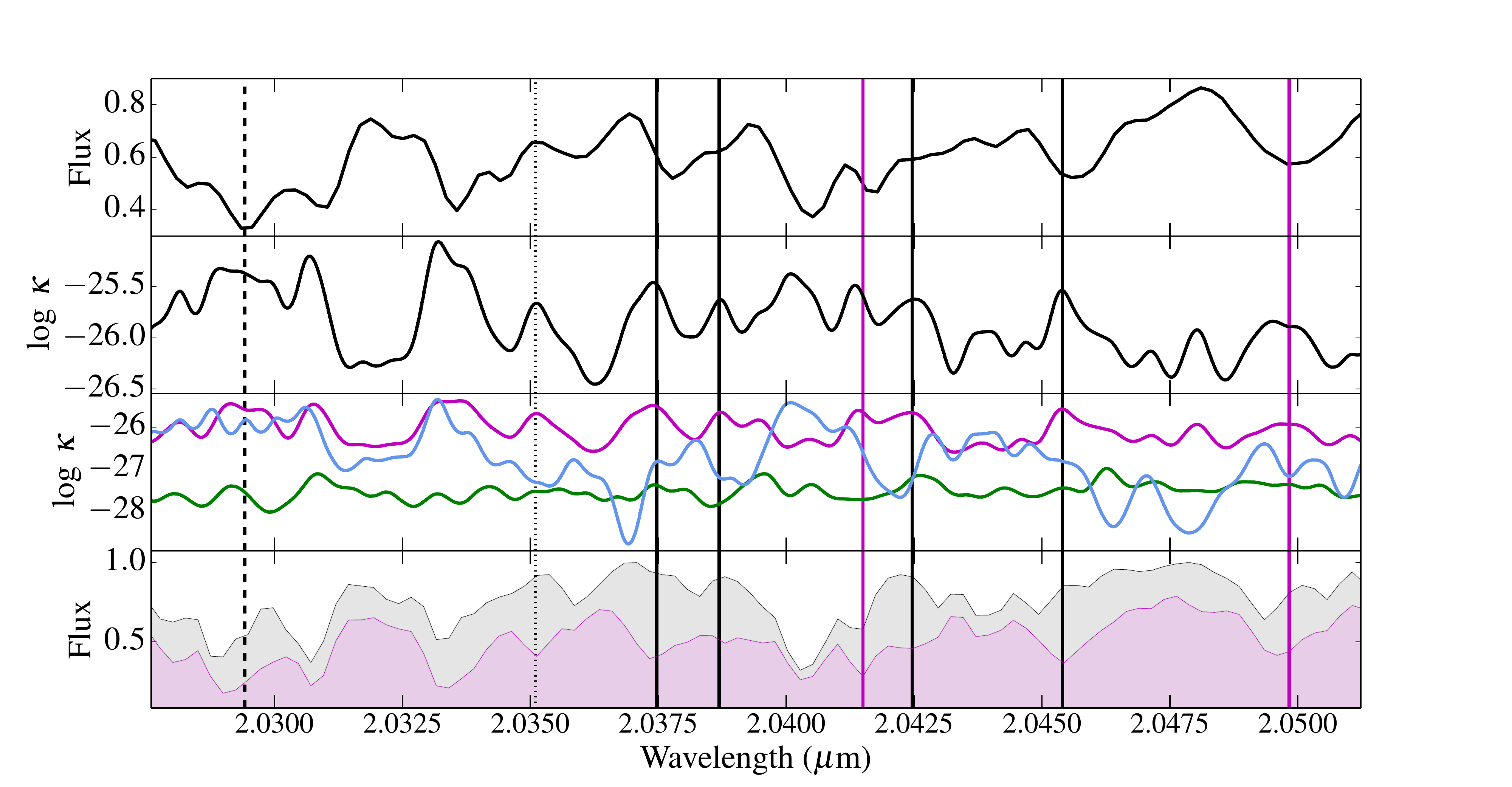} 
   \label{fig:subfig13}
                                 }
  \caption{NH$_{3}$ absorption features in the $K$-band spectrum of UGPS~0722 (see Table \ref{tab:7}). Solid black lines 
  are features in S12 which we can now confirm. These include seven features which were either missing 
  or ambiguous in S12. The black dotted lines are four S12 features which remain missing. The solid magenta lines 
  are features detected in this work. Scaled absorption cross-sections are calculated at 500~K for  CH$_{4}$ (green), 
  H$_{2}$O (blue) and NH$_{3}$ (magenta).}
  \label{fig:nh3_0722k}
\end{figure*}

\begin{figure*}
\centering  
\subfigure{
   \hspace*{0.5cm}\includegraphics[scale=0.355, angle=0]{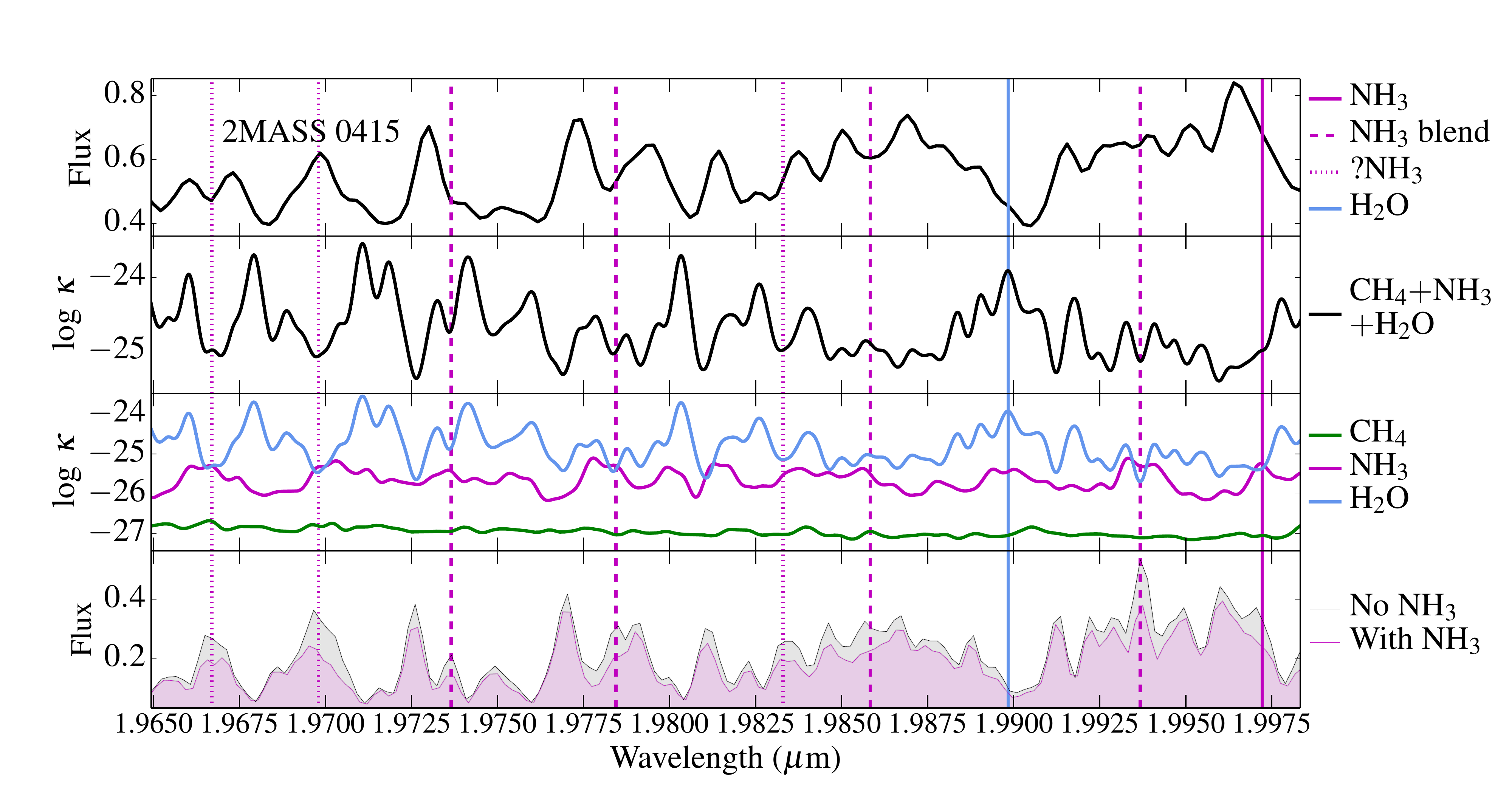} 
   \label{fig:subfig14}
                 }                                                           
\subfigure{
   \includegraphics[scale=0.34, angle=0]{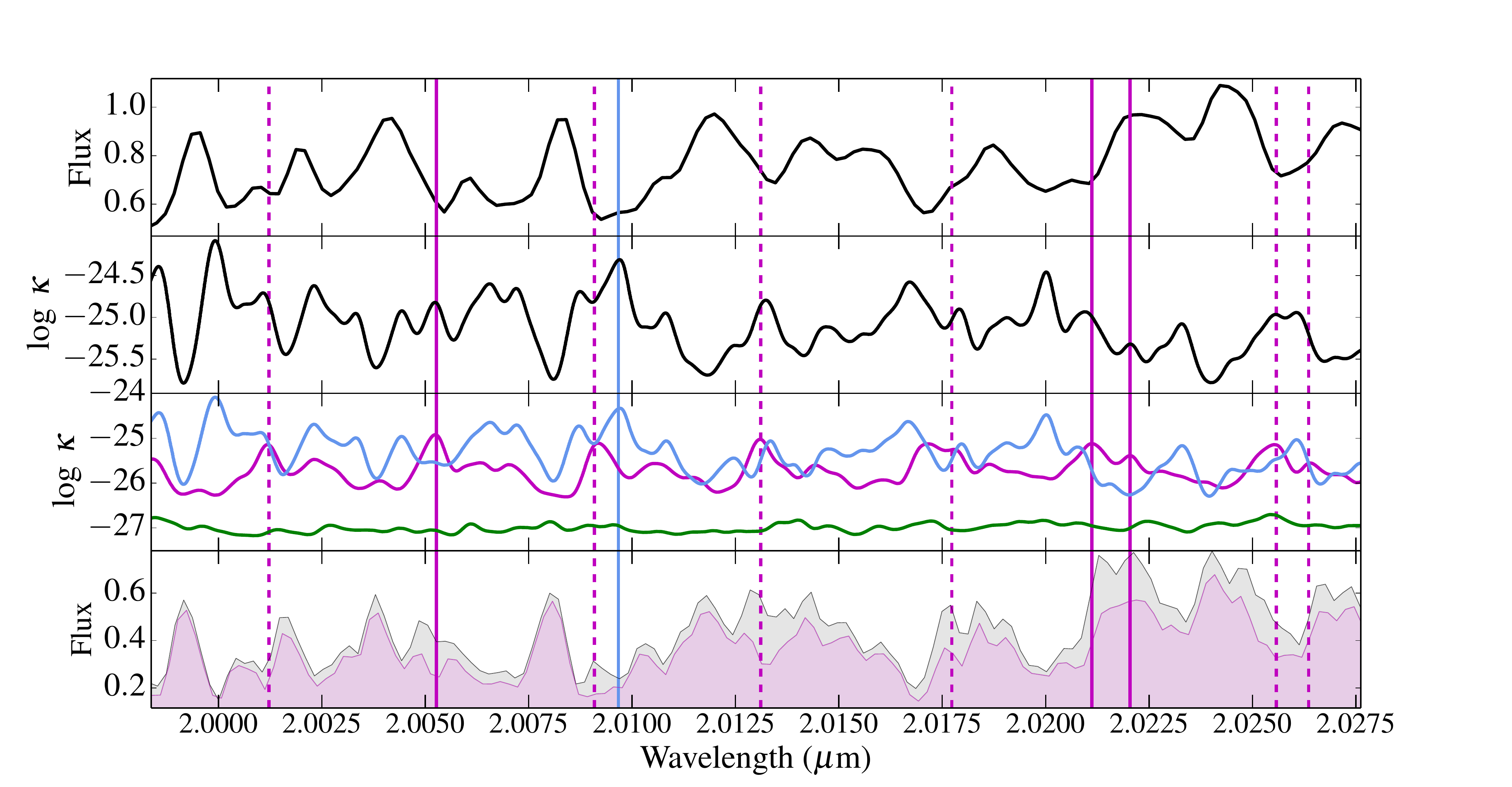} 
   \label{fig:subfig15}
                 }
                 \subfigure{
   \includegraphics[scale=0.34, angle=0]{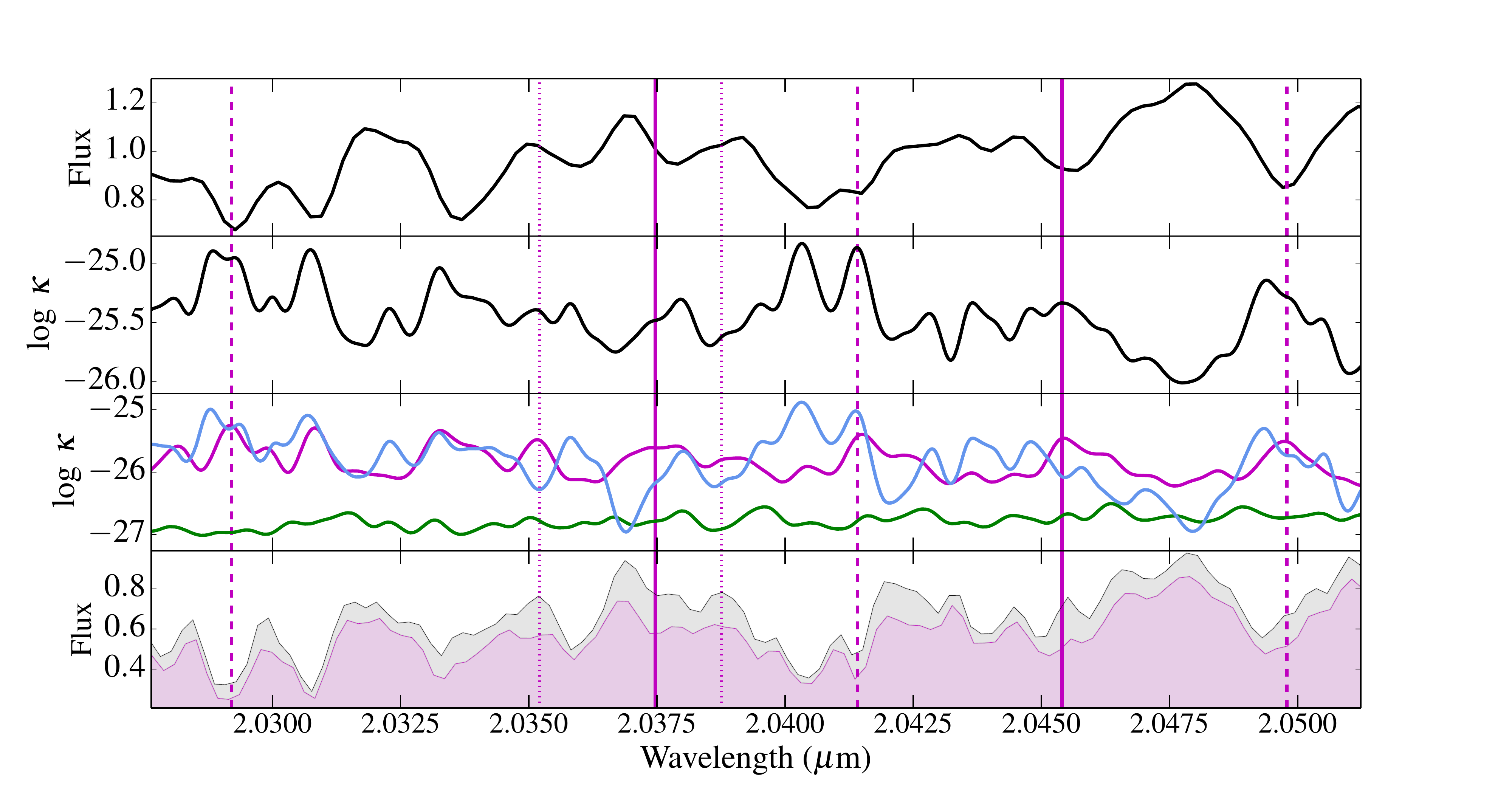} 
   \label{fig:subfig16}
                                 }
  \caption{NH$_{3}$ absorption features in the $K$-band spectrum of 2MASS~0415 (see Table \ref{tab:7}). Scaled absorption 
  cross-sections are calculated at 750~K for  CH$_{4}$ (green), 
  H$_{2}$O (blue) and NH$_{3}$ (magenta).}
  \label{fig:nh3_0415k}
\end{figure*}

In addition, we believe we have identified eight new NH$_{3}$ absorption features that correspond 
both to peaks in the NH$_{3}$ opacity and differences between the model spectra with and without 
NH$_{3}$. Among these, the feature at 1.5240~$\mu$m corresponds to peaks in the NH$_{3}$ 
opacity at 500~K and 750~K. While there is no obvious minimum in the spectrum of UGPS~0722 at this wavelength, 
the shape of the spectrum is most similar to the synthetic spectrum with NH$_{3}$ opacity at 500~K. This is an 
instance where we have identified an absorption feature based on the similarity of the shape of a T dwarf's spectrum 
to the shape of a synthetic spectrum, rather than the correspondence of a peak in molecular opacity with a trough in 
the T dwarf's flux.  The spectrum of 2MASS~0415 at this wavelength looks most similar to the synthetic spectrum without 
NH$_{3}$ opacity at 750~K. The feature at 1.5327~$\mu$m also corresponds to peaks in the NH$_{3}$ 
opacity at 500~K and 750~K and there appear to be absorption features at this wavelength in both T dwarf spectra. 
We note that at this wavelength there is a ``shoulder'' in the synthetic spectrum without NH$_{3}$ opacity at 500~K which 
is missing in the synthetic spectrum with NH$_{3}$ opacity. This feature is present in the synthetic spectra with and 
without NH$_{3}$ opacity calculated at 750~K.
Finally, the S12 models at 500~K predict a feature at 1.5382~$\mu$m. This feature is missing in the 
spectra of both T dwarfs.

We have found no significant differences in the identities of the ro-vibrational lines responsible for 
the absorption features in the two T dwarfs. The two blended NH$_{3}$ features at the shortest wavelengths, 
1.4996~$\mu$m and 1.5020~$\mu$m, are produced by R-branch line transitions. A weak Q-branch 
transition line is found in the feature at 1.4996~$\mu$m. The R-branch transition lines belong to the 
$\nu_{1}$+$\nu_{3}$ vibrational band, while the single Q-branch line arises from the the $\nu_{1}$+2$\nu_{4}$ band.  

NH$_{3}$ features in the $H$-band are produced by ro-vibrational transitions from the P-, Q-, and 
R-branches. Compare this with the CH$_{4}$ absorption features  in the $H$-band which are also 
produced by ro-vibrational transitions from all three branches. However, most of the NH$_{3}$ absorption 
features in the $H$-band are produced by P-branch transition lines, while the CH$_{4}$ absorption 
features in the $H$-band are almost equally distributed across the P-, Q-, and R-branches.
Q-branch transition lines are mostly from the $\nu_{1}$+$\nu_{3}$ vibrational band. 
However, of the four strongest transition lines responsible for the absorption feature at 1.5152~$\mu$m, 
while three lines are Q-branch transition lines, two of these lines include overtones. These two lines 
have approximately half the intensity of the Q-branch line from the $\nu_{1}$+$\nu_{3}$ vibrational 
band.  Another line responsible for the feature at 1.5152~$\mu$m is a P-branch transition line from 
the 2$\nu_{2}$+3$\nu_{4}$ vibrational band. This line has approximately twice the intensity of the 
strongest Q-branch line.
The absorption features at 1.5179~$\mu$m and 1.5201~$\mu$m also contain single P-branch transition 
lines with overtones. The feature at 1.5201~$\mu$m also contains a Q-branch transition line with an 
overtone. As expected, this line is approximately half the strength of the other Q-branch transition lines.
In the absorption features produced by P-branch line transitions (1.5224-1.5905~$\mu$m), there are a 
few Q-branch transition lines in the absorption features at longer wavelengths. Otherwise, absorption 
features are entirely due to P-branch transition lines. However, these lines belong to a larger assortment 
of vibrational bands than is the case for absorption features due to R- or Q-branch transition lines. 
Approximately half the absorption features are produced by transition lines belonging to the 
$\nu_{1}$+$\nu_{3}$ vibrational band, while somewhat less than half the features are from the 
$\nu_{1}$+2$\nu_{4}$ vibrational band. It does appear that absorption features at shorter wavelengths 
belong predominantly to the $\nu_{1}$+$\nu_{3}$ vibrational band, while those at longer wavelengths 
arise from the $\nu_{1}$+2$\nu_{4}$ vibrational band. One of the strongest transition lines producing the NH$_{3}$ absorption feature at the longest 
wavelength, 1.5905~$\mu$m, is the 4$\nu_{4}$ overtone. These results are described in more detail in the online tables A1-A5 and B1-B6.


\subsection{The $K$-Band}
In the $K$-band, S12 identified 19 absorption features corresponding to peaks in NH$_{3}$ opacity. 
Eight of these features matched absorption features in the Magellan/FIRE spectrum of UGPS~0722. 
A further six absorption features were tentatively identified. The remaining five features did not have any counterparts in the T dwarf spectrum. 

We have been able to confirm the eight absorption features first identified by S12. We are also able 
to confirm seven of the uncertain/missing detections, one of which appears to be an NH$_{3}$+H$_{2}$O blend. 
In addition, we have identified six new absorption features, including another NH$_{3}$+H$_{2}$O blend. 
There is a notch in the spectrum of UGPS 0722 at 1.9937 $\mu$m corresponding to 
where NH$_{3}$ opacity removes a large peak in the synthetic spectrum without NH$_{3}$ opacity. While there is no 
peak in NH$_{3}$ opacity at this wavelength, adding NH$_{3}$ removes a trough in the total opacity that would 
otherwise exist. We interpret this as an NH$_{3}$ signature. In 2MASS 0415, the observed spectrum is also a better fit to the 
750 K model spectrum with NH$_{3}$ opacity than without NH$_{3}$ opacity. In this case, ammonia opacity covers a gap in water opacity and 
removes a spike in the 750 K model spectrum.
These results are summarised in Figures \ref{fig:nh3_0722k}, \ref{fig:nh3_0415k}, and Table \ref{tab:7}. 

The absorption feature identified by B11 at 1.9900~$\mu$m as due to a combination of ammonia 
and water opacity does not correspond to a change in ammonia opacity in the S12 models. However, 
Figure \ref{fig:nh3_0722k} suggests that water opacity rather than ammonia opacity is the stronger
component in this feature. Indeed, in the spectrum of 2MASS~0415 the water opacity at this wavelength 
is an order of magnitude stronger.

In the ro-vibrational spectrum for this region of the $K$-band, there is a weak 4$\nu_{2}$ overtone 
among the transition lines contributing to the absorption feature at 2.0132~$\mu$m. Otherwise all 
the NH$_{3}$ absorption features in the $K$-band belong to the $\nu_{1}$+$\nu_{4}$ vibrational band. 
Approximately 71\% of the absorption features are produced by P-branch transition lines. There is a 
single feature at 1.9667~$\mu$m produced by R-branch transitions alone, and there is a very strong 
R-branch transition line among the Q-branch transition lines responsible for the absorption feature at 
1.9737~$\mu$m. The remaining $\sim$25\% absorption features are generated by Q-branch transitions. Note that 
NH$_{3}$ absorption features in the $K$-band spectra of these two T dwarfs are produced by 
transition lines from the P-, Q-, and R-branches, whereas we found only R-branch transition lines in 
the $K$-band absorption features due to CH$_{4}$.

\section{Discussion}
\label{sec:discussion}
In our analysis of CH$_{4}$ absorption features, we have assumed that the spectra of each 
T dwarf can be interpreted using an opacity spectrum computed at a single temperature. 
Absorption features tend to be produced in the higher, cooler parts of the photosphere. 
While it is possible that an absorption feature seen in the spectrum of 2MASS~0415 may be better 
modelled with a 500~K opacity spectrum than a 750~K opacity spectrum, we have not found any examples of this. 
This gives us confidence in the accuracy of the line lists and model spectra we have used 
in our analysis.
We have determined that the strongest absorption features on the long side of the $H$- and $K$-band 
flux peaks of both T dwarfs are due to CH$_{4}$ opacity (apart from mixed absorption features at 
2.0943~$\mu$m and 2.1017~$\mu$m in the spectra of both T dwarfs, and mixed features at 2.0971~$\mu$m, 
and  2.1129~$\mu$m in the spectrum of 2MASS~0415). There are significant differences between 
opacity sources in the spectra of the two T dwarfs on the long side of the $J$-band flux peak, where 
we identified a single methane feature in the spectrum of 2MASS~0415, compared to six features 
in the spectrum of UGPS~0722. There is a large number of methane blends in the $J$-band spectra 
of these objects compared to the pure methane features we have observed in 
these objects' $H$- and $K$-band spectra.

The disagreement between the 10to10 line list and the science data between 
1.6145~$\mu$m-1.6258~$\mu$m (see Figures \ref{fig:0722h} and \ref{fig:0415h}) cannot be due to faulty calibration of either the science spectra or the 10to10 line list, 
since elsewhere the correspondence between methane opacity and absorption feature is excellent. From our discussion 
in Section \ref{sec:ch4h}, it appears that the disparity is most probably due to a deficiency in the 10to10 line list in this wavelength region.
The differences in wavelength are slight. This, together with the better agreement obtained with 
the hybrid list including experimental data lead us to believe that the absorption features in the T dwarfs' spectra are 
due to methane.

The discrepancy between the 10to10 line list and the $H$-band data prompted us to ask whether some vibrational bands are more accurately represented in the 10to10 line list than others. To examine this, we looked at the vibrational band centres around 1.6 $\mu$m.
The errors in the 10to10 line list were assessed by comparing the experimental data from
HITRAN 2012 \citep{rothman13}. The errors are systematic within each
vibrational band, reflecting the quality (or flaws) of the underlying 
potential energy surfaces (PES), which in the case of 10to10 was obtained 
by refining an ab initio PES to available experimental energies.
We found that the band centres with the worst accuracy are for $\nu_{1}+2\nu_{2}$ and $\nu_{3}+2\nu_{4}$ where the accuracy is typically $\sim8\times$10$^{-3}$ $\mu$m. These are probably the weakest band centres, especially the former where the intensity of the associated line transitions is of the order of two magnitudes weaker than those due to the 2$\nu_{3}$ band. We have found that the band centre for 2$\nu_{3}$ is the most accurate, with accuracies varying between $\sim2\times$10$^{-6}$ $\mu$m and $\sim8\times$10$^{-5}$ $\mu$m. All the methane features we have identified in the $H$-band are from this vibrational band. We looked at other line transitions involving one quantum of $\nu_{3}$. These transitions are weaker than those from the 2$\nu_{3}$ band and are comparable in strength to line transitions arising from vibrational bands other than 2$\nu_{3}$ in this wavelength region. 
Our results are shown in Table \ref{tab:10to10}.  

\begin{table*}
\begin{minipage}{140mm}
\caption{Accuracy of the 10to10 line list in the $H$-band}  
\label{tab:10to10}
\begin{tabular}{ccccccccc} 
\hline\hline
\noalign{\vskip 2mm} 
Vibrational band           &            $\Delta\Gamma$                              & Accuracy, $\lambda$($\mu$m)\\ [0.5ex] 
\hline
\noalign{\vskip 2mm} 
2$\nu_{3}$                   &           F$_{1}$~$\rightarrow$~F$_{2}$        &  $\sim2\times$10$^{-6}$\\ 
2$\nu_{3}$                    &          E~$\rightarrow$~E                             &  $\sim8\times$10$^{-5}$\\  

\noalign{\vskip 2mm} 
2$\nu_{2}+\nu_{3}$       &          F$_{2}$~$\rightarrow$~F$_{1}$         & $\sim1\times$10$^{-4}$\\ 

\noalign{\vskip 2mm} 
$\nu_{3}+2\nu_{4}$       &         F$_{1}$~$\rightarrow$~F$_{2}$         &  $\sim6\times$10$^{-3}$\\ 
$\nu_{3}+2\nu_{4}$      &          F$_{2}$~$\rightarrow$~F$_{1}$         &  $\sim8\times$10$^{-3}$\\ 

\noalign{\vskip 2mm} 
$\nu_{1}+\nu_{3}$         &         F$_{1}$~$\rightarrow$~F$_{2}$         & $\sim6\times$10$^{-3}$\\ 

\noalign{\vskip 2mm} 
$\nu_{1}+2\nu_{2}$       &         A$_{2}$~$\rightarrow$~A$_{1}$        & $\sim8\times$10$^{-3}$\\ 
\noalign{\vskip 2mm}                      
\hline
\end{tabular}
\end{minipage}
\end{table*}

While several ammonia absorption features predicted by the S12 model spectra between 
$\sim$1.223-1.235~$\mu$m remain undetected, we have found a number of others. Although the 
S12 model spectra with and without NH$_{3}$ opacity show no significant differences at the wavelengths 
corresponding to several possible NH$_{3}$ absorption features between $\sim$1.239-1.266~$\mu$m, we find 
that NH$_{3}$ opacity is the strongest opacity source in this region (see Figures \ref{fig:nh3_0722j} and \ref{fig:nh3_0415j}). The failure of the S12 model spectra 
to predict these features may be because without a comprehensive, high-temperature methane line list, 
the models overestimate the importance of methane opacity in this region.

It has recently been suggested that the relatively red $J-€€€H$ and $J-K$ colours seen in T dwarfs of 
spectral types $\gtrsim$T8 in comparison with model predictions could be explained by the formation 
of sulphide clouds [M12]. The model spectra used in the analyses here and in S12 are cloudless. While 
the BYTe line list and the improved calculations of H$_{2}$ CIA have reddened the $J-K$ colours of 
model spectra for T dwarfs with T$_{eff}\gtrsim$600~K, the reddening is insufficient to match the observed
colours of the coolest T dwarfs [M12]. It is also possible that the inclusion of sulphide cloud opacity in a 
new set of synthetic spectra may be a better match to the data.

In UGPS~0722, NH$_{3}$ ro-vibrational transition lines in the $H$-band are an order of magnitude 
stronger than those in the $J$-band and NH$_{3}$ is the dominant source of opacity across the blue wing 
of the $H$-band (1.50-1.59~$\mu$m) (see Figure \ref{fig:nh3_0722h}).  This region of the $H$-band shows clear differences in the 
species responsible for the absorption features in the two T dwarfs' spectra. At 750~K, the scaled absorption 
cross-sections are of the same order of magnitude for all three of the main molecular opacity sources 
(H$_{2}$O, CH$_{4}$, NH$_{3}$) so that  features due to NH$_{3}$ opacity alone in the spectrum of 
UGPS~0722, are produced by a combination of opacities in the spectrum of 2MASS~0415 (see Figure \ref{fig:nh3_0415h}).

C11 detected absorption in the $H$-band spectrum of the Y dwarf WISEP J1738+2732 
which corresponded to the $\nu_{1}+\nu_{3}$ absorption band of NH$_{3}$ but were unable to confirm this 
owing to the low resolution of their data. Our results confirm C11's findings. Indeed, the intensity of the 
ro-vibrational transition lines responsible for these features at 300~K is up to twice the intensity at 500~K.

\section{Conclusions}
\label{sec:conclusions}
The BYTe and 10to10 line lists appear to be validated in 
that we have detected previously known features at the correct wavelengths. In addition, we have found 
new absorption features and corrected features which had previously been mis-identified.  The reasons 
for this are the high quality spectra used in this analysis, and the 10to10 line list, which is more complete 
at these temperatures than any previously available list. For example, the CH$_{4}$ laboratory line list 
used by B11 was made at 800~K and covered the spectral range 2000-5000~cm$^{-1}$ (wavelengths 
$\geq$2.00~$\mu$m) at a resolution of 0.02~cm$^{-1}$, and the spectral range 5000-6400~cm$^{-1}$ 
(wavelengths between 1.56-2.00~$\mu$m) at a resolution of 0.04~cm$^{-1}$~\citep{nassar03}. B11 used 
the HITRAN 2008 database~\citep{rothman09}, calculated at 296~K, to supplement the experimental line 
list. These facts may explain B11's mis-identifications, particularly in the $J$- and $H$-bands. The use of 
adaptive optics has enabled us to obtain data in the $H$- and $K$-bands with improved S/N ratios compared 
with the same passbands in B11. This has allowed us to add significantly to the number of detections of 
methane and ammonia absorption features in this region of the near-infrared.  Our near-infrared spectrum 
of UGPS~0722 and that of B11 appear very similar and we have not found any sign of variability of features 
between the two spectra.  As both data sets are independent, we are confident that these features are real. 

The  BYTe and 10to10 line lists indicates that NH$_{3}$ is the dominant opacity source between 
$\sim$1.233-1.266~$\mu$m in UGPS~0722, and we have tentatively identified several absorption features in this 
wavelength range in the T9's spectrum which may be due entirely to ammonia opacity.
Our analysis using the 10to10 line list suggests that water rather than methane is the 
dominant absorber in the red half of the $J$-band in 2MASS 0415, where water opacity is between 40-80\% 
stronger than methane opacity. In UGPS 0722, water is the major opacity source until $\sim$1.31~$\mu$m 
when methane opacity starts to dominate. This result can be examined when the 10to10 line list is included 
in a full model atmosphere.

The 10to10 line list has allowed us to accurately identify the opacity sources responsible 
for many of the T dwarf absorption features. We have also found that absorption features common to 
both T dwarf standards may have different opacity sources, or the relative strengths of the opacity sources 
producing these features may vary between them. This is particularly noticeable in the $J$-band, where the 
number of absorption features due solely to CH$_{4}$ opacity is fewer than previously thought. 

Using the high quality spectra of these T dwarfs, we have been able to confirm the presence of 15 of the 
19 NH$_{3}$ absorption features in the $K$-band predicted by the S12 models. In addition, we have identified 
six previously unknown absorption features. The ro-vibrational transition lines responsible for these features 
have up to twice the intensity of those in the $H$-band. The lines also appear to be more highly ordered than 
those in the $J$- or $H$-bands, making the identification of corresponding absorption features in the T dwarfs' 
spectra easier.
We think that the conspicuous peaks of NH$_{3}$ opacity in the $K$-band are due 
to the strong transition dipole of the stretching mode of the $\nu_{1}+\nu_{4}$ band.
The stretching mode ($\nu_{1}$) makes the strongest contribution to the ro-vibrational transition lines in this wavelength 
region. There are also no issues with forbidden bands, since the ammonia molecule has a permanent dipole. The 
$\nu_{1}+\nu_{4}$ band has a transition dipole moment of 0.017 Debye. This compares with a 
moment of 0.0005 Debye for the 3$\nu_{4}$ band, the band nearest in strength to the $\nu_{1}+\nu_{4}$ 
band at these wavelengths.

In this work we have looked at the most common isotopologues of methane and ammonia. 
In particular, the relative abundance of methane isotopologues is determined by the reaction rates of these 
isotopologues with methane sinks~\citep{rigby12}, and is therefore useful  in understanding the structure and 
evolution of atmospheres. 
The use of high-temperature line lists of other methane isotopologues in model spectra could be applied to 
the near-infrared spectra of T dwarfs and would allow a greater understanding of the evolution of their atmospheres.

\section*{Acknowledgements}
This paper is based on observations obtained in programmes GN-2010B-Q-18 and GN-2012B-Q-62 at the Gemini Observatory, which is operated by the Association of Universities for Research in Astronomy Inc., under a cooperative agreement with the NSF on behalf of the Gemini partnership: The National Science Foundation (USA), the Science and Technology Facilities Council (UK), the National Research Council (Canada), CONICYT (Chile), the Australian Research Council (Australia), CNPq (Brazil) and CONICET (Argentina).

J. I. Canty is supported by a University of Hertfordshire PhD studentship.
Sergei Yurchenko and Jonathan Tennyson acknowledge support by ERC Advanced Investigator Project 
267219 and the UK Science and Technology Research Council (STRC).

The authors are grateful to Didier Saumon for his numerous helpful comments.

\bibliographystyle{mn2e}

\label{lastpage}

\end{document}